# Measurements of differential and angle-integrated cross sections for the $^{10}$B(*n*, *α*)$^7$Li reaction in the neutron energy range from 1.0 eV to 2.5 MeV

Haoyu Jiang[1,a], Wei Jiang[2,3,a], Huaiyong Bai[1], Zengqi Cui[1], Guohui Zhang[1,*], Ruirui Fan[2,3,4], Han Yi[2,3], Changjun Ning[2,3], Liang Zhou[2,3], Jingyu Tang[2,3], Qi An[4,5], Jie Bao[6], Yu Bao[2,3], Ping Cao[4,5], Haolei Chen[4,5], Qiping Chen[7], Yonghao Chen[2,3], Yukai Chen[2,3], Zhen Chen[4,5], Changqing Feng[4,5], Keqing Gao[2,3], Minhao Gu[2,4], Changcai Han[8], Zijie Han[7], Guozhu He[6], Yongcheng He[2,3], Yang Hong[2,3,9], Hanxiong Huang[6], Weiling Huang[2,3], Xiru Huang[4,5], Xiaolu Ji[2,4], Xuyang Ji[4,10], Zhijie Jiang[4,5], Hantao Jing[2,3], Ling Kang[2,3], Mingtao Kang[2,3], Bo Li[2,3], Chao Li[4,5], Jiawen Li[4,10], Lun Li[2,3], Qiang Li[2,3], Xiao Li[2,3], Yang Li[2,3], Rong Liu[7], Shubin Liu[4,5], Xingyan Liu[7], Guangyuan Luan[6], Qili Mu[2,3], Binbin Qi[4,5], Jie Ren[6], Zhizhou Ren[7], Xichao Ruan[6], Zhaohui Song[8], Yingpeng Song[2,3], Hong Sun[2,3], Kang Sun[2,3,9], Xiaoyang Sun[2,3,9], Zhijia Sun[2,3,4], Zhixin Tan[2,3], Hongqing Tang[6], Xinyi Tang[4,5], Binbin Tian[2,3], Lijiao Wang[2,3,9], Pengcheng Wang[2,3], Qi Wang[6], Taofeng Wang[11], Zhaohui Wang[6], Jie Wen[7], Zhongwei Wen[7], Qingbiao Wu[2,3], Xiaoguang Wu[6], Xuan Wu[2,3], Likun Xie[4,10], Yiwei Yang[7], Li Yu[2,3], Tao Yu[4,5], Yongji Yu[2,3], Linhao Zhang[2,3,9], Qiwei Zhang[6], Xianpeng Zhang[8], Yuliang Zhang[2,3], Zhiyong Zhang[4,5], Yubin Zhao[2,3], Luping Zhou[2,3,9], Zuying Zhou[6], Danyang Zhu[4,5], Kejun Zhu[2,4,9], Peng Zhu[2,3]

[1] *State Key Laboratory of Nuclear Physics and Technology, School of Physics, Peking University, Beijing 100871, China*
[2] *Institute of High Energy Physics, Chinese Academy of Sciences (CAS), Beijing 100049, China*
[3] *Spallation Neutron Source Science Center, Dongguan 523803, China*
[4] *State Key Laboratory of Particle Detection and Electronics*
[5] *Department of Modern Physics, University of Science and Technology of China, Hefei 230026, China*
[6] *Key Laboratory of Nuclear Data, China Institute of Atomic Energy, Beijing 102413, China*
[7] *Institute of Nuclear Physics and Chemistry, China Academy of Engineering Physics, Mianyang 621900, China*
[8] *Northwest Institute of Nuclear Technology, Xi'an 710024, China*
[9] *University of Chinese Academy of Sciences, Beijing 100049, China*
[10] *Department of Engineering and Applied Physics, University of Science and Technology of China, Hefei 230026, China*
[11] *School of Physics, Beihang University, Beijing 100083, China*

**Abstract:** Differential and angle-integrated cross sections for the $^{10}$B(*n*, *α*)$^7$Li, $^{10}$B(*n*, *α*$_0$) $^7$Li and $^{10}$B(*n*, *α*$_1$) $^7$Li$^*$ reactions have been measured at CSNS Back-n white neutron source. Two enriched (90%) $^{10}$B samples 5.0 cm in diameter and ~85.0 μg/cm$^2$ in thickness each with an aluminum backing were prepared, and back-to-back mounted at the sample holder. The charged particles were detected using the silicon-detector array of the Light-charged Particle Detector Array (LPDA) system. The neutron energy $E_n$ was determined by TOF (time-of-flight) method, and the valid α events were extracted from the $E_n$-Amplitude two-dimensional spectrum. With 15 silicon detectors, the differential cross sections of α-particles were measured from 19.2° to 160.8°. Fitted with the Legendre polynomial series, the (*n*, *α*) cross sections were obtained through integration. The absolute cross sections were normalized using the standard cross sections of the $^{10}$B(*n*, *α*)$^7$Li reaction in the 0.3 − 0.5 MeV neutron energy region. The measurement neutron energy range for the $^{10}$B(*n*, *α*)$^7$Li reaction is 1.0 eV ≤ $E_n$ < 2.5 MeV (67 energy points), and for the $^{10}$B(*n*, *α*$_0$) $^7$Li and $^{10}$B(*n*, *α*$_1$) $^7$Li$^*$ reactions is 1.0 eV ≤ $E_n$ < 1.0 MeV (59 energy points). The present results have been analyzed by the resonance reaction mechanism and the level structure of the $^{11}$B compound system, and compared with existing measurements and evaluations.

**Keywords:** $^{10}$B(*n*, *α*)$^7$Li reaction; cross sections; LPDA; CSNS Back-n white neutron source

# I. INTRODUCTION

$^{10}$B, a stable isotope of boron with natural abundance of 19.9 %, is a crucial material in nuclear engineering, including radiation protection, neutron detection, reactor control, boron neutron capture therapy (BNCT), etc [1]. For neutron induced nuclear reactions of $^{10}$B, the $^{10}$B$(n, \alpha)^7$Li reaction is the dominate reaction channel for $E_n < 1.0$ MeV. In addition to various applications, the study of this reaction can enhance the understanding of nuclear reaction mechanism for light nuclei [2]. The $^{10}$B$(n, \alpha)^7$Li reaction has two main reaction channels, which are the $^{10}$B$(n, \alpha_0)^7$Li ($Q = 2.79$ MeV) and $^{10}$B$(n, \alpha_1)^7$Li$^*$ ($Q = 2.31$ MeV) reactions. Many measurements of the $^{10}$B$(n, \alpha)^7$Li reaction have been conducted since 1954 [3]. The cross sections of the $^{10}$B$(n, \alpha)^7$Li and $^{10}$B$(n, \alpha_1)^7$Li$^*$ reactions have been recommended as neutron cross section standard from thermal energy to 1.0 MeV region [4].

In the MeV region, however, due to the small cross sections and strong interference of background, discrepancies among different measurements and evaluations are apparent [3, 5]. Furthermore, existing measurements of the differential cross sections of the $^{10}$B$(n, \alpha)^7$Li reaction, as well those of as the $^{10}$B$(n, \alpha_0)^7$Li and $^{10}$B$(n, \alpha_1)^7$Li$^*$ reactions, are scarce. Only three measurements of angular distributions and differential cross sections (Sealock [6], Stelts [7] and Hambsch [2]) can be found in EXFOR for $E_n \leq 1.2$ MeV and there is no data in the 1.2 MeV $< E_n <$ 2.5 MeV region. Taking these factors into consideration, accurate measurements of differential and angle-integrated cross sections for the $^{10}$B$(n, \alpha)^7$Li reaction are demanded.

In the present work, a LPDA (Light-charged Particle Detector Array) system, which mainly consisted with a silicon detector array in a vacuum chamber, was built to study the neutron induced charged particle emission reaction at CSNS (China Spallation Neutron Source) Back-n white neutron source [8]. With 15 silicon detectors distributed from 19.2° to 160.8°, the differential and angle-integrated cross sections were obtained for the $^{10}$B$(n, \alpha)^7$Li reaction in the 1.0 eV $\leq E_n <$ 2.5 MeV region (67 energy points), as well as the two reaction channels, $^{10}$B$(n, \alpha_0)^7$Li and $^{10}$B$(n, \alpha_1)^7$Li$^*$, in the 1.0 eV $\leq E_n <$ 1.0 MeV region (59 energy points). The present results have been analyzed with the resonance reaction mechanism and the level structure of the $^{11}$B compound system, and compared with existing measurements and evaluations.

# II. EXPERIMENTAL DETAILS

## A. Neutron source

The neutrons were produced by double bunched proton beam (1.6 GeV, ~ 20 kW) bombarding a tungsten target at CSNS Back-n white neutron source [9]. The repetition rate of the beam pulse was 25 Hz and the pulse width was ~ 41 ns. The interval between the two proton bunches was 410 ns [10]. The experiment was conducted at Endstation #1, where the length of the flight path was 57.99 m, and the neutron flux was ~ $3.5 \times 10^6$ n/(cm$^2 \cdot$s). The beam spot size at Endstation #1 was determined by the apertures of the shutter and Collimator-1 [11]. The diameter of the collimation aperture for the shutter was 50 mm and that for Collimator-1 was also 50 mm in the present work. The full width at half maximum (FWHM) of the neutron beam spot was 54 – 58 mm [12]. The relative neutron intensity could be monitored by the number of protons in the beam and a Li-Si detector array mounted in the beamline. Using the single-bunch operation mode, the neutron energy spectrum was measured by a multi-layer $^{235}$U fission chamber at Endstation #2, where the length of flight was 75.76 m [13, 14]. The details of the neutron energy spectrum could be found in Refs. [8] and [14].

The neutron energy spectrum and the neutron energy bins in the present work are shown in Fig. 1. The error bar in Fig. 1 represents the uncertainty of relative neutron fluence ($\varphi_{E\_bin}$), which is (0.5 - 21.4) % (for 49 of the 67 energy points, this uncertainty is less than 5 %). Compared with the origin neutron energy spectrum in Ref. [14], the new wider energy bins of the spectrum were defined in the present work. Sixty-seven the energy points called $E\_bin$ were specified from 1.0 eV to 2.5 MeV, and each $E\_bin$ was correlated with a neutron energy bin. The energy points ($E\_bin$) were specified as the follows: 49 equally spaced points were defined in the logarithmic coordinate between 1.0 eV and 0.1 MeV, 8 points were defined with equal interval of 0.1 MeV in the linear coordinate between 0.1 MeV and 1.0 MeV. Above 1.0 MeV, the interval was 0.2 MeV up to 2.5 MeV. Next, the neutron energy corresponding to each event was obtained from TOF, then the linearly nearest $E\_bin$ was searched, and the event was counted into the corresponding neutron energy bin.

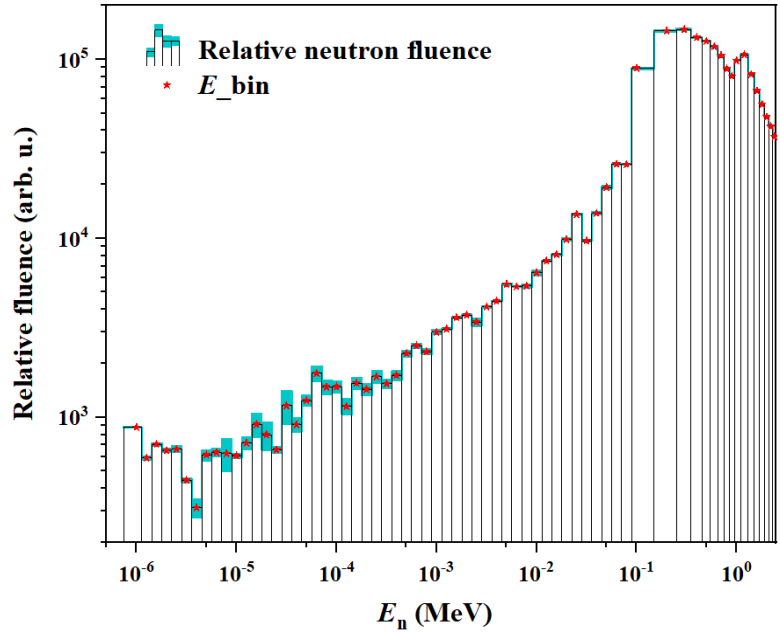

**FIG. 1.** (Color online) The neutron energy spectrum with uncertainty presented by blue bars (below 2.5 MeV).

### B. Samples

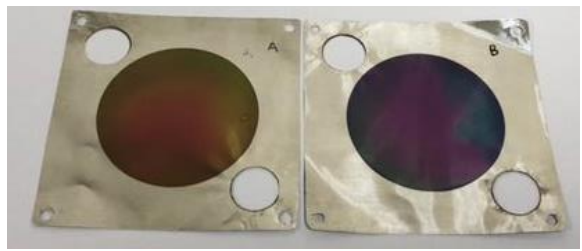

**FIG. 2.** (Color online) The $^{10}$B samples.

Two enriched (90%) $^{10}$B samples were prepared as shown in Fig. 2. Each $^{10}$B sample was evaporated on an aluminum sheet 50 $\mu$m in thickness. The two samples were 5.0 cm in diameter both, and 82.59 and 85.05 $\mu$g/cm$^2$ in thickness, respectively. The two $^{10}$B samples were back-to-back mounted at one of the four sample positions of the sample holder as shown in Fig. 3. At other sample positions, two back-to-back $^{241}$Am $\alpha$ sources and two aluminum sheets 50 $\mu$m each in thickness were mounted. The $^{241}$Am $\alpha$ sources were used to calibrate the detectors and the DAQ (Data Acquisition) system, and the aluminum sheets were used for the background measurement. The angle between the normal of the samples and the neutron beam line was 60° as shown in Fig.4 (a) so that the energy loss of $\alpha$- particles from the samples could be minimized.

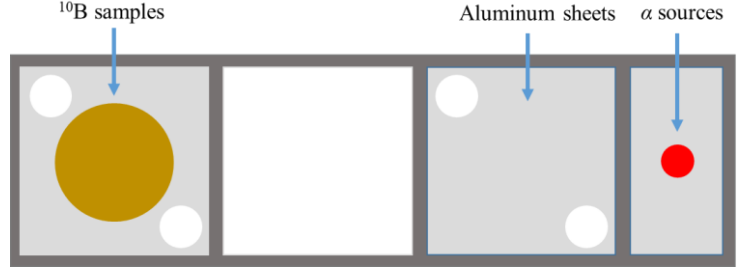

**FIG. 3.** (Color online) The diagram of the sample holder.

## C. Detectors

The charged particles were detected by the LPDA system, which mainly consisted of a silicon detector array and a vacuum chamber as shown in Fig. 4. Apart from the silicon detectors, other detectors such as a gridded ionization chamber (GIC) and three $\Delta E$-$E$ detectors were installed and tested. Fifteen rectangular (2.0 cm × 2.5 cm) silicon detectors 500 $\mu$m in thickness could cover the emission angle of the particles from 19.2° to 160.8°, and their solid angles were (0.0123 - 0.0125) ($\pm$0.3%) sr according to Monte Carlo simulation. The distance between the center of the silicon detector and that of the $^{10}$B sample was 20.0 cm. The angle between the normal of the silicon detectors and the horizon was 16° in order to avoid shielding $\Delta E$-$E$ detectors.

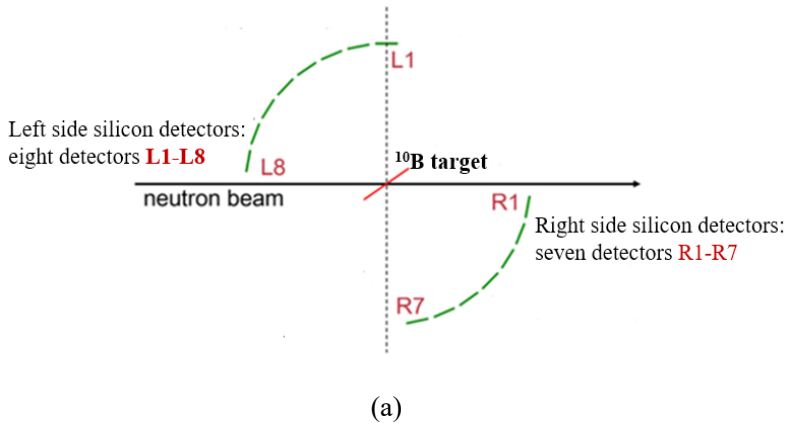
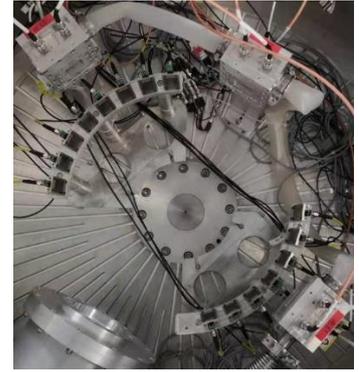

(a)          (b)

**FIG. 4.** (Color online) (a) The sketch of the silicon detectors. (b) The photo of the detectors.

## D. DAQ system

The DAQ system was designed based on PXIe platform [15]. The sampling rate of the DAQ system was 1 GHz with the resolution of 12 bits. When the signal amplitude exceeded the predefined threshold of the corresponding channel, the full signal waveform with a time window 15 $\mu$s would be recorded. In order to obtain the starting time of the signal, the original signal was filtered and differentiated, and the starting time was determined by the position of the one-tenth maximum height of differentiated signal. The TOF of the neutron could be calculated by

$$\mathrm{TOF} = T_{\mathrm{event}} - T_0 + \frac{L}{c} \quad , \tag{1}$$

where $T_{\mathrm{event}}$ is the signal starting time of the corresponding event, $T_0$ is the generation moment of the related neutrons determined using the starting time of $\gamma$-flash events, $L$ is the length of flight path, $c$ is the velocity of light. The neutron energy distribution due to the double-bunched operation mode would be unfolded as described in section III.C.

**E. Experimental process**

In the experiment, the 15 silicon detectors and the DAQ system were firstly calibrated using the $^{241}$Am $\alpha$ sources. Then, the $^{10}$B samples and the Aluminum backing sheets were measured in turns (~ 16 h for measurement foreground and ~ 8 h measurement for background for each turn). The total beam duration was ~ 357 h.

**III. DATA ANALYSIS AND RESULTS**

With the recorded signal waveforms and the corresponding TOFs, the $E_n$-Amplitude two-dimensional spectrum could be obtained, and the valid area of $\alpha$ events could be determined. Next, the events were counted into the corresponding neutron energy bins as described in section II. A, and then the background was subtracted to obtain the net events. After that, the neutron energy distribution caused by the neutron energy bin width and by the double proton bunches, and the spread of the detection angle were unfolded using the iterative method. Next, the relative differential cross sections of the $^{10}$B($n$, $\alpha$)$^7$Li reaction were obtained, then the relative angle-integrated cross sections were calculated via integration. The results were normalized using the standard cross sections of the $^{10}$B($n$, $\alpha$)$^7$Li reaction in the 0.3 − 0.5 MeV region. After that, the ratios of the $^{10}$B($n$, $\alpha_0$)$^7$Li and $^{10}$B($n$, $\alpha_1$)$^7$Li$^*$ reactions was calculated by the unfolded spectrum of net $\alpha$ events, and the differential and angle-integrated cross sections for these two reaction channels were obtained. The process of data analysis is shown in Fig. 5.

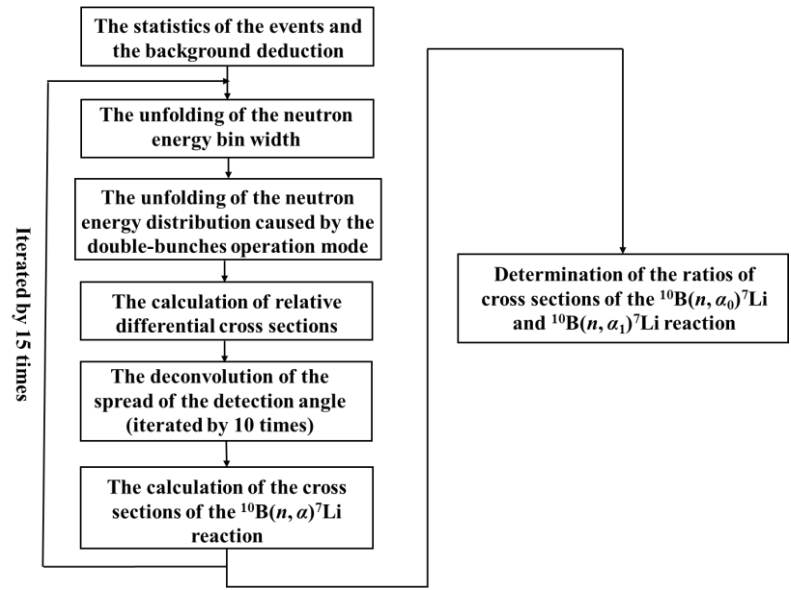

**FIG. 5.** The flow chart of data analysis.

**A. The statistics of the events and the background subtraction**

The measurement data have been sorted into two-dimensional distributions ($E_n$-Amplitude) at every detection angle. The amplitude of each event could be obtained from the recorded waveform, and the corresponding $E_n$ could be calculated from its TOF. An $E_n$-Amplitude two-dimensional spectrum is shown in Fig. 6 as an example, in which the areas of $\alpha_0$ and $\alpha_1$ events, as well as the Li and Li$^*$ events, and recoil proton events are labeled. From the two-dimensional spectrum, the valid-event-area of $\alpha$ events could be decided. Then, the events in the valid-event-area were projected into their corresponding neutron energy bins which was described in section II.A.

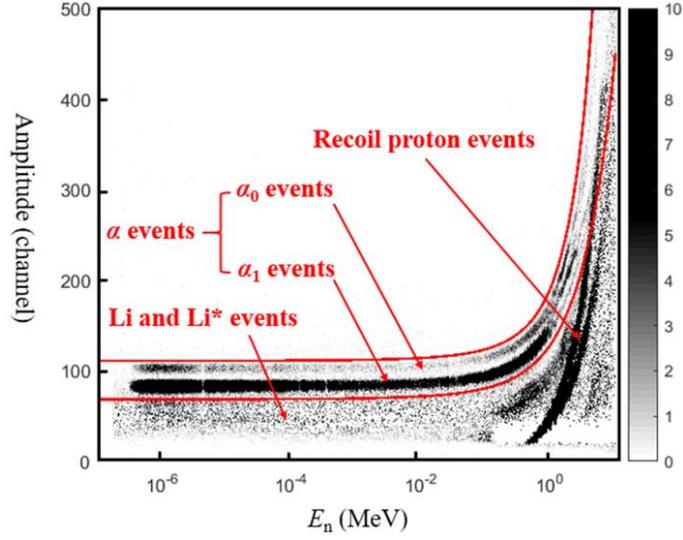

**FIG. 6.** (Color online) The $E_n$-Amplitude two-dimensional spectrum at the detection angle near 26.9°.

The net events in different energy bins at each detection angle could be obtained after the background subtraction shown in Fig. 7 as an example. The normalization factor was decided by the ratio of the number of the protons in the beam during the foreground measurement over that during the background measurement. Although the background from the sample itself, such as the charge particles from the $^{11}$B$(n, p)$ ($Q = 0.23$ MeV) and $^{10}$B$(n, t2\alpha)$ ($Q = 0.32$ MeV) reactions, could not be subtracted, these interferences can be ignored in $E_n < 1$ MeV region because of their fairly small cross sections [2]. In the 1.0 MeV $\leq E_n <$ 2.5 MeV region, the valid-event-area could be separated from background area because the charged particles from the background reactions with small $Q$-values have quite low energies. For $E_n \geq 2.5$ MeV, the energies of the emitted background particles are high enough to interfere with the valid-event-area. Besides, the recoil protons from hydrogen adsorbed in the samples would be another notable source of the background. Therefore only the cross sections of the $^{10}$B$(n, \alpha)^7$Li reaction below 2.5 MeV region were obtained in the present measurement.

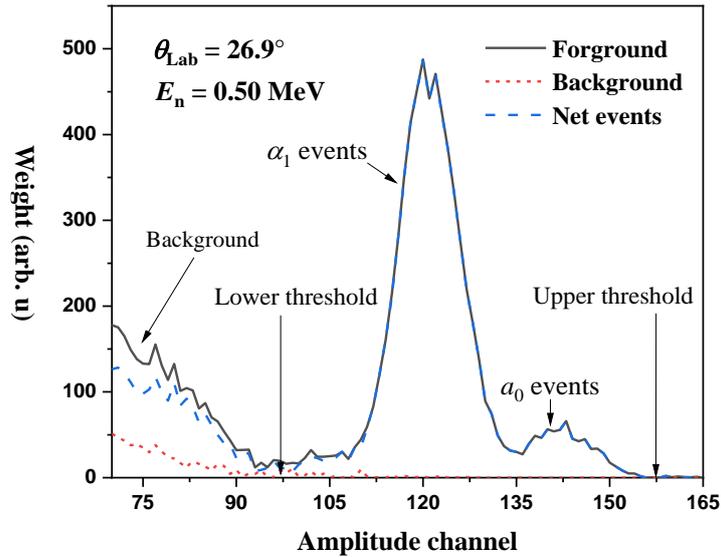

**FIG. 7.** (Color online) The measured $\alpha$ events at the detection angle 26.9° and at the neutron energy 0.50 MeV.

**B. The unfolding of the neutron energy bin width**

The unfolding is necessary due to the influence of the width of neutron energy bin. Each event was weighted as

$$w_{E,\theta} = \frac{\sigma_{E\_\text{bin},\theta}}{\sigma_{E,\theta}} , \qquad (2)$$

where $w$ is the weight of a single event, $E$ is the energy of the neutron calculated from TOF, $\theta$ is the detection angle, $E\_\text{bin}$ is the linearly nearest neighbor energy point described in section II.A, and $\sigma_{E\_\text{bin},\theta}$ and $\sigma_{E,\theta}$ are the $^{10}$B$(n, \alpha)^7$Li differential reaction cross sections at the neutron energy of $E\_\text{bin}$ and $E$, respectively. The calculation of $\sigma_{E\_\text{bin},\theta}$ and $\sigma_{E,\theta}$ were described in section III.F. In the first step of iteration, the weight of every event was set as 1, and the $^{10}$B$(n, \alpha)^7$Li reaction cross sections could be calculated. And in all the following steps, the weight of events at every neutron energy bin was calculated by the cross sections obtained from the last step, then the cross sections were recalculated. Through the unfolding, the uncertainty of the neutron energy could be decreased from (4.2 − 50.0) % to (0.4 − 17.2) %. However, the uncertainty of the differential cross section would be introduced, and the magnitude is (0.1 − 51.1) % (for 944 of the 1005 $\sigma_{E\_\text{bin},\theta}$ results, this uncertainty is less than 5 %).

**C. The unfolding of the neutron energy distribution caused by the double-bunched operation mode**

The interval (410 ns) of the double bunched proton beams would lead to fairly big uncertainty of the neutron energy especially in high energy region. For $E_n \geq 0.02$ MeV, the uncertainty of the neutron energy is (1.5 − 17.2) % which cannot be not negligible, and the unfolding is thus needed. The details of the unfolding method could be found in Ref. [16], which will be described briefly here.

At each detection angle, every event was split into two child events and each of them was weighted as

$$\begin{cases} w_{E_{n1},\theta} = \dfrac{I_{E_{n1}} \sigma^{\text{last\_itera}}_{E_{n1},\theta}}{I_{E_{n1}} \sigma^{\text{last\_itera}}_{E_{n1},\theta} + I_{E_{n2}} \sigma^{\text{last\_itera}}_{E_{n2},\theta}} \\[2mm] w_{E_{n2},\theta} = \dfrac{I_{E_{n2}} \sigma^{\text{last\_itera}}_{E_{n2},\theta}}{I_{E_{n1}} \sigma^{\text{last\_itera}}_{E_{n1},\theta} + I_{E_{n2}} \sigma^{\text{last\_itera}}_{E_{n2},\theta}} \end{cases} , \qquad (3)$$

where the subscripts of $E_{n1}$ and $E_{n2}$ are the neutron energies determined by the TOF of the event plus 205 ns and minus 205 ns, respectively. $I_{E_{n1}}$ and $I_{E_{n1}}$ are the unit neutron fluence (i.e. n/ns) (the neutron spectrum is transformed into the number of neutrons per unit time of TOF), $\sigma^{\text{last\_itera}}_{E_{n1},\theta}$ and $\sigma^{\text{last\_itera}}_{E_{n2},\theta}$ are the differential cross sections obtained from the last iteration. Using Eq. (2), the two child events would be counted into related bins, and the new differential cross sections described in in section III.F could be obtained. The unfolding could decrease the uncertainty of the neutron energy to (0.4 − 1.5) %. However, the unfolding would lead to the uncertainty of the differential cross sections. In the present work, the correction was processed for $E_n \geq 0.02$ MeV and the corresponding uncertainty is (1.2 − 11.9) % (for 305 of the 360 $\sigma_{E\_\text{bin},\theta}$ results above 0.02 MeV region, this uncertainty is less than 5 %).

**D. The calculation of the relative differential cross sections**

The relative differential cross section $\sigma^{re}_{E\_\text{bin},\theta}$ can be obtained using

$$\sigma^{re}_{E\_\text{bin},\theta} = \frac{W_{E\_\text{bin},\theta}}{\varphi_{E\_\text{bin}} \Omega_\theta N_B \varepsilon_{E\_\text{bin},\theta}} , \qquad (4)$$

where

$$W_{E\_\text{bin},\theta} = \sum_{E \in E\_\text{bin}} w_{E\_\text{bin},\theta} \qquad (5)$$

is the total weight of the net events, $\varphi_{E\_bin}$ is the relative neutron fluence shown in Fig. 1, $\Omega_\theta$ is the detection solid angle of the corresponding silicon detector, $N_B$ is the number of the $^{10}$B atoms in the sample (the two $^{10}$B samples are not the same), $\varepsilon_{E\_bin,\theta}$ is the detection efficiency for α-particles (~ 100 %). Sources of the uncertainty of $W_{E\_bin,\theta}$ include the errors of statistics, background subtraction and uncertainty of the valid-event-area determination with the magnitudes of (0.7 − 6.9) %, (0.1 − 9.0) % and (0.1 − 6.5) %, respectively. The uncertainties of $\Omega_\theta$, and $N_B$ are 0.3 % and 1.0 %, respectively.

### E. The deconvolution of the spread of the detection angle

The detection angle of each silicon detector was obtained from the Monte Carlo simulation. In the simulation, the particles were assumed to be emitted isotropically from the random position in the sample and reach the detector. According to the simulation, the spread of the detection angle for each detector was expected to be 3.8° − 4.0° as shown in Fig. 8, which lead to the uncertainty of the detection angle. The iterative method was used to perform a correction for the spread of the detection angle.

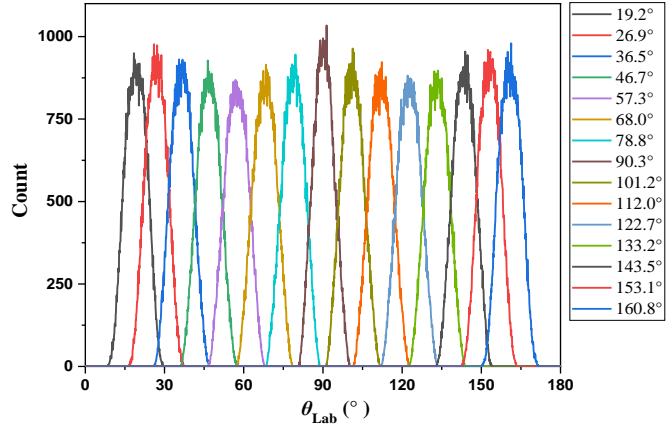

**FIG. 8.** (Color online) The simulated distributions of the receiving angles of the silicon detectors.

For each neutron energy bin, the 15 measured $\sigma^{re}_{E\_bin,\theta}$ were fitted using the Legendre polynomial series, then the fitting curve were convoluted with the simulated distribution shown in Fig. 8. The process of deconvolution and iteration would take the anisotropy into account. After that, the new corrected relative differential cross section $C\_\sigma^{re}_{E\_bin,\theta}$ was obtained. Next, the deconvolution cross section $cor\_\sigma^{re}_{E\_bin,\theta}$ was calculated by

$$\mathrm{cor}\_\sigma^{re}_{E\_bin,\theta} = \sigma^{re}_{E\_bin,\theta} \frac{\sigma^{re}_{E\_bin,\theta}}{C\_\sigma^{re}_{E\_bin,\theta}} . \quad (6)$$

This deconvolution process was iterated until the variation of $cor\_\sigma^{re}_{E\_bin,\theta}$ was less than 0.1 % (usually 10 times). This correction could reduce the uncertainty of the detection angles to ~ 0.01°. The uncertainty of the differential cross section introduced from the angle deconvolution is (0.1 − 0.9) %.

### F. The calculation of the cross sections of the $^{10}$B(n, α)$^7$Li reaction

It is commonly accepted that the relative differential cross section could be represented by the Legendre polynomial series

$$f^{re}_{E\_bin}(\cos(\theta)) = \sum_{i=0}^{M} A_i P_i(\cos(\theta)) \quad (7)$$

where M is the maximum Legendre polynomial order (M = 1 in the 1.0 eV ≤ $E_n$ < 1.0 keV region, M = 2 in the 1.0 keV ≤ $E_n$ < 0.1 MeV region, M = 3 in the 0.1 MeV ≤ $E_n$ < 2.5 MeV region), $A_i$ is the $i^{th}$ coefficient which is determined by fitting the $cor\_\sigma^{re}_{E\_bin,\theta}$ at each $E\_bin$, $P_i(\cos(\theta))$ is the $i^{th}$-order Legendre polynomial. Then through the integration of $f^{re}_{E\_bin}$, the relative angle-integrated cross sections of the $^{10}B(n, \alpha)^7Li$ reaction $\sigma^{re}_{E\_bin}$ could be obtained, and the uncertainty of the fitting is (0.2 − 9.5) %. Since the absolute neutron flux was not measured, the measured cross sections were normalized according to standard cross section $\sigma^{Standard}_{E\_bin}$ from ENDF/B-VIII.0 in 0.3 − 0.5 MeV region, where the results obtained in the present work have relatively small uncertainties [5]. The uncertainty of normalization is 1.3 % deduced from the information of standard library [4, 5]. Then, the cross sections were iterated using the method presented in sections III.A − E until the variation was less than 1.0 % (usually 15 times). Finally, the absolute differential cross section $\sigma_{E\_bin,\theta}$ and the angle-integrated cross section $\sigma_{E\_bin}$ of the $^{10}B(n, \alpha)^7Li$ reaction can be obtained. The total uncertainty of the differential cross section $\sigma_{E\_bin,\theta}$ is (2.6 − 53.0) % (for 868 of the 1005 $\sigma_{E\_bin,\theta}$ results, this uncertainty is less than 10 %), and those of the angle-integrated cross section $\sigma_{E\_bin}$ is (2.1 − 21.5) % (for 43 of the 67 $\sigma_{E\_bin}$ results, this uncertainty is less than 5 %). Sources of uncertainty and their magnitudes were shown in Table. I. The uncertainty of the $^{10}B(n, \alpha)^7Li$ reaction is fairly big around $E_n$ = 1.0 MeV due to few counts of $\alpha$ events near the valley in the excitation function. The results are presented in Table. II of Appendix A. Selected results of differential cross sections are shown in Fig. 9, and those of angle-integrated cross sections are shown in Fig. 10. The variation trend of the differential cross sections will be discussed in section V.

**Table I.** Sources of uncertainty and their magnitudes.

| Sources of uncertainty | Magnitude (%) | |
|---|---|---|
| | Differential cross sections | Angle-integrated cross sections |
| Relative neutron fluence ($\varphi_{E\_bin}$) | 0.5 − 21.4[a], 0.6 − 1.9[b], 0.7 − 0.8[c] | 0.5 − 21.4[a], 0.6 − 1.9[b], 0.7 − 0.8[c] |
| Unfolding of the neutron energy bin width ($w_{E\_bin,\theta}$) | 0.1 − 4.2[a], 0.1 − 7.8[b], 0.3 − 51.1[c] | 0.1 − 3.8[a], 0.1 − 3.4[b], 0.9 − 7.4[c] |
| Unfolding of the expanded neutron energy due to the double-bunched operation mode ($w_{E\_bin,\theta}$) | 1.2 − 10.9[b], 2.9 − 11.9[c] | 0.4 − 1.3[b], 1.0 − 1.6[c] |
| Uncertainty of neutron energy ($E\_bin$, lateral error) | 0.4 − 1.4[a], 0.6 − 1.0[b], 1.0 − 1.5[c] | 0.4 − 1.4[a], 0.6 − 1.0[b], 1.0 − 1.5[c] |
| Statistical error of the valid $\alpha$ events ($W_{E\_bin,\theta}$) | 0.7 − 3.1[a], 1.0 − 5.6[b], 2.6 − 6.9[c] | 0.2 − 0.8[a], 0.3 − 1.0[b], 0.8 − 1.5[c] |
| Background subtraction ($W_{E\_bin,\theta}$) | 0.1 − 3.4[a], 0.1 − 9.0[b], 0.1 − 8.7[c] | < 0.6[a], < 0.7[b], < 0.7[c] |
| Determination the valid-even-area ($W_{E\_bin,\theta}$) | 0.1 − 3.8[a], 0.3 − 6.5[b], 0.4 − 4.4[c] | < 0.4[a], < 0.3[b], < 0.2[c] |
| Detection solid angle ($\Omega_\theta$) | 0.3 % | 0.3 % |
| Number of the $^{10}B$ atoms ($N_B$) | 1.0 % | 1.0 % |
| Deconvolution of the spread of the detection angle ($cor\_\sigma^{re}_{E\_bin,\theta}$) | < 1.0 % | < 0.1 % |
| Fitting using the Legendre polynomial series ($f^{re}_{E\_bin}$) | ——— | 0.2 − 1.6[a], 0.6 − 2.0[b], 1.8 − 9.5[c] |
| Normalization using the standard library ($\sigma^{Standard}_{E\_bin}$) | 1.3 % | 1.3 % |
| Ratios of the $^{10}B(n, \alpha_0)^7Li$ reaction ($R^0_{E\_bin,\theta}$) | 2.9 − 25.8[a], 3.6 − 36.4[b] | 1.7 − 4.3[a], 1.9 − 4.6[b] |
| Ratios of the $^{10}B(n, \alpha_1)^7Li$ reaction ($R^1_{E\_bin,\theta}$) | 1.0 − 6.6[a], 1.7 − 21.6[b] | 0.4 − 2.0[a], 0.9 − 4.3[b] |
| Total uncertainty of the $^{10}$B$(n, \alpha)^7$Li reaction ($\sigma_{E\_bin,\theta}$ and $\sigma_{E\_bin}$) | 2.8 − 21.6$^a$, 2.6 − 17.3$^b$, 4.5 − 53.0$^c$ | 2.2 − 21.5$^a$, 2.1 − 4.4$^b$, 3.1 − 12.4$^c$ |
| Total uncertainty of the $^{10}$B$(n, \alpha_0)^7$Li reaction ($\sigma^0_{E\_bin,\theta}$ and $\sigma^0_{E\_bin}$) | 4.7 − 26.3$^a$, 4.7 − 36.7$^b$ | 3.2 − 21.6$^a$, 2.8 − 6.3$^b$ |
| Total uncertainty of the $^{10}$B$(n, \alpha_1)^7$Li reaction ($\sigma^1_{E\_bin,\theta}$ and $\sigma^1_{E\_bin}$) | 3.1 − 21.8$^a$, 3.3 − 27.7$^b$ | 2.5 − 21.5$^a$, 2.3 − 5.7$^b$ |

$^a$ Uncertainties for 1.0 eV ≤ $E_n$ < 0.02 MeV.

$^b$ Uncertainties for 0.02 MeV ≤ $E_n$ < 1.0 MeV.

$^c$ Uncertainties for 1.0 MeV ≤ $E_n$ < 2.5 MeV

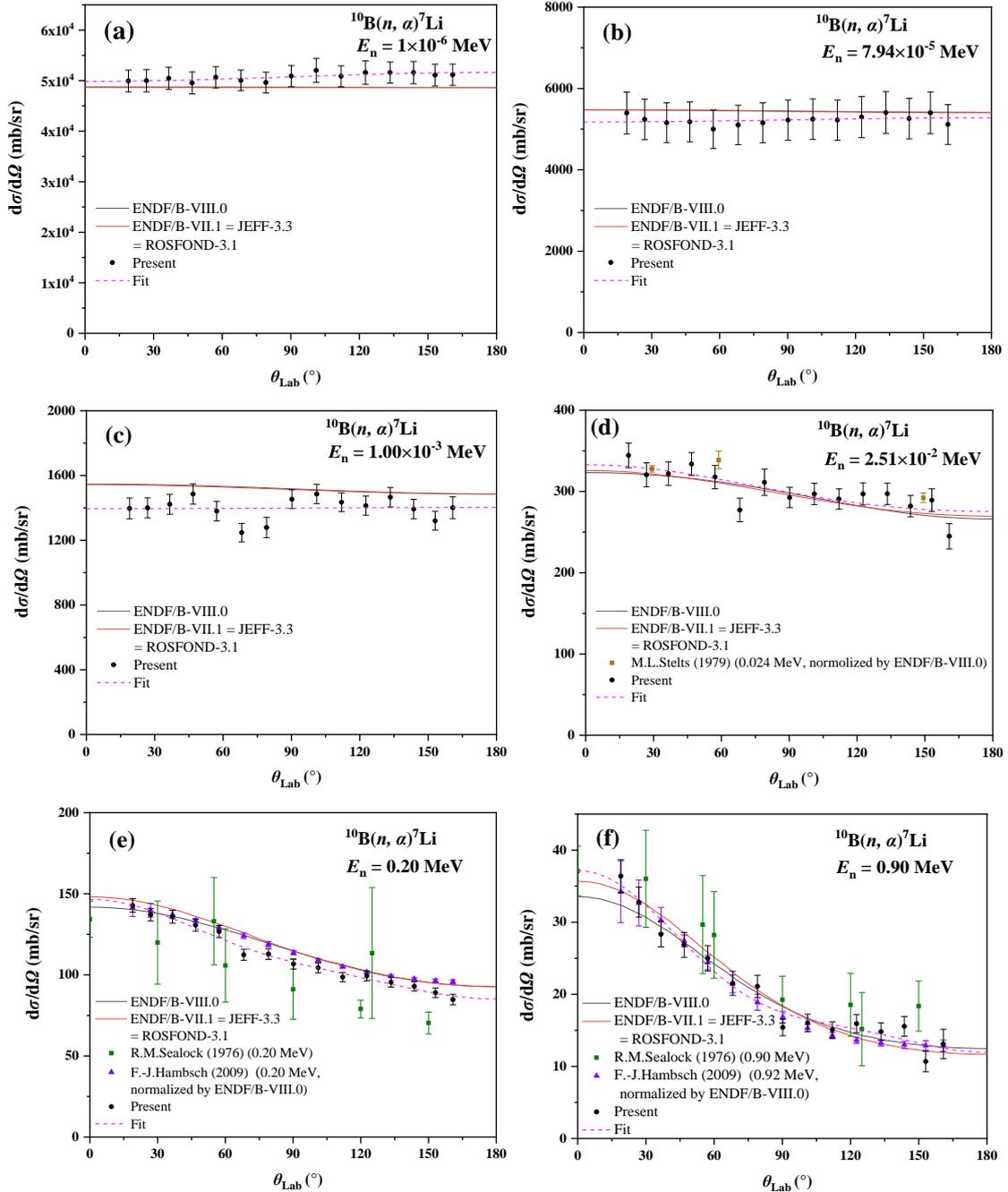

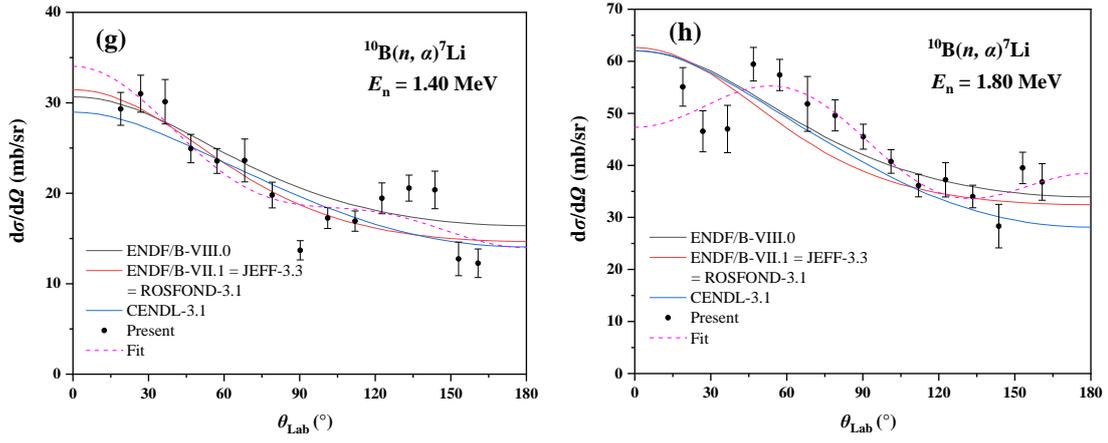

**FIG. 9.** (Color online) The present differential cross sections of the $^{10}$B(n, α)$^7$Li reaction at selected energy points as a function of the $\theta_{Lab}$ compared with existing results of evaluations and measurements [3, 5].

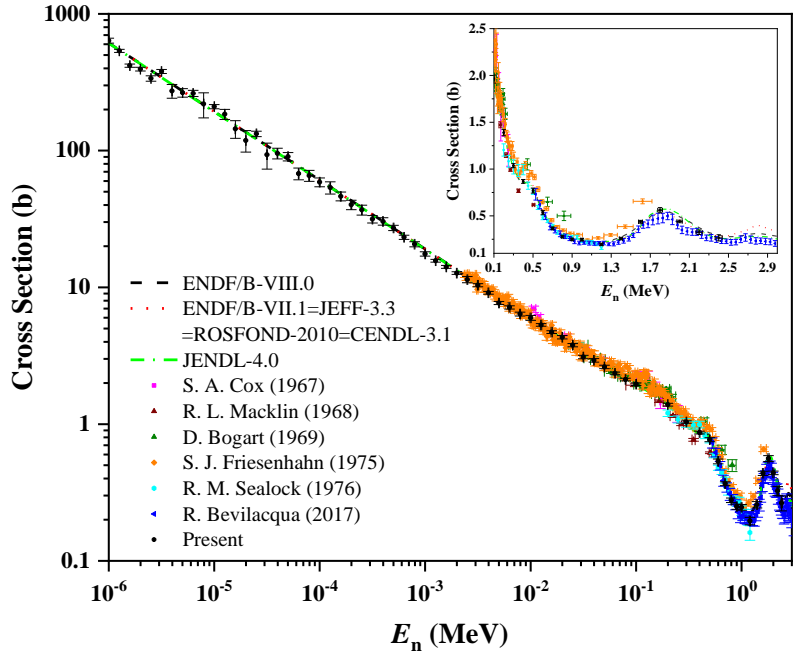

**FIG. 10.** (Color online) The angle-integrated cross sections of the $^{10}$B(n, α)$^7$Li reaction compared with existing results of evaluations and measurements since 1965 [3, 5, 17].

## G. Determination of the ratios of the cross sections of the $^{10}$B(n, α$_0$)$^7$Li and $^{10}$B(n, α$_1$)$^7$Li* reactions

After the processes presented in sections III.A − F, the differential and angle-integrated cross sections the $^{10}$B(n, α)$^7$Li reaction have been obtained. The $^{10}$B(n, α)$^7$Li reaction has two reaction channels which are the $^{10}$B(n, α$_0$)$^7$Li and $^{10}$B(n, α$_1$)$^7$Li* reactions. Their differential cross sections, $\sigma^0_{E\_bin,\theta}$ and $\sigma^1_{E\_bin,\theta}$, can be obtained by

$$\sigma^0_{E\_bin,\theta} = R^0_{E\_bin,\theta} \sigma_{E\_bin,\theta} \tag{8}$$

and

$$\sigma^1_{E\_bin,\theta} = R^1_{E\_bin,\theta} \sigma_{E\_bin,\theta}, \tag{9}$$

where $R^0_{E\_bin,\theta}$ and $R^1_{E\_bin,\theta}$ are the ratios of the differential cross sections of the $^{10}$B(n, α$_0$)$^7$Li and $^{10}$B(n, α$_1$)$^7$Li* reactions over that of the $^{10}$B(n, α)$^7$Li reaction, respectively. After the processes presented in sections III.A − F, the weight of the events has been corrected at the corresponding neutron energy bin and the detection angle, and the ratios can be obtained from the spectrum of the net α events. The peaks of $α_0$ and $α_1$ could be separated directly from the spectrum in the low neutron energy region from 1.0 eV to 0.01 MeV except for several detection angles, such as 19.2° and 160.8°. However, as the neutron energy increases, overlap occurs between the two peaks. The cross sections of the two reaction channels were obtained from 0.01 MeV to 1.0 MeV using the double Gaussian functions fitting method; above 1.0 MeV, the two peaks overlap almost completely. The calculation detail of $R^0_{E\_bin,\theta}$ and $R^1_{E\_bin,\theta}$ is described as fellows.

Firstly, from the spectrum at the corresponding neutron energy bin and the detection angle, the valley between the two peaks of $α_0$ and $α_1$ could be found. If the minimum height of the valley was less than the one-tenth maximum height of the lower peak (usually $α_0$ peak), a threshold at the least-count position would separate the two peaks directly as an example shown in Fig. 11 (a). In this condition, the area of the overlap is less than 2 % of the $α_0$ or $α_1$ peak area, and the overlap is thus negligible. Below 0.01 MeV, the $α_0$ and $α_1$ peaks of most spectra could be separated directly, except for the spectra of several detection angles, such as 19.2° and 160.8°.

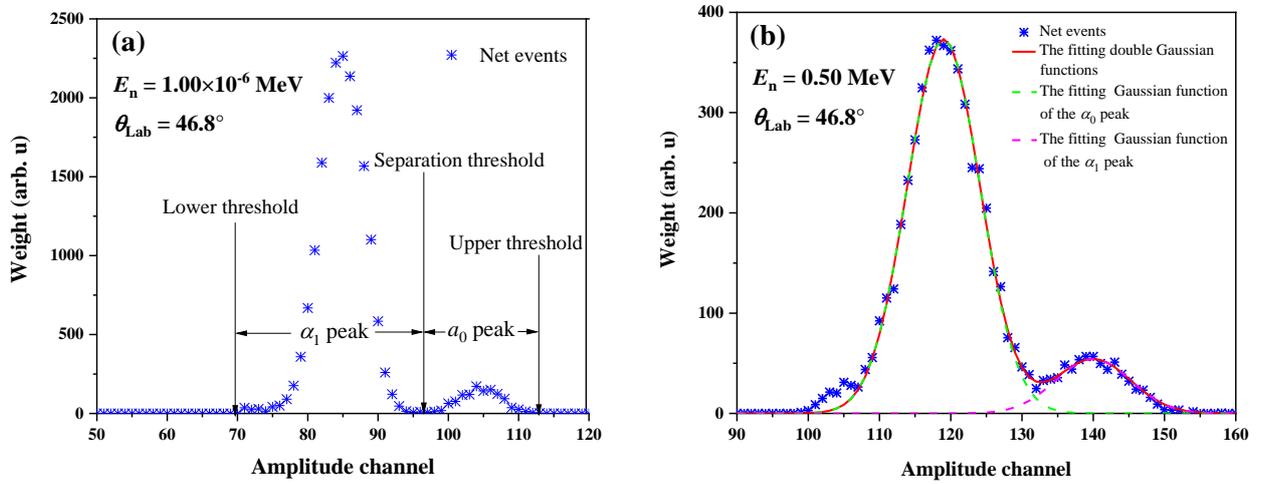

**FIG. 11.** (Color online) The separation of the $α_0$ and $α_1$ peaks by (a) threshold at the detection angle 46.8° and at the neutron energy $1.00×10^{-6}$ MeV and (b) the fitting method at the detection angle 46.8° and at the neutron energy 0.50 MeV.

Secondly, if the overlap between the two peaks in the spectrum of net α events cannot be ignored, then double Gaussian functions were used to fit them. Compared with other fitting functions, such as the exponential function and the polynomial function, the double Gaussian functions agree better with the measurement spectrum. Then the area of the overlapping region would be separated by the fitting result. A typical example is shown in Fig. 11 (b).

After the calculations of $R^0_{E\_bin,\theta}$ and $R^1_{E\_bin,\theta}$, $\sigma^0_{E\_bin,\theta}$ and $\sigma^1_{E\_bin,\theta}$ would be obtained from Eqs. (6) and (7). The uncertainties of the calculations of $R^0_{E\_bin,\theta}$ and $R^1_{E\_bin,\theta}$ are (2.9 − 36.4) % and (1.0 − 21.6) % (for 503 of the 885 $\sigma^0_{E\_bin,\theta}$ results, this uncertainty is less than 10 %; for 864 of the 885 $\sigma^1_{E\_bin,\theta}$ results, this uncertainty is less than 10 %).

The differential cross sections would be fitted by the Legendre polynomial series, then the ratios of the angle-integrated cross sections of the two reaction channel would be calculated using fitting results. Finally, the angle-integrated cross sections of the $^{10}$B(n, $α_0$)$^7$Li and $^{10}$B(n, $α_1$)$^7$Li* reactions, $\sigma^0_{E\_bin}$ and $\sigma^1_{E\_bin}$, were obtained. The total uncertainties of the differential cross sections $\sigma^0_{E\_bin,\theta}$ and $\sigma^1_{E\_bin,\theta}$ are (4.7 − 36.7) % and (3.1 − 27.7) % (for 271 of the 885 $\sigma^0_{E\_bin,\theta}$ results, this uncertainty is less than 10 %; for 738 of the 885 $\sigma^1_{E\_bin,\theta}$ results, this uncertainty is less than 10 %), and those of the angle-integrated cross sections $\sigma^0_{E\_bin}$ and $\sigma^1_{E\_bin}$ are (2.8 − 21.6) % and (2.3 − 21.5) % (for 31 of the 59 $\sigma^0_{E\_bin}$ results, this uncertainty is less than 5 %; for 37 of the 59 $\sigma^1_{E\_bin}$ results, this uncertainty is less than 5 %). Sources of uncertainty and their magnitudes were shown in Table. I. The uncertainty of the $^{10}$B(n, $\alpha_0$)$^7$Li reaction is quite large for $E_n$ < 0.1 MeV because of the small counts of $\alpha_0$-particles. Besides, the uncertainties of the ratios, $R^0_{E\_bin,\theta}$ and $R^1_{E\_bin,\theta}$, are quite large at the detection angles of 19.2° and 160.8°, because the emitted $\alpha$-particles at these two emission angles would have longer track in the sample, which lead to more energy loss and more overlap of the two peaks. The final results are presented in Table. III and Table. IV of Appendix A. Selected results of differential cross sections are shown in Figs. 12 and 13, and the angle-integrated cross sections are shown in Figs. 14 and 15. The differential cross sections of the $^{10}$B(n, $\alpha_1$)$^7$Li* reaction are very similar to that of the $^{10}$B(n, $\alpha$)$^7$Li* reaction below 1.0 MeV, because the cross section of $^{10}$B(n, $\alpha_1$)$^7$Li* reaction take a large proportion ( > 75 % below 0.5 MeV region and > 55 % in the 0.5 MeV ≤ $E_n$ < 1.0 MeV region) of the total cross section. The differential cross sections will be discussed in section V.

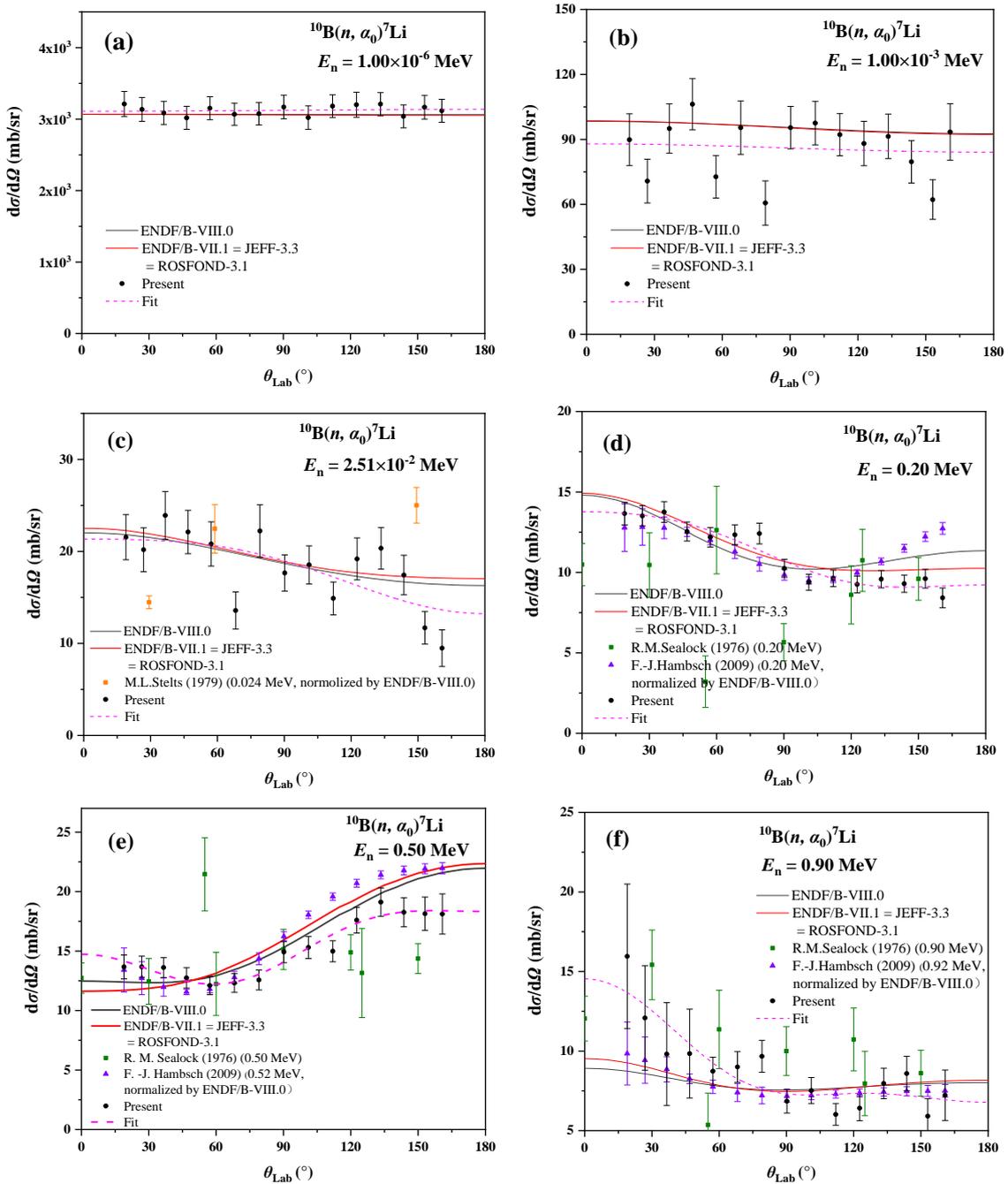

**FIG. 12.** (Color online) The present differential cross sections of the $^{10}$B(n, $\alpha_0$)$^7$Li reaction at selected energy points as a function of the $\theta_{Lab}$ compared with existing results of evaluations and measurements [3, 5].

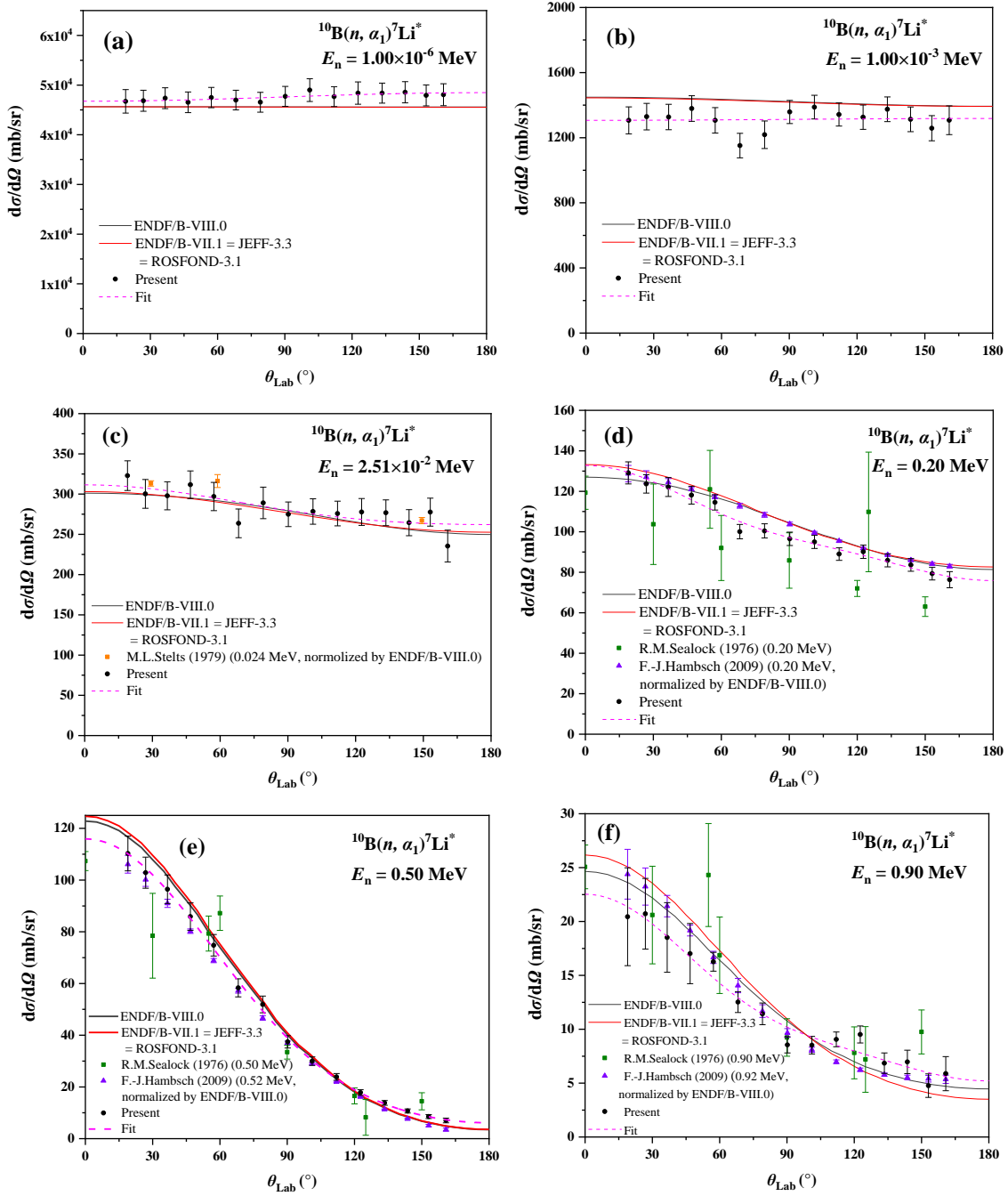

**FIG. 13.** (Color online) The present differential cross sections of the $^{10}$B$(n, \alpha_1)^7$Li$^*$ reaction at selected energy points as a function of the $\theta_{Lab}$ compared with existing results of evaluations and measurements [3, 5].

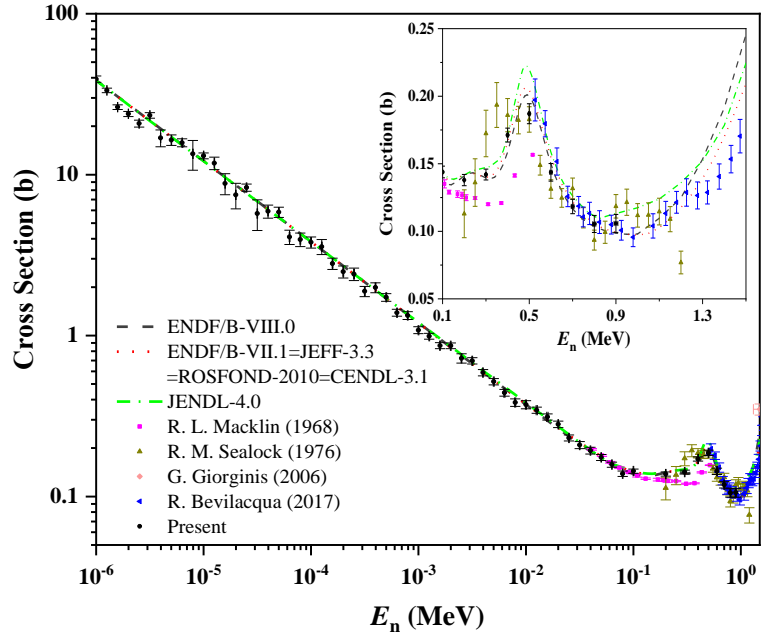

**FIG. 14.** (Color online) The angle-integrated cross sections of the $^{10}$B$(n, \alpha_0)^7$Li reaction compared with existing results of evaluations and measurements since 1965 [3, 5, 17].

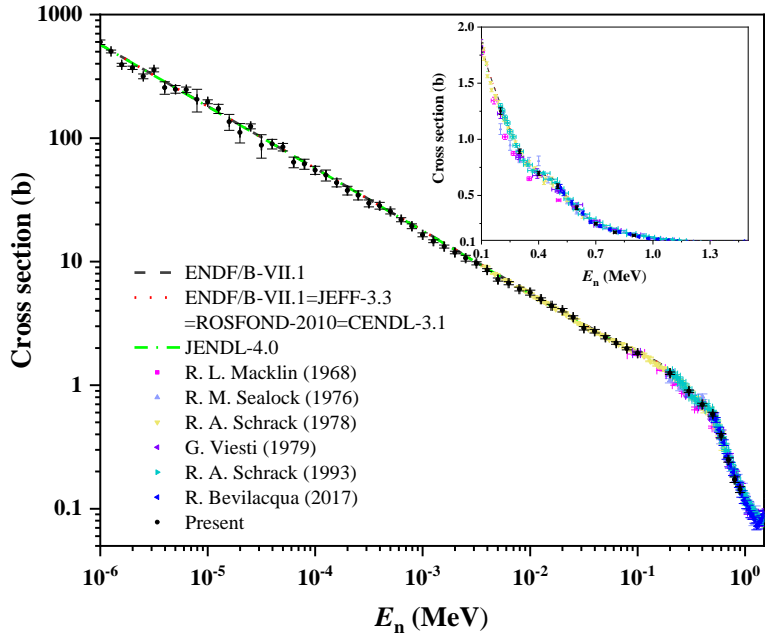

**FIG. 15.** (Color online) The present cross sections of the $^{10}$B$(n, \alpha_1)^7$Li$^*$ reaction compared with existing results of evaluations and measurements since 1965 [3, 5, 17].

## IV. Discussions

### A. Comparison of the results with different measurements and evaluations

The present differential cross sections have been compared with existing measurement data and evaluations [3, 5]:

1) For $E_n$ < 0.1 MeV, the present differential cross sections of the $^{10}$B$(n, \alpha)^7$Li, $^{10}$B$(n, \alpha_0)^7$Li and $^{10}$B$(n, \alpha_1)^7$Li$^*$ reactions agree well with the measurement data of Hambsch (2009, 0.40 keV − 1.20 MeV, normalized by ENDF/B-VIII.0) and Stelts [7] (1979, 2 keV − 24 keV, normalized by ENDF/B-VIII.0) relatively. The present differential cross sections of the $^{10}$B(*n*, *α*)$^7$Li, $^{10}$B(*n*, *α*$_0$)$^7$Li and $^{10}$B(*n*, *α*$_1$)$^7$Li$^*$ reactions show that the differential cross sections are almost isotropic in the center-of-mass system and slightly forward-peaked in the laboratory system, as well as those of different evaluations.

2) In the 0.1 MeV ≤ $E_n$ < 1.0 MeV region, the present differential cross sections of the $^{10}$B(*n*, *α*)$^7$Li and $^{10}$B(*n*, *α*$_1$)$^7$Li$^*$ reactions agree with the measurement data of Hambsch (2009, 0.40 keV − 0.98 MeV, normalized by ENDF/B-VIII.0) [2], as well as the ENDF/B-VIII.0 and ENDF/B-VII.1 library.

For the present differential cross sections of the $^{10}$B(*n*, *α*$_0$)$^7$Li reaction, there are differences between the present data and the measurement data of Hambsch, as well as those of ENDF/B-VIII.0 and ENDF/B-VII.1 library. In this neutron energy region, compared with the measurement data of Hambsch and evaluation data, the present differential cross sections are commonly higher in the forward emission angles, and lower in the backward angles. However, due to the large uncertainty of the present results of the $^{10}$B(*n*, *α*$_0$)$^7$Li reaction, more research is needed to clarified these differences.

Besides, the present differential cross sections of the $^{10}$B(*n*, *α*$_0$)$^7$Li and $^{10}$B(*n*, *α*$_1$)$^7$Li$^*$ don't agree with the results of Sealock (1976, 0.20 MeV − 1.20 MeV) [6]. Compared with the results of Sealock, the uncertainty of the present differential cross sections is smaller. The average uncertainty of the results of Sealock is 22.4 %, while that of the present results is 8.6 % in this neutron energy region.

3) In the 1.0 MeV ≤ $E_n$ < 2.5 MeV region, the discrepancies between the present differential cross sections of the $^{10}$B(*n*, *α*)$^7$Li reaction and evaluations exist. There is no other data in the 1.0 MeV ≤ $E_n$ < 2.5 MeV region except the results of Sealock (1976, 0.20 MeV − 1.20 MeV), which has fairly big uncertainty [6]. So, further researches are therefore demanded in the MeV neutron energy region.

The present cross sections have been also compared with existing measurement data from EXFOR library since 1965 and evaluations [3, 5]：

1) For the $^{10}$B(*n*, *α*)$^7$Li reaction, the present cross sections agree well with the measurement data of Friesenhahn (1975, 2.35 keV − 1.72 MeV) [18] for $E_n$ ≤ 0.3 MeV, Sealock (1976, 0.20 MeV − 1.20 MeV) [6] in the 0.2 MeV ≤ $E_n$ ≤ 1.2 MeV region and Bevilacqu (2017, 0.50 MeV – 3.00 MeV) [17] in the 0.5 MeV < $E_n$ < 1.0 MeV region. The present cross sections of the $^{10}$B(*n*, *α*)$^7$Li reaction are 17.7 % lower than those of Friesenhahn for $E_n$ > 0.3 MeV, and 12.6 % higher than those of Bevilacqu for $E_n$ ≥ 1.0 MeV.

2) For the $^{10}$B(*n*, *α*$_0$)$^7$Li reaction, the present cross sections agree well with the measurement data of Macklin (1968, 0.04 MeV − 0.52 MeV) [19] in the 0.04 MeV ≤ $E_n$ ≤ 0.1 MeV region, Sealock (1976, 0.20 MeV − 1.20 MeV) [6] in the 0.2 MeV ≤ $E_n$ < 1.0 MeV region, and Bevilacqu (2017, 0.50 MeV – 3.00 MeV) [17] in the 0.8 MeV ≤ $E_n$ < 1.0 MeV region. The present cross sections of the $^{10}$B(*n*, *α*$_0$)$^7$Li reaction are 19.5 % higher than those of Macklin for $E_n$ > 0.1 MeV and 8.1 % lower than those of Bevilacqu for $E_n$ < 0.8 MeV.

3) For the $^{10}$B(*n*, *α*$_1$)$^7$Li$^*$ reaction, the present cross sections agree well with the measurement data of Schrack (1978, 3.82 keV − 0.63 MeV) [20], Viesti (1979, 0.10 MeV − 2.20 MeV) [21], Sealock (1976, 0.20 MeV − 1.20 MeV) [6], Schrack (1993, 0.20 − 4.06 MeV) [22] and Bevilacqu (2017, 0.50 MeV – 3.00 MeV) [17].

Compared with different evaluations, including ENDF/B-VIII.0, ENDF/B-VII.1, JENDL-4.0, ROSFOND-2010 and CENDL-3.1 libraries, the present cross sections of the $^{10}$B(*n*, *α*$_0$)$^7$Li and $^{10}$B(*n*, *α*$_1$)$^7$Li$^*$ reactions generally agree better with ENDF/B-VIII.0 library; those of the $^{10}$B(*n*, *α*)$^7$Li reaction agree well with ENDF/B-VIII.0 library for $E_n$

≤ 1.8 MeV, and well with ENDF/B-VII.1 library for $E_n >$ 1.8 MeV [5].

**B. The future experimental plan**

As mentioned in section II.A, the experiment was performed at Endstation #1 (the length of flight path was 57.99 m) while the neutron spectrum was measured at Endstation #2 (the length of flight path was 75.78 m), so the actual neutron energy spectrum should be a little different between the two positions. The absolute cross sections are affected by the uncertainty of the neutron energy spectrum, while the relative angular distributions are not. The results of simulation by Fluka code was used to determine this effect [10]. According to the simulation, the deviations between the spectrum of the two position were expected to be less than 2 % in the 0.3 − 0.5 MeV region which was chosen for normalization. In the 0.1 − 2.5 MeV region, the deviations were less than 4 %. Besides, one could notice that the neutron energy spectrum has fairly big uncertainty for $E_n <$ 1 keV because of the resonance of the cross section of the $^{235}$U(*n, f*) reaction below 2.5 keV region and the uncertainty of the moderation length of the neutron source. In the future, precise neutron energy spectrum at Endstation #1 should be measured.

The present measurement results are limited in $E_n <$ 2.5 MeV region; more work should be done in the higher neutron energy region. Besides, the peaks of $\alpha_0$ and $\alpha_1$ could not be separated for $E_n >$ 1.0 MeV in the present work; thinner sample and the detectors with higher resolution should be used in the future. The unfolding of the expand neutron energy due to the double-bunched operation mode would introduce fairly big uncertainty for $E_n >$ 2.0 MeV, so the single bunched proton beam should be used in the coming work.

**V. Theoretical analysis**

As shown in Figs. 10, 14 and 15, the cross sections of the $^{10}$B(*n, α*) $^7$Li and $^{10}$B(*n, α*$_1$) $^7$Li* reactions are smooth and obey the 1/*v* law for $E_n <$ 0.1 MeV, as well as those of the $^{10}$B(*n, α*$_0$) $^7$Li reaction for $E_n <$ 0.01 MeV. The $^{10}$B(*n, α*$_1$) $^7$Li* reaction dominates below 0.01 MeV region where the $^{10}$B(*n, α*$_1$) / $^{10}$B(*n, α*) cross-section ratio is (93.77± 0.92) % according to the present results. For the low neutron energy region ($E_n <$ 0.1 MeV), the present differential cross sections of the $^{10}$B(*n, α*)$^7$Li, $^{10}$B(*n, α*$_0$)$^7$Li and $^{10}$B(*n, α*$_1$)$^7$Li* reactions are almost isotropic in the center-of-mass system and slightly forward-peaked in the laboratory system as shown in Figs. 9 (a) − (d), Figs. 12 (a) − (c) and Figs. 13 (a) − (c).

The big $^{10}$B(*n, α*$_1$) / $^{10}$B(*n, α*) cross-section ratio for the low neutron energy region could be explained by the resonance reaction mechanism and the level structure of the $^{11}$B compound system [23]. When the $^{10}$B target nucleus interacts with the low-energy neutrons ($l_n = 0$), the scattering and bound states with $J^\pi = 7/2^+$ of the compound nucleus $^{11}$B would be formed, then $^{11}$B would decay to $^7$Li or $^7$Li*. The *s*-wave scattering state of $^{11}$B has the resonance energy of $E_n =$ 0.37 MeV, and the excitation energy of $^7$Li* is 0.48 MeV. With the broad *s*-wave state ($\Gamma_{n(C.M.)} =$ 0.77 MeV and $\Gamma_{\alpha_1(C.M.)} =$ 0.113 MeV), the $^{10}$B(*n, α*$_1$) $^7$Li* reaction would have large cross section in the low neutron energy region. The $\alpha_0$ partial width ($\Gamma_{\alpha_0(C.M.)} =$ 0.001 MeV) is negligible compared to the $\alpha_1$ partial width, so it is very difficult for the $^{10}$B(*n, α*$_0$) $^7$Li reaction to have the resonance with this state of $^{11}$B. Besides, the two $7/2^+$ *s*-wave states are primarily responsible for the 1/*v* law of the excitation function for the low neutron energy region [24].

In the neutron energy range from 0.1 MeV to 1.0 MeV, the differential cross sections of the $^{10}$B(*n, α*$_1$) $^7$Li* reaction are forward-peaked as shown in Figs. 13 (d) − (f) which are also mainly contributed by the two $7/2^+$ *s*-wave states of the $^{11}$B [24]. For the $^{10}$B(*n, α*$_0$) $^7$Li reaction, the differential cross sections are backward-peaked in 0.3 − 0.6 MeV region as shown in Fig. 12 (e), which are caused by the resonance of $5/2^-$ *p*-wave state of the $^{11}$B which would occur as $E_n \approx 0.52$ MeV. As for the excitation function, the resonance of $5/2^-$ $p$-wave state would lead to a peak of the $^{10}$B$(n, \alpha_0)$ $^7$Li reaction around $E_n = 0.50$ MeV. Because of the competition of the $^{10}$B$(n, \alpha_0)$ $^7$Li reaction, the cross section of the $^{10}$B$(n, \alpha_1)$ $^7$Li$^*$ reaction would decrease rapidly above 0.50 MeV region [2].

For 1.0 MeV $\leq E_n <$ 2.5 MeV, the differential cross sections of the $^{10}$B$(n, \alpha)^7$Li reaction are approximately forward-peaked in the laboratory (Lab) system, while the deviations exist between the measurements and evaluations. For example, the anomaly was observed as shown in Fig. 9 (h) around $E_n = 1.80$ MeV. It might be due to the contribution of the resonance of the $9/2^-$ $p$-wave state at $E_n = 1.83$ MeV and $5/2^+$ or $7/2^+$ $d$-wave state at $E_n = 1.88$ MeV of $^{11}$B [25]. These resonance states would also lead to the peak of the excitation function of the $^{10}$B$(n, \alpha)$ $^7$Li reaction near $E_n \approx 1.80$ MeV as shown in Fig. 10.

## VI. Conclusions

In the present work, with 15 silicon detectors distributed from 19.2° to 160.8°, differential and angle-integrated cross sections for the $^{10}$B$(n, \alpha)^7$Li, $^{10}$B$(n, \alpha_0)$ $^7$Li and $^{10}$B$(n, \alpha_1)$ $^7$Li$^*$ reactions have been measured using the LPDA detector system at CSNS Back-n white neutron source. Compared with existing measurements, the present results have been obtained in the wider neutron energy range. The differential and angle-integrated cross sections were obtained for the $^{10}$B$(n, \alpha)^7$Li reaction for 1.0 eV $\leq E_n <$ 2.5 MeV (67 energy points), as well as the two reaction channels, $^{10}$B$(n, \alpha_0)$ $^7$Li and $^{10}$B$(n, \alpha_1)$ $^7$Li$^*$, for 1.0 eV $\leq E_n <$ 1.0 MeV (59 energy points). There is no previous measurement datum of the differential cross section of the $^{10}$B$(n, \alpha)^7$Li reaction existing in EXFOR library in the 1.2 MeV $< E_n <$ 2.5 MeV region; the present data are the first measurement results in this region. The measurement results have been analyzed by the resonance reaction mechanism and the level structure of the $^{11}$B compound system. Compared with existing evaluations and measurements, the present results show a good agreement except for several energy points around 1.0 MeV and above 2.0 MeV, which need further research especially in the MeV neutron energy region.

**ACKNOWLEDGEMENTS**

The authors are indebted to the operation crew of the CSNS Back-n white neutron source. Dr. Qiwen Fan from China Institute of Atomic Energy is appreciated for preparing the $^{10}$B samples. Prof. Zhenpeng Chen from Tsinghua University is appreciated for the beneficial discussions. The present work is financially supported by the National Key R&D Program of China (No. 2016YFA0401604) and the National Natural Science Foundation of China (No. 11775006).

# Appendix A: Results of differential and angle-integrated cross sections of the $^{10}$B$(n, \alpha)^7$Li, $^{10}$B$(n, \alpha_0)^7$Li and $^{10}$B$(n, \alpha_1)^7$Li$^*$ reactions

**Table II (a).** The differential cross sections of the $^{10}$B$(n, \alpha)^7$Li reaction in the laboratory reference system.

| $E_n$ (MeV) | $\sigma_{E\_bin,\theta}$ (mb/sr) | | | | | |
|---|---|---|---|---|---|---|
| | 19.2° | 26.9° | 36.5° | 46.7° | 57.3° | 68.0° |
| 1.00×10$^{-6}$±4.2×10$^{-9}$ | 4.99×10$^4$±2.1×10$^3$ | 5.00×10$^4$±2.2×10$^3$ | 5.05×10$^4$±2.2×10$^3$ | 4.96×10$^4$±2.2×10$^3$ | 5.06×10$^4$±2.1×10$^3$ | 5.00×10$^4$±2.0×10$^3$ |
| 1.26×10$^{-6}$±5.4×10$^{-9}$ | 4.13×10$^4$±1.2×10$^3$ | 4.18×10$^4$±1.2×10$^3$ | 4.22×10$^4$±1.2×10$^3$ | 4.22×10$^4$±1.2×10$^3$ | 4.24×10$^4$±1.2×10$^3$ | 4.16×10$^4$±1.2×10$^3$ |
| 1.58×10$^{-6}$±6.8×10$^{-9}$ | 3.30×10$^4$±9.7×10$^2$ | 3.24×10$^4$±9.4×10$^2$ | 3.27×10$^4$±9.4×10$^2$ | 3.27×10$^4$±9.5×10$^2$ | 3.29×10$^4$±9.5×10$^2$ | 3.29×10$^4$±9.5×10$^2$ |
| 2.00×10$^{-6}$±8.7×10$^{-9}$ | 3.07×10$^4$±1.0×10$^3$ | 3.10×10$^4$±1.0×10$^3$ | 3.08×10$^4$±1.0×10$^3$ | 3.06×10$^4$±1.0×10$^3$ | 3.13×10$^4$±1.0×10$^3$ | 3.13×10$^4$±1.0×10$^3$ |
| 2.51×10$^{-6}$±1.1×10$^{-8}$ | 2.64×10$^4$±1.2×10$^3$ | 2.65×10$^4$±1.2×10$^3$ | 2.63×10$^4$±1.2×10$^3$ | 2.67×10$^4$±1.2×10$^3$ | 2.61×10$^4$±1.2×10$^3$ | 2.61×10$^4$±1.2×10$^3$ |
| 3.16×10$^{-6}$±1.4×10$^{-8}$ | 2.94×10$^4$±1.1×10$^3$ | 2.96×10$^4$±1.1×10$^3$ | 2.92×10$^4$±1.1×10$^3$ | 3.04×10$^4$±1.1×10$^3$ | 2.98×10$^4$±1.1×10$^3$ | 3.01×10$^4$±1.1×10$^3$ |
| 3.98×10$^{-6}$±1.8×10$^{-8}$ | 2.09×10$^4$±2.5×10$^3$ | 2.16×10$^4$±2.5×10$^3$ | 2.12×10$^4$±2.5×10$^3$ | 2.13×10$^4$±2.5×10$^3$ | 2.14×10$^4$±2.5×10$^3$ | 2.09×10$^4$±2.5×10$^3$ |
| 5.01×10$^{-6}$±2.3×10$^{-8}$ | 2.07×10$^4$±1.5×10$^3$ | 2.07×10$^4$±1.5×10$^3$ | 2.06×10$^4$±1.5×10$^3$ | 2.12×10$^4$±1.6×10$^3$ | 2.07×10$^4$±1.5×10$^3$ | 2.13×10$^4$±1.6×10$^3$ |
| 6.31×10$^{-6}$±2.9×10$^{-8}$ | 2.03×10$^4$±1.1×10$^3$ | 2.11×10$^4$±1.1×10$^3$ | 2.03×10$^4$±1.1×10$^3$ | 2.03×10$^4$±1.1×10$^3$ | 2.07×10$^4$±1.1×10$^3$ | 2.05×10$^4$±1.1×10$^3$ |
| 7.94×10$^{-6}$±3.7×10$^{-8}$ | 1.70×10$^4$±3.5×10$^3$ | 1.69×10$^4$±3.5×10$^3$ | 1.70×10$^4$±3.6×10$^3$ | 1.73×10$^4$±3.6×10$^3$ | 1.75×10$^4$±3.7×10$^3$ | 1.75×10$^4$±3.7×10$^3$ |
| 1.00×10$^{-5}$±4.7×10$^{-8}$ | 1.68×10$^4$±6.5×10$^2$ | 1.61×10$^4$±6.3×10$^2$ | 1.66×10$^4$±6.4×10$^2$ | 1.64×10$^4$±6.3×10$^2$ | 1.62×10$^4$±6.3×10$^2$ | 1.65×10$^4$±6.4×10$^2$ |
| 1.26×10$^{-5}$±6.4×10$^{-8}$ | 1.45×10$^4$±1.3×10$^3$ | 1.50×10$^4$±1.3×10$^3$ | 1.44×10$^4$±1.3×10$^3$ | 1.44×10$^4$±1.3×10$^3$ | 1.45×10$^4$±1.3×10$^3$ | 1.45×10$^4$±1.3×10$^3$ |
| 1.58×10$^{-5}$±8.6×10$^{-8}$ | 1.11×10$^4$±1.7×10$^3$ | 1.14×10$^4$±1.7×10$^3$ | 1.12×10$^4$±1.7×10$^3$ | 1.13×10$^4$±1.7×10$^3$ | 1.16×10$^4$±1.7×10$^3$ | 1.12×10$^4$±1.7×10$^3$ |
| 2.00×10$^{-5}$±1.2×10$^{-7}$ | 9.50×10$^3$±1.7×10$^3$ | 9.39×10$^3$±1.7×10$^3$ | 9.45×10$^3$±1.7×10$^3$ | 9.42×10$^3$±1.7×10$^3$ | 9.46×10$^3$±1.7×10$^3$ | 9.19×10$^3$±1.7×10$^3$ |
| 2.51×10$^{-5}$±1.6×10$^{-7}$ | 1.04×10$^4$±5.1×10$^2$ | 1.03×10$^4$±5.1×10$^2$ | 1.07×10$^4$±5.2×10$^2$ | 1.05×10$^4$±5.1×10$^2$ | 1.02×10$^4$±5.0×10$^2$ | 1.03×10$^4$±5.1×10$^2$ |
| 3.16×10$^{-5}$±2.1×10$^{-7}$ | 7.29×10$^3$±1.6×10$^3$ | 7.35×10$^3$±1.6×10$^3$ | 7.40×10$^3$±1.6×10$^3$ | 7.53×10$^3$±1.6×10$^3$ | 7.49×10$^3$±1.6×10$^3$ | 7.19×10$^3$±1.6×10$^3$ |
| 3.98×10$^{-5}$±2.7×10$^{-7}$ | 7.73×10$^3$±7.1×10$^2$ | 7.36×10$^3$±6.7×10$^2$ | 7.55×10$^3$±6.9×10$^2$ | 7.45×10$^3$±6.8×10$^2$ | 7.60×10$^3$±6.9×10$^2$ | 7.49×10$^3$±6.8×10$^2$ |
| 5.01×10$^{-5}$±3.6×10$^{-7}$ | 7.24×10$^3$±5.2×10$^2$ | 7.22×10$^3$±5.2×10$^2$ | 7.00×10$^3$±5.0×10$^2$ | 6.92×10$^3$±5.0×10$^2$ | 7.25×10$^3$±5.2×10$^2$ | 7.07×10$^3$±5.1×10$^2$ |
| 6.31×10$^{-5}$±4.8×10$^{-7}$ | 5.40×10$^3$±5.4×10$^2$ | 5.42×10$^3$±5.4×10$^2$ | 5.37×10$^3$±5.4×10$^2$ | 5.34×10$^3$±5.3×10$^2$ | 5.37×10$^3$±5.4×10$^2$ | 5.08×10$^3$±5.1×10$^2$ |
| 7.94×10$^{-5}$±6.4×10$^{-7}$ | 5.40×10$^3$±5.1×10$^2$ | 5.24×10$^3$±5.0×10$^2$ | 5.16×10$^3$±4.9×10$^2$ | 5.18×10$^3$±4.9×10$^2$ | 5.00×10$^3$±4.8×10$^2$ | 5.10×10$^3$±4.9×10$^2$ |
| 1.00×10$^{-4}$±8.4×10$^{-7}$ | 4.76×10$^3$±3.7×10$^2$ | 4.78×10$^3$±3.7×10$^2$ | 4.62×10$^3$±3.6×10$^2$ | 4.80×10$^3$±3.7×10$^2$ | 4.74×10$^3$±3.7×10$^2$ | 4.55×10$^3$±3.5×10$^2$ |
| 1.26×10$^{-4}$±1.0×10$^{-6}$ | 4.48×10$^3$±4.7×10$^2$ | 4.00×10$^3$±4.3×10$^2$ | 4.23×10$^3$±4.4×10$^2$ | 4.32×10$^3$±4.5×10$^2$ | 4.41×10$^3$±4.6×10$^2$ | 4.13×10$^3$±4.3×10$^2$ |
| 1.58×10$^{-4}$±1.3×10$^{-6}$ | 3.72×10$^3$±2.9×10$^2$ | 3.72×10$^3$±2.9×10$^2$ | 3.57×10$^3$±2.8×10$^2$ | 3.62×10$^3$±2.8×10$^2$ | 3.86×10$^3$±3.0×10$^2$ | 3.61×10$^3$±2.8×10$^2$ |
| 2.00×10$^{-4}$±1.6×10$^{-6}$ | 3.37×10$^3$±2.8×10$^2$ | 3.07×10$^3$±2.6×10$^2$ | 3.08×10$^3$±2.6×10$^2$ | 3.15×10$^3$±2.6×10$^2$ | 3.14×10$^3$±2.6×10$^2$ | 3.04×10$^3$±2.5×10$^2$ |
| 2.51×10$^{-4}$±2.0×10$^{-6}$ | 3.08×10$^3$±2.7×10$^2$ | 2.87×10$^3$±2.5×10$^2$ | 2.91×10$^3$±2.5×10$^2$ | 2.93×10$^3$±2.5×10$^2$ | 2.85×10$^3$±2.5×10$^2$ | 2.86×10$^3$±2.5×10$^2$ |
| 3.16×10$^{-4}$±2.5×10$^{-6}$ | 2.78×10$^3$±1.9×10$^2$ | 2.41×10$^3$±1.7×10$^2$ | 2.47×10$^3$±1.7×10$^2$ | 2.29×10$^3$±1.6×10$^2$ | 2.55×10$^3$±1.7×10$^2$ | 2.36×10$^3$±1.6×10$^2$ |
| 3.98×10$^{-4}$±3.1×10$^{-6}$ | 2.63×10$^3$±1.8×10$^2$ | 2.49×10$^3$±1.7×10$^2$ | 2.28×10$^3$±1.6×10$^2$ | 2.40×10$^3$±1.6×10$^2$ | 2.40×10$^3$±1.7×10$^2$ | 2.27×10$^3$±1.6×10$^2$ |
| 5.01×10$^{-4}$±3.9×10$^{-6}$ | 2.28×10$^3$±1.2×10$^2$ | 2.04×10$^3$±1.1×10$^2$ | 2.04×10$^3$±1.0×10$^2$ | 2.30×10$^3$±1.1×10$^2$ | 2.19×10$^3$±1.1×10$^2$ | 2.10×10$^3$±1.1×10$^2$ |
| 6.31×10$^{-4}$±4.8×10$^{-6}$ | 1.98×10$^3$±8.4×10$^1$ | 1.74×10$^3$±7.7×10$^1$ | 1.87×10$^3$±8.1×10$^1$ | 1.81×10$^3$±7.8×10$^1$ | 1.89×10$^3$±8.2×10$^1$ | 1.73×10$^3$±7.9×10$^1$ |
| 7.94×10$^{-4}$±6.1×10$^{-6}$ | 1.79×10$^3$±9.4×10$^1$ | 1.62×10$^3$±8.8×10$^1$ | 1.67×10$^3$±8.6×10$^1$ | 1.64×10$^3$±8.2×10$^1$ | 1.60×10$^3$±8.9×10$^1$ | 1.62×10$^3$±8.4×10$^1$ |
| 1.00×10$^{-3}$±7.7×10$^{-6}$ | 1.40×10$^3$±6.4×10$^1$ | 1.40×10$^3$±6.2×10$^1$ | 1.42×10$^3$±6.1×10$^1$ | 1.49×10$^3$±6.3×10$^1$ | 1.38×10$^3$±6.0×10$^1$ | 1.25×10$^3$±5.7×10$^1$ |
| 1.26×10$^{-3}$±9.8×10$^{-6}$ | 1.24×10$^3$±4.8×10$^1$ | 1.19×10$^3$±4.6×10$^1$ | 1.20×10$^3$±4.3×10$^1$ | 1.32×10$^3$±4.4×10$^1$ | 1.24×10$^3$±4.4×10$^1$ | 1.13×10$^3$±4.6×10$^1$ |
| 1.58×10$^{-3}$±1.3×10$^{-5}$ | 1.17×10$^3$±4.3×10$^1$ | 1.08×10$^3$±5.0×10$^1$ | 1.18×10$^3$±4.0×10$^1$ | 1.11×10$^3$±4.0×10$^1$ | 1.13×10$^3$±4.0×10$^1$ | 1.08×10$^3$±4.2×10$^1$ |
| 2.00×10$^{-3}$±1.6×10$^{-5}$ | 1.06×10$^3$±3.9×10$^1$ | 1.04×10$^3$±3.8×10$^1$ | 1.03×10$^3$±3.6×10$^1$ | 1.03×10$^3$±3.6×10$^1$ | 1.04×10$^3$±3.6×10$^1$ | 9.41×10$^2$±4.0×10$^1$ |
| 2.51×10$^{-3}$±2.1×10$^{-5}$ | 8.72×10$^2$±5.7×10$^1$ | 8.82×10$^2$±5.5×10$^1$ | 9.22×10$^2$±5.6×10$^1$ | 9.55×10$^2$±5.8×10$^1$ | 9.03×10$^2$±5.8×10$^1$ | 7.98×10$^2$±5.5×10$^1$ |
| 3.16×10$^{-3}$±2.7×10$^{-5}$ | 8.74×10$^2$±3.6×10$^1$ | 8.29×10$^2$±3.3×10$^1$ | 7.96×10$^2$±3.0×10$^1$ | 8.36×10$^2$±3.1×10$^1$ | 7.91×10$^2$±3.0×10$^1$ | 7.39×10$^2$±3.5×10$^1$ |
| 3.98×10$^{-3}$±3.6×10$^{-5}$ | 7.05×10$^2$±3.1×10$^1$ | 7.71×10$^2$±3.3×10$^1$ | 7.62×10$^2$±2.8×10$^1$ | 7.54×10$^2$±2.7×10$^1$ | 7.60×10$^2$±2.8×10$^1$ | 6.85×10$^2$±3.0×10$^1$ |
| 5.01×10$^{-3}$±4.8×10$^{-5}$ | 5.90×10$^2$±2.5×10$^1$ | 6.07×10$^2$±2.5×10$^1$ | 6.39×10$^2$±2.4×10$^1$ | 6.32×10$^2$±2.3×10$^1$ | 6.21×10$^2$±2.4×10$^1$ | 5.44×10$^2$±2.5×10$^1$ |
| 6.31×10$^{-3}$±6.4×10$^{-5}$ | 6.05×10$^2$±2.5×10$^1$ | 5.94×10$^2$±2.4×10$^1$ | 5.63×10$^2$±2.3×10$^1$ | 5.91×10$^2$±2.2×10$^1$ | 5.66×10$^2$±2.2×10$^1$ | 5.53×10$^2$±2.3×10$^1$ |
| 7.94×10$^{-3}$±8.6×10$^{-5}$ | 5.11×10$^2$±2.7×10$^1$ | 5.40×10$^2$±2.5×10$^1$ | 5.06×10$^2$±2.4×10$^1$ | 5.35×10$^2$±2.3×10$^1$ | 5.29×10$^2$±2.4×10$^1$ | 4.81×10$^2$±2.4×10$^1$ |
| 1.00×10$^{-2}$±1.2×10$^{-4}$ | 5.14×10$^2$±2.7×10$^1$ | 4.89×10$^2$±2.6×10$^1$ | 5.04×10$^2$±2.6×10$^1$ | 5.01×10$^2$±2.6×10$^1$ | 4.69×10$^2$±2.5×10$^1$ | 4.71×10$^2$±2.6×10$^1$ |
| 1.26×10$^{-2}$±1.6×10$^{-4}$ | 4.01×10$^2$±1.6×10$^1$ | 4.33×10$^2$±1.6×10$^1$ | 4.46×10$^2$±1.6×10$^1$ | 4.49×10$^2$±1.6×10$^1$ | 4.22×10$^2$±1.6×10$^1$ | 4.18×10$^2$±1.7×10$^1$ |
| 1.58×10$^{-2}$±2.2×10$^{-4}$ | 3.81×10$^2$±1.6×10$^1$ | 3.83×10$^2$±1.5×10$^1$ | 3.99×10$^2$±1.5×10$^1$ | 3.81×10$^2$±1.5×10$^1$ | 3.94×10$^2$±1.6×10$^1$ | 3.54×10$^2$±1.5×10$^1$ |
| 2.00×10$^{-2}$±1.2×10$^{-4}$ | 3.61×10$^2$±1.9×10$^1$ | 3.56×10$^2$±1.9×10$^1$ | 3.62×10$^2$±1.8×10$^1$ | 3.49×10$^2$±1.7×10$^1$ | 3.61×10$^2$±1.8×10$^1$ | 3.10×10$^2$±1.8×10$^1$ |

| $E_n$ (MeV) | | | | | | |
|---|---|---|---|---|---|---|
| 2.51×10⁻²±1.5×10⁻⁴ | 3.44×10²±1.5×10¹ | 3.21×10²±1.5×10¹ | 3.22×10²±1.4×10¹ | 3.34×10²±1.4×10¹ | 3.18×10²±1.4×10¹ | 2.77×10²±1.5×10¹ |
| 3.16×10⁻²±1.9×10⁻⁴ | 2.70×10²±1.6×10¹ | 2.68×10²±1.5×10¹ | 2.61×10²±1.5×10¹ | 2.46×10²±1.5×10¹ | 2.63×10²±1.6×10¹ | 2.33×10²±1.6×10¹ |
| 3.98×10⁻²±2.4×10⁻⁴ | 2.66×10²±1.4×10¹ | 2.42×10²±1.3×10¹ | 2.52×10²±1.3×10¹ | 2.45×10²±1.3×10¹ | 2.56×10²±1.3×10¹ | 2.22×10²±1.3×10¹ |
| 5.01×10⁻²±3.0×10⁻⁴ | 2.30×10²±1.1×10¹ | 2.21×10²±1.1×10¹ | 2.34×10²±1.0×10¹ | 2.30×10²±1.0×10¹ | 2.18×10²±1.0×10¹ | 2.02×10²±1.0×10¹ |
| 6.31×10⁻²±3.8×10⁻⁴ | 2.15×10²±8.1×10⁰ | 2.08×10²±8.0×10⁰ | 2.17×10²±7.9×10⁰ | 2.11×10²±7.9×10⁰ | 2.03×10²±7.9×10⁰ | 1.92×10²±8.0×10⁰ |
| 7.94×10⁻²±4.8×10⁻⁴ | 1.91×10²±7.6×10⁰ | 1.97×10²±7.6×10⁰ | 1.93×10²±7.4×10⁰ | 1.87×10²±7.3×10⁰ | 1.90×10²±7.5×10⁰ | 1.66×10²±7.2×10⁰ |
| 1.00×10⁻¹±6.1×10⁻⁴ | 1.86×10²±5.1×10⁰ | 1.84×10²±5.1×10⁰ | 1.79×10²±5.0×10⁰ | 1.75×10²±4.9×10⁰ | 1.73×10²±4.8×10⁰ | 1.52×10²±4.7×10⁰ |
| 2.00×10⁻¹±1.3×10⁻³ | 1.43×10²±4.1×10⁰ | 1.37×10²±3.9×10⁰ | 1.36×10²±4.0×10⁰ | 1.31×10²±3.7×10⁰ | 1.27×10²±3.6×10⁰ | 1.12×10²±3.4×10⁰ |
| 3.00×10⁻¹±2.0×10⁻³ | 1.11×10²±3.1×10⁰ | 1.08×10²±2.8×10⁰ | 1.05×10²±2.8×10⁰ | 1.02×10²±2.7×10⁰ | 9.82×10¹±2.7×10⁰ | 8.38×10¹±2.4×10⁰ |
| 4.00×10⁻¹±2.9×10⁻³ | 1.17×10²±3.7×10⁰ | 1.11×10²±3.6×10⁰ | 1.08×10²±3.5×10⁰ | 9.67×10¹±3.2×10⁰ | 9.03×10¹±2.9×10⁰ | 7.38×10¹±2.4×10⁰ |
| 5.00×10⁻¹±3.9×10⁻³ | 1.24×10²±5.7×10⁰ | 1.17×10²±5.5×10⁰ | 1.10×10²±5.0×10⁰ | 9.86×10¹±4.7×10⁰ | 8.69×10¹±4.0×10⁰ | 7.06×10¹±3.2×10⁰ |
| 6.00×10⁻¹±5.0×10⁻³ | 8.34×10¹±4.5×10⁰ | 8.00×10¹±4.2×10⁰ | 7.22×10¹±3.6×10⁰ | 6.63×10¹±3.6×10⁰ | 6.04×10¹±3.1×10⁰ | 4.83×10¹±2.7×10⁰ |
| 7.00×10⁻¹±6.1×10⁻³ | 4.93×10¹±2.6×10⁰ | 5.09×10¹±2.3×10⁰ | 4.79×10¹±2.1×10⁰ | 4.36×10¹±2.0×10⁰ | 3.96×10¹±1.7×10⁰ | 3.09×10¹±1.8×10⁰ |
| 8.00×10⁻¹±7.4×10⁻³ | 3.94×10¹±1.8×10⁰ | 3.81×10¹±1.7×10⁰ | 3.61×10¹±1.6×10⁰ | 3.19×10¹±1.5×10⁰ | 2.63×10¹±1.9×10⁰ | 2.32×10¹±1.3×10⁰ |
| 9.00×10⁻¹±8.7×10⁻³ | 3.64×10¹±2.3×10⁰ | 3.28×10¹±2.1×10⁰ | 2.83×10¹±1.8×10⁰ | 2.69×10¹±1.7×10⁰ | 2.50×10¹±1.7×10⁰ | 2.15×10¹±1.7×10⁰ |
| 1.00×10⁰±1.0×10⁻² | 2.97×10¹±1.3×10⁰ | 2.94×10¹±1.4×10⁰ | 2.61×10¹±1.5×10⁰ | 2.52×10¹±1.3×10⁰ | 2.74×10¹±1.7×10⁰ | 1.80×10¹±1.2×10⁰ |
| 1.20×10⁰±1.3×10⁻² | 2.73×10¹±1.4×10⁰ | 2.44×10¹±1.2×10⁰ | 2.49×10¹±1.2×10⁰ | 2.07×10¹±1.5×10⁰ | 1.85×10¹±1.7×10⁰ | 1.74×10¹±1.2×10⁰ |
| 1.40×10⁰±1.7×10⁻² | 2.93×10¹±1.8×10⁰ | 3.10×10¹±2.0×10⁰ | 3.01×10¹±2.5×10⁰ | 2.49×10¹±1.6×10⁰ | 2.36×10¹±1.4×10⁰ | 2.36×10¹±2.4×10⁰ |
| 1.60×10⁰±2.0×10⁻² | 3.79×10¹±2.1×10⁰ | 3.49×10¹±2.1×10⁰ | 3.76×10¹±2.0×10⁰ | 4.09×10¹±2.5×10⁰ | 4.25×10¹±2.6×10⁰ | 3.68×10¹±1.8×10⁰ |
| 1.80×10⁰±2.4×10⁻² | 5.51×10¹±3.7×10⁰ | 4.65×10¹±3.9×10⁰ | 4.70×10¹±4.5×10⁰ | 5.94×10¹±3.2×10⁰ | 5.74×10¹±3.0×10⁰ | 5.18×10¹±5.3×10⁰ |
| 2.00×10⁰±2.8×10⁻² | 5.71×10¹±3.4×10⁰ | 6.12×10¹±3.1×10⁰ | 5.83×10¹±2.7×10⁰ | 5.44×10¹±3.0×10⁰ | 4.94×10¹±2.8×10⁰ | 4.16×10¹±2.5×10⁰ |
| 2.20×10⁰±3.2×10⁻² | 4.73×10¹±7.0×10⁰ | 5.37×10¹±5.3×10⁰ | 5.64×10¹±4.2×10⁰ | 3.34×10¹±2.6×10⁰ | 2.94×10¹±1.9×10⁰ | 2.84×10¹±3.6×10⁰ |
| 2.40×10⁰±3.7×10⁻² | 1.43×10¹±1.3×10⁰ | 3.39×10¹±4.0×10⁰ | 3.74×10¹±4.4×10⁰ | 1.89×10¹±1.9×10⁰ | 2.22×10¹±2.2×10⁰ | 2.83×10¹±1.4×10¹ |

**Table II (b).** The differential cross sections of the $^{10}$B($n, \alpha$)$^7$Li reaction in the laboratory reference system.

| $E_n$ (MeV) | $\sigma_{E\_bin,\theta}$ (mb/sr) | | | | |
|---|---|---|---|---|---|
| | 78.8° | 90.3° | 101.2° | 112.0° | 122.7° |
| 1.00×10⁻⁶±4.2×10⁻⁹ | 4.96×10⁴±2.1×10³ | 5.09×10⁴±2.1×10³ | 5.20×10⁴±2.4×10³ | 5.09×10⁴±2.1×10³ | 5.16×10⁴±2.3×10³ |
| 1.26×10⁻⁶±5.4×10⁻⁹ | 4.24×10⁴±1.2×10³ | 4.39×10⁴±1.3×10³ | 4.34×10⁴±1.2×10³ | 4.32×10⁴±1.2×10³ | 4.34×10⁴±1.2×10³ |
| 1.58×10⁻⁶±6.8×10⁻⁹ | 3.33×10⁴±9.6×10² | 3.38×10⁴±9.8×10² | 3.38×10⁴±9.8×10² | 3.35×10⁴±9.6×10² | 3.34×10⁴±9.7×10² |
| 2.00×10⁻⁶±8.7×10⁻⁹ | 3.08×10⁴±1.0×10³ | 3.17×10⁴±1.1×10³ | 3.15×10⁴±1.0×10³ | 3.18×10⁴±1.1×10³ | 3.18×10⁴±1.0×10³ |
| 2.51×10⁻⁶±1.1×10⁻⁸ | 2.69×10⁴±1.3×10³ | 2.70×10⁴±1.3×10³ | 2.77×10⁴±1.3×10³ | 2.69×10⁴±1.3×10³ | 2.77×10⁴±1.3×10³ |
| 3.16×10⁻⁶±1.4×10⁻⁸ | 2.99×10⁴±1.1×10³ | 3.11×10⁴±1.1×10³ | 3.07×10⁴±1.1×10³ | 3.11×10⁴±1.1×10³ | 3.03×10⁴±1.1×10³ |
| 3.98×10⁻⁶±1.8×10⁻⁸ | 2.15×10⁴±2.5×10³ | 2.25×10⁴±2.6×10³ | 2.19×10⁴±2.6×10³ | 2.22×10⁴±2.6×10³ | 2.20×10⁴±2.6×10³ |
| 5.01×10⁻⁶±2.3×10⁻⁸ | 2.05×10⁴±1.5×10³ | 2.10×10⁴±1.5×10³ | 2.11×10⁴±1.5×10³ | 2.13×10⁴±1.6×10³ | 2.12×10⁴±1.6×10³ |
| 6.31×10⁻⁶±2.9×10⁻⁸ | 2.08×10⁴±1.1×10³ | 2.10×10⁴±1.1×10³ | 2.12×10⁴±1.1×10³ | 2.09×10⁴±1.1×10³ | 2.16×10⁴±1.2×10³ |
| 7.94×10⁻⁶±3.7×10⁻⁸ | 1.74×10⁴±3.6×10³ | 1.75×10⁴±3.6×10³ | 1.82×10⁴±3.8×10³ | 1.76×10⁴±3.7×10³ | 1.76×10⁴±3.7×10³ |
| 1.00×10⁻⁵±4.7×10⁻⁸ | 1.68×10⁴±6.5×10² | 1.68×10⁴±6.5×10² | 1.69×10⁴±6.5×10² | 1.66×10⁴±6.4×10² | 1.72×10⁴±6.7×10² |
| 1.26×10⁻⁵±6.4×10⁻⁸ | 1.46×10⁴±1.3×10³ | 1.46×10⁴±1.3×10³ | 1.46×10⁴±1.3×10³ | 1.50×10⁴±1.3×10³ | 1.53×10⁴±1.3×10³ |
| 1.58×10⁻⁵±8.6×10⁻⁸ | 1.13×10⁴±1.7×10³ | 1.17×10⁴±1.7×10³ | 1.16×10⁴±1.7×10³ | 1.16×10⁴±1.7×10³ | 1.20×10⁴±1.8×10³ |
| 2.00×10⁻⁵±1.2×10⁻⁷ | 9.29×10³±1.7×10³ | 9.78×10³±1.8×10³ | 9.56×10³±1.7×10³ | 9.23×10³±1.7×10³ | 9.74×10³±1.8×10³ |
| 2.51×10⁻⁵±1.6×10⁻⁷ | 1.06×10⁴±5.2×10² | 1.08×10⁴±5.3×10² | 1.07×10⁴±5.2×10² | 1.06×10⁴±5.2×10² | 1.07×10⁴±5.3×10² |
| 3.16×10⁻⁵±2.1×10⁻⁷ | 7.31×10³±1.6×10³ | 7.41×10³±1.6×10³ | 7.39×10³±1.6×10³ | 7.59×10³±1.6×10³ | 7.48×10³±1.6×10³ |
| 3.98×10⁻⁵±2.7×10⁻⁷ | 7.36×10³±6.8×10² | 7.92×10³±7.2×10² | 7.85×10³±7.2×10² | 7.73×10³±7.0×10² | 7.55×10³±6.9×10² |
| 5.01×10⁻⁵±3.6×10⁻⁷ | 7.09×10³±5.1×10² | 7.26×10³±5.2×10² | 7.36×10³±5.3×10² | 7.31×10³±5.2×10² | 7.22×10³±5.2×10² |
| 6.31×10⁻⁵±4.8×10⁻⁷ | 5.32×10³±5.3×10² | 5.52×10³±5.5×10² | 5.62×10³±5.6×10² | 5.46×10³±5.4×10² | 5.53×10³±5.5×10² |
| 7.94×10⁻⁵±6.4×10⁻⁷ | 5.16×10³±4.9×10² | 5.22×10³±5.0×10² | 5.25×10³±5.0×10² | 5.22×10³±4.9×10² | 5.30×10³±5.0×10² |
| 1.00×10⁻⁴±8.4×10⁻⁷ | 4.55×10³±3.5×10² | 4.69×10³±3.6×10² | 4.75×10³±3.6×10² | 4.81×10³±3.7×10² | 4.78×10³±3.7×10² |
| $1.26\times10^{-4}\pm1.0\times10^{-6}$ | $4.19\times10^{3}\pm4.4\times10^{2}$ | $4.35\times10^{3}\pm4.5\times10^{2}$ | $4.52\times10^{3}\pm4.7\times10^{2}$ | $4.24\times10^{3}\pm4.4\times10^{2}$ | $4.48\times10^{3}\pm4.7\times10^{2}$ |
| $1.58\times10^{-4}\pm1.3\times10^{-6}$ | $3.54\times10^{3}\pm2.8\times10^{2}$ | $3.60\times10^{3}\pm2.8\times10^{2}$ | $3.93\times10^{3}\pm3.0\times10^{2}$ | $3.65\times10^{3}\pm2.8\times10^{2}$ | $3.54\times10^{3}\pm2.7\times10^{2}$ |
| $2.00\times10^{-4}\pm1.6\times10^{-6}$ | $3.24\times10^{3}\pm2.7\times10^{2}$ | $3.25\times10^{3}\pm2.7\times10^{2}$ | $3.27\times10^{3}\pm2.7\times10^{2}$ | $3.35\times10^{3}\pm2.7\times10^{2}$ | $3.16\times10^{3}\pm2.6\times10^{2}$ |
| $2.51\times10^{-4}\pm2.0\times10^{-6}$ | $2.81\times10^{3}\pm2.5\times10^{2}$ | $3.01\times10^{3}\pm2.6\times10^{2}$ | $2.87\times10^{3}\pm2.5\times10^{2}$ | $3.12\times10^{3}\pm2.7\times10^{2}$ | $3.08\times10^{3}\pm2.7\times10^{2}$ |
| $3.16\times10^{-4}\pm2.5\times10^{-6}$ | $2.58\times10^{3}\pm1.8\times10^{2}$ | $2.56\times10^{3}\pm1.7\times10^{2}$ | $2.49\times10^{3}\pm1.6\times10^{2}$ | $2.49\times10^{3}\pm1.6\times10^{2}$ | $2.72\times10^{3}\pm1.8\times10^{2}$ |
| $3.98\times10^{-4}\pm3.1\times10^{-6}$ | $2.36\times10^{3}\pm1.6\times10^{2}$ | $2.47\times10^{3}\pm1.6\times10^{2}$ | $2.52\times10^{3}\pm1.7\times10^{2}$ | $2.44\times10^{3}\pm1.6\times10^{2}$ | $2.51\times10^{3}\pm1.7\times10^{2}$ |
| $5.01\times10^{-4}\pm3.9\times10^{-6}$ | $2.06\times10^{3}\pm1.1\times10^{2}$ | $2.17\times10^{3}\pm1.1\times10^{2}$ | $2.24\times10^{3}\pm1.1\times10^{2}$ | $2.13\times10^{3}\pm1.0\times10^{2}$ | $2.16\times10^{3}\pm1.1\times10^{2}$ |
| $6.31\times10^{-4}\pm4.8\times10^{-6}$ | $1.72\times10^{3}\pm8.2\times10^{1}$ | $1.88\times10^{3}\pm7.7\times10^{1}$ | $1.81\times10^{3}\pm7.4\times10^{1}$ | $1.81\times10^{3}\pm7.4\times10^{1}$ | $1.93\times10^{3}\pm8.0\times10^{1}$ |
| $7.94\times10^{-4}\pm6.1\times10^{-6}$ | $1.51\times10^{3}\pm8.4\times10^{1}$ | $1.67\times10^{3}\pm8.1\times10^{1}$ | $1.72\times10^{3}\pm8.3\times10^{1}$ | $1.71\times10^{3}\pm8.2\times10^{1}$ | $1.65\times10^{3}\pm8.3\times10^{1}$ |
| $1.00\times10^{-3}\pm7.7\times10^{-6}$ | $1.28\times10^{3}\pm6.3\times10^{1}$ | $1.45\times10^{3}\pm5.8\times10^{1}$ | $1.49\times10^{3}\pm5.9\times10^{1}$ | $1.43\times10^{3}\pm5.8\times10^{1}$ | $1.41\times10^{3}\pm5.9\times10^{1}$ |
| $1.26\times10^{-3}\pm9.8\times10^{-6}$ | $1.14\times10^{3}\pm5.1\times10^{1}$ | $1.33\times10^{3}\pm4.4\times10^{1}$ | $1.30\times10^{3}\pm4.1\times10^{1}$ | $1.31\times10^{3}\pm4.0\times10^{1}$ | $1.25\times10^{3}\pm4.2\times10^{1}$ |
| $1.58\times10^{-3}\pm1.3\times10^{-5}$ | $1.06\times10^{3}\pm4.3\times10^{1}$ | $1.18\times10^{3}\pm3.9\times10^{1}$ | $1.14\times10^{3}\pm3.7\times10^{1}$ | $1.14\times10^{3}\pm3.6\times10^{1}$ | $1.17\times10^{3}\pm4.1\times10^{1}$ |
| $2.00\times10^{-3}\pm1.6\times10^{-5}$ | $9.74\times10^{2}\pm4.1\times10^{1}$ | $1.03\times10^{3}\pm3.3\times10^{1}$ | $1.04\times10^{3}\pm3.2\times10^{1}$ | $9.88\times10^{2}\pm3.1\times10^{1}$ | $1.05\times10^{3}\pm3.5\times10^{1}$ |
| $2.51\times10^{-3}\pm2.1\times10^{-5}$ | $9.26\times10^{2}\pm6.0\times10^{1}$ | $9.55\times10^{2}\pm5.4\times10^{1}$ | $9.48\times10^{2}\pm5.4\times10^{1}$ | $9.53\times10^{2}\pm5.4\times10^{1}$ | $9.24\times10^{2}\pm5.4\times10^{1}$ |
| $3.16\times10^{-3}\pm2.7\times10^{-5}$ | $7.90\times10^{2}\pm3.5\times10^{1}$ | $8.05\times10^{2}\pm2.9\times10^{1}$ | $8.67\times10^{2}\pm2.9\times10^{1}$ | $8.83\times10^{2}\pm2.9\times10^{1}$ | $8.82\times10^{2}\pm3.1\times10^{1}$ |
| $3.98\times10^{-3}\pm3.6\times10^{-5}$ | $6.68\times10^{2}\pm3.1\times10^{1}$ | $7.17\times10^{2}\pm2.6\times10^{1}$ | $7.63\times10^{2}\pm2.5\times10^{1}$ | $7.14\times10^{2}\pm2.4\times10^{1}$ | $7.21\times10^{2}\pm2.6\times10^{1}$ |
| $5.01\times10^{-3}\pm4.8\times10^{-5}$ | $6.09\times10^{2}\pm2.6\times10^{1}$ | $6.04\times10^{2}\pm2.2\times10^{1}$ | $6.57\times10^{2}\pm2.2\times10^{1}$ | $6.17\times10^{2}\pm2.1\times10^{1}$ | $6.41\times10^{2}\pm2.3\times10^{1}$ |
| $6.31\times10^{-3}\pm6.4\times10^{-5}$ | $5.23\times10^{2}\pm2.9\times10^{1}$ | $5.91\times10^{2}\pm2.0\times10^{1}$ | $6.01\times10^{2}\pm2.1\times10^{1}$ | $5.86\times10^{2}\pm2.0\times10^{1}$ | $5.60\times10^{2}\pm2.1\times10^{1}$ |
| $7.94\times10^{-3}\pm8.6\times10^{-5}$ | $4.69\times10^{2}\pm2.5\times10^{1}$ | $5.23\times10^{2}\pm2.2\times10^{1}$ | $5.27\times10^{2}\pm2.2\times10^{1}$ | $5.14\times10^{2}\pm2.1\times10^{1}$ | $5.18\times10^{2}\pm2.3\times10^{1}$ |
| $1.00\times10^{-2}\pm1.2\times10^{-4}$ | $4.27\times10^{2}\pm2.5\times10^{1}$ | $5.04\times10^{2}\pm2.5\times10^{1}$ | $4.96\times10^{2}\pm2.5\times10^{1}$ | $4.75\times10^{2}\pm2.3\times10^{1}$ | $4.80\times10^{2}\pm2.4\times10^{1}$ |
| $1.26\times10^{-2}\pm1.6\times10^{-4}$ | $4.27\times10^{2}\pm1.8\times10^{1}$ | $4.39\times10^{2}\pm1.6\times10^{1}$ | $4.12\times10^{2}\pm1.4\times10^{1}$ | $4.19\times10^{2}\pm1.4\times10^{1}$ | $4.07\times10^{2}\pm1.5\times10^{1}$ |
| $1.58\times10^{-2}\pm2.2\times10^{-4}$ | $3.84\times10^{2}\pm1.6\times10^{1}$ | $3.67\times10^{2}\pm1.4\times10^{1}$ | $3.76\times10^{2}\pm1.3\times10^{1}$ | $3.78\times10^{2}\pm1.3\times10^{1}$ | $3.63\times10^{2}\pm1.4\times10^{1}$ |
| $2.00\times10^{-2}\pm1.2\times10^{-4}$ | $3.40\times10^{2}\pm2.0\times10^{1}$ | $3.52\times10^{2}\pm1.7\times10^{1}$ | $3.71\times10^{2}\pm1.6\times10^{1}$ | $3.26\times10^{2}\pm1.5\times10^{1}$ | $3.48\times10^{2}\pm1.7\times10^{1}$ |
| $2.51\times10^{-2}\pm1.5\times10^{-4}$ | $3.11\times10^{2}\pm1.6\times10^{1}$ | $2.93\times10^{2}\pm1.3\times10^{1}$ | $2.97\times10^{2}\pm1.3\times10^{1}$ | $2.91\times10^{2}\pm1.2\times10^{1}$ | $2.97\times10^{2}\pm1.4\times10^{1}$ |
| $3.16\times10^{-2}\pm1.9\times10^{-4}$ | $2.40\times10^{2}\pm1.7\times10^{1}$ | $2.60\times10^{2}\pm1.3\times10^{1}$ | $2.47\times10^{2}\pm1.3\times10^{1}$ | $2.51\times10^{2}\pm1.3\times10^{1}$ | $2.57\times10^{2}\pm1.4\times10^{1}$ |
| $3.98\times10^{-2}\pm2.4\times10^{-4}$ | $2.29\times10^{2}\pm1.4\times10^{1}$ | $2.47\times10^{2}\pm1.2\times10^{1}$ | $2.37\times10^{2}\pm1.1\times10^{1}$ | $2.26\times10^{2}\pm1.1\times10^{1}$ | $2.31\times10^{2}\pm1.2\times10^{1}$ |
| $5.01\times10^{-2}\pm3.0\times10^{-4}$ | $2.12\times10^{2}\pm1.1\times10^{1}$ | $2.08\times10^{2}\pm9.2\times10^{0}$ | $2.12\times10^{2}\pm9.5\times10^{0}$ | $2.01\times10^{2}\pm9.0\times10^{0}$ | $2.01\times10^{2}\pm9.7\times10^{0}$ |
| $6.31\times10^{-2}\pm3.8\times10^{-4}$ | $1.77\times10^{2}\pm8.3\times10^{0}$ | $1.80\times10^{2}\pm6.9\times10^{0}$ | $1.91\times10^{2}\pm7.2\times10^{0}$ | $1.79\times10^{2}\pm6.7\times10^{0}$ | $1.78\times10^{2}\pm7.3\times10^{0}$ |
| $7.94\times10^{-2}\pm4.8\times10^{-4}$ | $1.65\times10^{2}\pm7.8\times10^{0}$ | $1.69\times10^{2}\pm6.6\times10^{0}$ | $1.64\times10^{2}\pm6.6\times10^{0}$ | $1.65\times10^{2}\pm6.5\times10^{0}$ | $1.57\times10^{2}\pm7.0\times10^{0}$ |
| $1.00\times10^{-1}\pm6.1\times10^{-4}$ | $1.61\times10^{2}\pm4.8\times10^{0}$ | $1.58\times10^{2}\pm4.5\times10^{0}$ | $1.48\times10^{2}\pm4.4\times10^{0}$ | $1.47\times10^{2}\pm4.1\times10^{0}$ | $1.43\times10^{2}\pm4.4\times10^{0}$ |
| $2.00\times10^{-1}\pm1.3\times10^{-3}$ | $1.13\times10^{2}\pm3.3\times10^{0}$ | $1.07\times10^{2}\pm3.2\times10^{0}$ | $1.04\times10^{2}\pm3.0\times10^{0}$ | $9.86\times10^{1}\pm2.9\times10^{0}$ | $9.94\times10^{1}\pm3.0\times10^{0}$ |
| $3.00\times10^{-1}\pm2.0\times10^{-3}$ | $8.46\times10^{1}\pm2.5\times10^{0}$ | $7.65\times10^{1}\pm2.3\times10^{0}$ | $7.77\times10^{1}\pm2.3\times10^{0}$ | $7.06\times10^{1}\pm2.1\times10^{0}$ | $7.12\times10^{1}\pm2.3\times10^{0}$ |
| $4.00\times10^{-1}\pm2.9\times10^{-3}$ | $6.96\times10^{1}\pm2.4\times10^{0}$ | $6.33\times10^{1}\pm2.1\times10^{0}$ | $5.69\times10^{1}\pm2.0\times10^{0}$ | $5.18\times10^{1}\pm1.8\times10^{0}$ | $5.08\times10^{1}\pm2.0\times10^{0}$ |
| $5.00\times10^{-1}\pm3.9\times10^{-3}$ | $6.45\times10^{1}\pm3.0\times10^{0}$ | $5.24\times10^{1}\pm2.1\times10^{0}$ | $4.53\times10^{1}\pm2.0\times10^{0}$ | $3.89\times10^{1}\pm1.6\times10^{0}$ | $3.54\times10^{1}\pm1.5\times10^{0}$ |
| $6.00\times10^{-1}\pm5.0\times10^{-3}$ | $4.60\times10^{1}\pm2.6\times10^{0}$ | $3.54\times10^{1}\pm1.7\times10^{0}$ | $3.25\times10^{1}\pm1.7\times10^{0}$ | $2.81\times10^{1}\pm1.4\times10^{0}$ | $2.57\times10^{1}\pm1.3\times10^{0}$ |
| $7.00\times10^{-1}\pm6.1\times10^{-3}$ | $3.09\times10^{1}\pm1.6\times10^{0}$ | $2.43\times10^{1}\pm1.2\times10^{0}$ | $2.42\times10^{1}\pm1.3\times10^{0}$ | $2.02\times10^{1}\pm1.1\times10^{0}$ | $1.98\times10^{1}\pm1.2\times10^{0}$ |
| $8.00\times10^{-1}\pm7.4\times10^{-3}$ | $2.22\times10^{1}\pm1.6\times10^{0}$ | $1.74\times10^{1}\pm1.5\times10^{0}$ | $1.90\times10^{1}\pm1.1\times10^{0}$ | $1.71\times10^{1}\pm1.4\times10^{0}$ | $1.65\times10^{1}\pm1.2\times10^{0}$ |
| $9.00\times10^{-1}\pm8.7\times10^{-3}$ | $2.11\times10^{1}\pm1.5\times10^{0}$ | $1.54\times10^{1}\pm1.2\times10^{0}$ | $1.60\times10^{1}\pm1.2\times10^{0}$ | $1.51\times10^{1}\pm1.1\times10^{0}$ | $1.59\times10^{1}\pm1.3\times10^{0}$ |
| $1.00\times10^{0}\pm1.0\times10^{-2}$ | $2.08\times10^{1}\pm1.9\times10^{0}$ | $2.05\times10^{1}\pm1.9\times10^{0}$ | $1.59\times10^{1}\pm9.6\times10^{-1}$ | $1.68\times10^{1}\pm1.8\times10^{0}$ | $1.49\times10^{1}\pm1.3\times10^{0}$ |
| $1.20\times10^{0}\pm1.3\times10^{-2}$ | $1.31\times10^{1}\pm1.6\times10^{0}$ | $1.30\times10^{1}\pm1.5\times10^{0}$ | $1.48\times10^{1}\pm9.2\times10^{-1}$ | $1.16\times10^{1}\pm1.5\times10^{0}$ | $1.28\times10^{1}\pm1.2\times10^{0}$ |
| $1.40\times10^{0}\pm1.7\times10^{-2}$ | $1.98\times10^{1}\pm1.4\times10^{0}$ | $1.37\times10^{1}\pm1.1\times10^{0}$ | $1.73\times10^{1}\pm1.2\times10^{0}$ | $1.69\times10^{1}\pm1.1\times10^{0}$ | $1.94\times10^{1}\pm1.7\times10^{0}$ |
| $1.60\times10^{0}\pm2.0\times10^{-2}$ | $3.85\times10^{1}\pm2.6\times10^{0}$ | $3.58\times10^{1}\pm2.0\times10^{0}$ | $3.00\times10^{1}\pm1.7\times10^{0}$ | $3.32\times10^{1}\pm2.3\times10^{0}$ | $3.32\times10^{1}\pm2.0\times10^{0}$ |
| $1.80\times10^{0}\pm2.4\times10^{-2}$ | $4.96\times10^{1}\pm3.0\times10^{0}$ | $4.55\times10^{1}\pm2.4\times10^{0}$ | $4.08\times10^{1}\pm2.3\times10^{0}$ | $3.61\times10^{1}\pm2.2\times10^{0}$ | $3.72\times10^{1}\pm3.3\times10^{0}$ |
| $2.00\times10^{0}\pm2.8\times10^{-2}$ | $3.71\times10^{1}\pm2.3\times10^{0}$ | $2.87\times10^{1}\pm2.0\times10^{0}$ | $2.72\times10^{1}\pm2.3\times10^{0}$ | $2.11\times10^{1}\pm1.6\times10^{0}$ | $2.24\times10^{1}\pm2.5\times10^{0}$ |
| $2.20\times10^{0}\pm3.2\times10^{-2}$ | $2.27\times10^{1}\pm2.2\times10^{0}$ | $1.39\times10^{1}\pm1.5\times10^{0}$ | $1.71\times10^{1}\pm1.8\times10^{0}$ | $1.79\times10^{1}\pm1.7\times10^{0}$ | $1.85\times10^{1}\pm3.6\times10^{0}$ |
| 2.40×10⁰±3.7×10⁻² | 1.92×10¹±1.8×10⁰ | 1.22×10¹±1.4×10⁰ | 1.39×10¹±1.4×10⁰ | 2.31×10¹±2.0×10⁰ | 1.69×10¹±6.4×10⁰ |

**Table II (c).** The differential and angle-integrated cross sections of the $^{10}$B$(n, \alpha)^7$Li reaction in the laboratory reference system.

| $E_n$ (MeV) | $\sigma_{E\_bin,\theta}$ (mb/sr) | | | | $\sigma_{E\_bin}$ (mb) |
|---|---|---|---|---|---|
| | 133.2° | 143.5° | 153.1° | 160.8° | |
| 1.00×10⁻⁶±4.2×10⁻⁹ | 5.16×10⁴±2.1×10³ | 5.16×10⁴±2.2×10³ | 5.11×10⁴±2.2×10³ | 5.12×10⁴±2.1×10³ | 6.38×10⁵±2.7×10⁴ |
| 1.26×10⁻⁶±5.4×10⁻⁹ | 4.36×10⁴±1.2×10³ | 4.29×10⁴±1.2×10³ | 4.34×10⁴±1.2×10³ | 4.30×10⁴±1.2×10³ | 5.37×10⁵±1.4×10⁴ |
| 1.58×10⁻⁶±6.8×10⁻⁹ | 3.42×10⁴±9.9×10² | 3.33×10⁴±9.7×10² | 3.35×10⁴±9.9×10² | 3.39×10⁴±1.0×10³ | 4.19×10⁵±1.1×10⁴ |
| 2.00×10⁻⁶±8.7×10⁻⁹ | 3.24×10⁴±1.1×10³ | 3.21×10⁴±1.1×10³ | 3.18×10⁴±1.0×10³ | 3.17×10⁴±1.1×10³ | 3.95×10⁵±1.2×10⁴ |
| 2.51×10⁻⁶±1.1×10⁻⁸ | 2.73×10⁴±1.3×10³ | 2.68×10⁴±1.3×10³ | 2.71×10⁴±1.3×10³ | 2.71×10⁴±1.3×10³ | 3.37×10⁵±1.5×10⁴ |
| 3.16×10⁻⁶±1.4×10⁻⁸ | 3.10×10⁴±1.1×10³ | 3.07×10⁴±1.1×10³ | 3.04×10⁴±1.1×10³ | 2.96×10⁴±1.1×10³ | 3.80×10⁵±1.3×10⁴ |
| 3.98×10⁻⁶±1.8×10⁻⁸ | 2.16×10⁴±2.6×10³ | 2.26×10⁴±2.7×10³ | 2.21×10⁴±2.6×10³ | 2.21×10⁴±2.6×10³ | 2.73×10⁵±3.2×10⁴ |
| 5.01×10⁻⁶±2.3×10⁻⁸ | 2.12×10⁴±1.6×10³ | 2.15×10⁴±1.6×10³ | 2.21×10⁴±1.6×10³ | 2.11×10⁴±1.6×10³ | 2.65×10⁵±1.9×10⁴ |
| 6.31×10⁻⁶±2.9×10⁻⁸ | 2.11×10⁴±1.1×10³ | 2.11×10⁴±1.1×10³ | 2.08×10⁴±1.1×10³ | 2.09×10⁴±1.1×10³ | 2.62×10⁵±1.4×10⁴ |
| 7.94×10⁻⁶±3.7×10⁻⁸ | 1.73×10⁴±3.6×10³ | 1.78×10⁴±3.7×10³ | 1.76×10⁴±3.7×10³ | 1.76×10⁴±3.7×10³ | 2.19×10⁵±4.6×10⁴ |
| 1.00×10⁻⁵±4.7×10⁻⁸ | 1.61×10⁴±6.3×10² | 1.68×10⁴±6.6×10² | 1.74×10⁴±6.8×10² | 1.72×10⁴±6.8×10² | 2.10×10⁵±7.5×10³ |
| 1.26×10⁻⁵±6.4×10⁻⁸ | 1.49×10⁴±1.3×10³ | 1.49×10⁴±1.3×10³ | 1.48×10⁴±1.3×10³ | 1.43×10⁴±1.2×10³ | 1.85×10⁵±1.6×10⁴ |
| 1.58×10⁻⁵±8.6×10⁻⁸ | 1.16×10⁴±1.7×10³ | 1.18×10⁴±1.8×10³ | 1.16×10⁴±1.7×10³ | 1.13×10⁴±1.7×10³ | 1.44×10⁵±2.1×10⁴ |
| 2.00×10⁻⁵±1.2×10⁻⁷ | 9.35×10³±1.7×10³ | 9.67×10³±1.7×10³ | 9.48×10³±1.7×10³ | 9.19×10³±1.7×10³ | 1.19×10⁵±2.1×10⁴ |
| 2.51×10⁻⁵±1.6×10⁻⁷ | 1.07×10⁴±5.3×10² | 1.07×10⁴±5.3×10² | 1.07×10⁴±5.3×10² | 1.06×10⁴±5.4×10² | 1.33×10⁵±6.0×10³ |
| 3.16×10⁻⁵±2.1×10⁻⁷ | 7.60×10³±1.6×10³ | 7.66×10³±1.7×10³ | 7.39×10³±1.6×10³ | 7.35×10³±1.6×10³ | 9.34×10⁴±2.0×10⁴ |
| 3.98×10⁻⁵±2.7×10⁻⁷ | 7.80×10³±7.1×10² | 7.81×10³±7.2×10² | 7.60×10³±7.0×10² | 7.37×10³±6.8×10² | 9.57×10⁴±8.6×10³ |
| 5.01×10⁻⁵±3.6×10⁻⁷ | 7.43×10³±5.3×10² | 7.26×10³±5.2×10² | 7.30×10³±5.2×10² | 7.06×10³±5.1×10² | 9.05×10⁴±6.3×10³ |
| 6.31×10⁻⁵±4.8×10⁻⁷ | 5.30×10³±5.3×10² | 5.39×10³±5.4×10² | 5.56×10³±5.5×10² | 5.43×10³±5.4×10² | 6.80×10⁴±6.7×10³ |
| 7.94×10⁻⁵±6.4×10⁻⁷ | 5.41×10³±5.1×10² | 5.26×10³±5.0×10² | 5.40×10³±5.1×10² | 5.11×10³±4.9×10² | 6.57×10⁴±6.1×10³ |
| 1.00×10⁻⁴±8.4×10⁻⁷ | 4.59×10³±3.5×10² | 4.87×10³±3.7×10² | 4.58×10³±3.6×10² | 4.33×10³±3.5×10² | 5.88×10⁴±4.4×10³ |
| 1.26×10⁻⁴±1.0×10⁻⁶ | 4.34×10³±4.5×10² | 4.20×10³±4.4×10² | 4.13×10³±4.3×10² | 4.01×10³±4.3×10² | 5.36×10⁴±5.5×10³ |
| 1.58×10⁻⁴±1.3×10⁻⁶ | 3.72×10³±2.9×10² | 3.81×10³±2.9×10² | 3.71×10³±2.9×10² | 3.70×10³±2.9×10² | 4.63×10⁴±3.5×10³ |
| 2.00×10⁻⁴±1.6×10⁻⁶ | 3.34×10³±2.8×10² | 3.20×10³±2.7×10² | 3.05×10³±2.6×10² | 3.24×10³±2.8×10² | 4.02×10⁴±3.2×10³ |
| 2.51×10⁻⁴±2.0×10⁻⁶ | 3.01×10³±2.6×10² | 2.88×10³±2.5×10² | 2.96×10³±2.6×10² | 2.75×10³±2.5×10² | 3.69×10⁴±3.1×10³ |
| 3.16×10⁻⁴±2.5×10⁻⁶ | 2.53×10³±1.7×10² | 2.60×10³±1.7×10² | 2.61×10³±1.8×10² | 2.21×10³±1.6×10² | 3.15×10⁴±2.0×10³ |
| 3.98×10⁻⁴±3.1×10⁻⁶ | 2.49×10³±1.7×10² | 2.35×10³±1.6×10² | 2.35×10³±1.6×10² | 2.29×10³±1.7×10² | 3.04×10⁴±1.9×10³ |
| 5.01×10⁻⁴±3.9×10⁻⁶ | 2.23×10³±1.1×10² | 2.20×10³±1.1×10² | 2.20×10³±1.1×10² | 2.17×10³±1.2×10² | 2.72×10⁴±1.2×10³ |
| 6.31×10⁻⁴±4.8×10⁻⁶ | 2.01×10³±8.4×10¹ | 1.93×10³±8.0×10¹ | 1.86×10³±8.0×10¹ | 1.77×10³±9.0×10¹ | 2.33×10⁴±8.4×10² |
| 7.94×10⁻⁴±6.1×10⁻⁶ | 1.70×10³±8.4×10¹ | 1.65×10³±8.3×10¹ | 1.65×10³±8.5×10¹ | 1.47×10³±8.5×10¹ | 2.07×10⁴±9.0×10² |
| 1.00×10⁻³±7.7×10⁻⁶ | 1.47×10³±6.0×10¹ | 1.39×10³±6.0×10¹ | 1.32×10³±5.8×10¹ | 1.40×10³±6.8×10¹ | 1.76×10⁴±6.2×10² |
| 1.26×10⁻³±9.8×10⁻⁶ | 1.24×10³±4.1×10¹ | 1.27×10³±4.3×10¹ | 1.24×10³±4.4×10¹ | 1.21×10³±5.1×10¹ | 1.56×10⁴±3.7×10² |
| 1.58×10⁻³±1.3×10⁻⁵ | 1.13×10³±3.8×10¹ | 1.15×10³±4.0×10¹ | 1.07×10³±3.9×10¹ | 9.97×10²±4.3×10¹ | 1.41×10⁴±3.5×10² |
| 2.00×10⁻³±1.6×10⁻⁵ | 1.00×10³±3.4×10¹ | 9.97×10²±3.5×10¹ | 1.03×10³±3.5×10¹ | 9.91×10²±4.0×10¹ | 1.27×10⁴±2.7×10² |
| 2.51×10⁻³±2.1×10⁻⁵ | 9.23×10²±5.4×10¹ | 8.39×10²±5.0×10¹ | 8.29×10²±5.2×10¹ | 8.79×10²±6.1×10¹ | 1.14×10⁴±6.0×10² |
| 3.16×10⁻³±2.7×10⁻⁵ | 8.66×10²±3.0×10¹ | 8.71×10²±3.1×10¹ | 7.61×10²±3.1×10¹ | 8.17×10²±3.7×10¹ | 1.04×10⁴±2.7×10² |
| 3.98×10⁻³±3.6×10⁻⁵ | 7.54×10²±2.6×10¹ | 6.95×10²±2.5×10¹ | 7.41×10²±2.9×10¹ | 7.40×10²±3.1×10¹ | 9.13×10³±2.1×10² |
| 5.01×10⁻³±4.8×10⁻⁵ | 6.05×10²±2.2×10¹ | 6.16×10²±2.3×10¹ | 5.91×10²±2.4×10¹ | 5.65×10²±2.6×10¹ | 7.69×10³±2.0×10² |
| 6.31×10⁻³±6.4×10⁻⁵ | 5.74×10²±2.1×10¹ | 5.88×10²±2.3×10¹ | 4.75×10²±2.1×10¹ | 5.18×10²±2.5×10¹ | 7.14×10³±1.9×10² |
| 7.94×10⁻³±8.6×10⁻⁵ | 5.23×10²±2.2×10¹ | 4.79×10²±2.2×10¹ | 4.88×10²±2.3×10¹ | 4.59×10²±3.0×10¹ | 6.38×10³±2.1×10² |
| 1.00×10⁻²±1.2×10⁻⁴ | 4.56×10²±2.4×10¹ | 4.57×10²±2.4×10¹ | 4.74×10²±2.5×10¹ | 4.43×10²±2.6×10¹ | 5.99×10³±2.6×10² |
| 1.26×10⁻²±1.6×10⁻⁴ | 4.10×10²±1.5×10¹ | 4.11×10²±1.5×10¹ | 4.32×10²±1.7×10¹ | 4.03×10²±1.8×10¹ | 5.31×10³±1.2×10² |
| 1.58×10⁻²±2.2×10⁻⁴ | 3.71×10²±1.4×10¹ | 3.78×10²±1.4×10¹ | 3.64×10²±1.6×10¹ | 3.34×10²±1.7×10¹ | 4.71×10³±1.1×10² |
| 2.00×10⁻²±1.2×10⁻⁴ | 3.50×10²±1.6×10¹ | 3.35×10²±1.7×10¹ | 3.37×10²±1.7×10¹ | 2.92×10²±2.0×10¹ | 4.33×10³±1.2×10² |
| 2.51×10⁻²±1.5×10⁻⁴ | 2.97×10²±1.3×10¹ | 2.82×10²±1.3×10¹ | 2.89×10²±1.4×10¹ | 2.45×10²±1.6×10¹ | 3.78×10³±1.0×10² |
| 3.16×10⁻²±1.9×10⁻⁴ | 2.45×10²±1.4×10¹ | 2.26×10²±1.4×10¹ | 2.10×10²±1.5×10¹ | 1.99×10²±1.8×10¹ | 3.10×10³±8.7×10¹ |

| $E_n$ (MeV) | | | | | |
|---|---|---|---|---|---|
| 3.98×10⁻²±2.4×10⁻⁴ | 2.21×10²±1.2×10¹ | 2.14×10²±1.2×10¹ | 1.98×10²±1.2×10¹ | 1.88×10²±1.4×10¹ | 2.93×10³±9.0×10¹ |
| 5.01×10⁻²±3.0×10⁻⁴ | 1.93×10²±9.2×10⁰ | 1.96×10²±9.6×10⁰ | 1.90×10²±9.9×10⁰ | 1.72×10²±1.1×10¹ | 2.62×10³±7.5×10¹ |
| 6.31×10⁻²±3.8×10⁻⁴ | 1.72×10²±7.2×10⁰ | 1.70×10²±7.1×10⁰ | 1.59×10²±7.3×10⁰ | 1.63×10²±8.4×10⁰ | 2.35×10³±5.1×10¹ |
| 7.94×10⁻²±4.8×10⁻⁴ | 1.50×10²±6.9×10⁰ | 1.54×10²±6.9×10⁰ | 1.46×10²±7.0×10⁰ | 1.43×10²±8.0×10⁰ | 2.12×10³±4.7×10¹ |
| 1.00×10⁻¹±6.1×10⁻⁴ | 1.41×10²±4.3×10⁰ | 1.37×10²±4.4×10⁰ | 1.34×10²±4.4×10⁰ | 1.32×10²±4.4×10⁰ | 1.96×10³±4.3×10¹ |
| 2.00×10⁻¹±1.3×10⁻³ | 9.55×10¹±3.0×10⁰ | 9.29×10¹±2.9×10⁰ | 8.89×10¹±2.9×10⁰ | 8.48×10¹±3.2×10⁰ | 1.39×10³±3.3×10¹ |
| 3.00×10⁻¹±2.0×10⁻³ | 6.75×10¹±2.1×10⁰ | 6.81×10¹±2.3×10⁰ | 6.57×10¹±2.2×10⁰ | 6.17×10¹±2.3×10⁰ | 1.04×10³±2.2×10¹ |
| 4.00×10⁻¹±2.9×10⁻³ | 4.67×10¹±1.9×10⁰ | 4.41×10¹±1.9×10⁰ | 4.39×10¹±1.9×10⁰ | 4.14×10¹±2.2×10⁰ | 8.68×10²±2.0×10¹ |
| 5.00×10⁻¹±3.9×10⁻³ | 3.30×10¹±1.5×10⁰ | 2.90×10¹±1.4×10⁰ | 2.67×10¹±1.5×10⁰ | 2.51×10¹±1.7×10⁰ | 7.70×10²±2.7×10¹ |
| 6.00×10⁻¹±5.0×10⁻³ | 2.29×10¹±1.2×10⁰ | 2.28×10¹±1.3×10⁰ | 1.86×10¹±1.2×10⁰ | 1.88×10¹±1.7×10⁰ | 5.34×10²±2.2×10¹ |
| 7.00×10⁻¹±6.1×10⁻³ | 1.92×10¹±1.2×10⁰ | 1.68×10¹±1.2×10⁰ | 1.91×10¹±1.5×10⁰ | 1.57×10¹±1.7×10⁰ | 3.67×10²±1.2×10¹ |
| 8.00×10⁻¹±7.4×10⁻³ | 1.71×10¹±1.2×10⁰ | 1.59×10¹±1.2×10⁰ | 1.36×10¹±1.6×10⁰ | 8.73×10⁰±1.5×10⁰ | 2.77×10²±1.1×10¹ |
| 9.00×10⁻¹±8.7×10⁻³ | 1.48×10¹±1.2×10⁰ | 1.56×10¹±1.4×10⁰ | 1.07×10¹±1.4×10⁰ | 1.31×10¹±2.0×10⁰ | 2.49×10²±1.1×10¹ |
| 1.00×10⁰±1.0×10⁻² | 1.27×10¹±9.8×10⁻¹ | 1.16×10¹±1.1×10⁰ | 1.73×10¹±2.6×10⁰ | 1.34×10¹±2.1×10⁰ | 2.46×10²±1.2×10¹ |
| 1.20×10⁰±1.3×10⁻² | 1.26×10¹±1.0×10⁰ | 1.26×10¹±1.0×10⁰ | 9.46×10⁰±1.7×10⁰ | 8.25×10⁰±1.7×10⁰ | 1.96×10²±1.2×10¹ |
| 1.40×10⁰±1.7×10⁻² | 2.06×10¹±1.4×10⁰ | 2.04×10¹±2.1×10⁰ | 1.27×10¹±1.9×10⁰ | 1.23×10¹±1.6×10⁰ | 2.58×10²±1.4×10¹ |
| 1.60×10⁰±2.0×10⁻² | 3.04×10¹±2.0×10⁰ | 2.60×10¹±2.7×10⁰ | 3.28×10¹±3.3×10⁰ | 2.39×10¹±2.5×10⁰ | 4.37×10²±1.5×10¹ |
| 1.80×10⁰±2.4×10⁻² | 3.40×10¹±2.2×10⁰ | 2.83×10¹±4.2×10⁰ | 3.95×10¹±3.0×10⁰ | 3.68×10¹±3.5×10⁰ | 5.61×10²±2.8×10¹ |
| 2.00×10⁰±2.8×10⁻² | 2.18×10¹±2.9×10⁰ | 2.22×10¹±2.3×10⁰ | 1.66×10¹±2.2×10⁰ | 2.45×10¹±2.9×10⁰ | 4.40×10²±1.3×10¹ |
| 2.20×10⁰±3.2×10⁻² | 2.54×10¹±3.0×10⁰ | 2.19×10¹±5.6×10⁰ | 1.50×10¹±6.0×10⁰ | 1.99×10¹±4.4×10⁰ | 3.25×10²±2.4×10¹ |
| 2.40×10⁰±3.7×10⁻² | 2.47×10¹±6.2×10⁰ | 2.84×10¹±5.1×10⁰ | 1.46×10¹±7.7×10⁰ | 1.58×10¹±3.3×10⁰ | 2.65×10²±3.3×10¹ |

**Table III (a).** The differential cross sections of the $^{10}$B$(n, \alpha_0)^7$Li reaction in the laboratory reference system.

| $E_n$ (MeV) | $\sigma^0_{E\_bin,\theta}$ (mb/sr) | | | | | |
|---|---|---|---|---|---|---|
| | 19.2° | 26.9° | 36.5° | 46.7° | 57.3° | 68.0° |
| 1.00×10⁻⁶±4.2×10⁻⁹ | 3.21×10³±1.8×10² | 3.14×10³±1.7×10² | 3.09×10³±1.6×10² | 3.02×10³±1.6×10² | 3.15×10³±1.6×10² | 3.07×10³±1.5×10² |
| 1.26×10⁻⁶±5.4×10⁻⁹ | 2.68×10³±1.4×10² | 2.51×10³±1.2×10² | 2.54×10³±1.2×10² | 2.46×10³±1.2×10² | 2.80×10³±1.3×10² | 2.72×10³±1.3×10² |
| 1.58×10⁻⁶±6.8×10⁻⁹ | 1.93×10³±1.0×10² | 2.08×10³±1.0×10² | 2.07×10³±1.0×10² | 2.13×10³±1.1×10² | 2.07×10³±1.0×10² | 2.37×10³±1.1×10² |
| 2.00×10⁻⁶±8.7×10⁻⁹ | 1.88×10³±1.1×10² | 1.92×10³±1.1×10² | 1.91×10³±1.0×10² | 1.87×10³±1.0×10² | 1.85×10³±1.0×10² | 1.92×10³±1.0×10² |
| 2.51×10⁻⁶±1.1×10⁻⁸ | 1.69×10³±1.2×10² | 1.58×10³±1.1×10² | 1.66×10³±1.1×10² | 1.75×10³±1.1×10² | 1.74×10³±1.1×10² | 1.68×10³±1.1×10² |
| 3.16×10⁻⁶±1.4×10⁻⁸ | 1.89×10³±1.3×10² | 1.87×10³±1.2×10² | 1.76×10³±1.2×10² | 1.88×10³±1.2×10² | 1.94×10³±1.3×10² | 1.89×10³±1.2×10² |
| 3.98×10⁻⁶±1.8×10⁻⁸ | 1.54×10³±2.2×10² | 1.23×10³±1.8×10² | 1.33×10³±1.9×10² | 1.41×10³±2.0×10² | 1.31×10³±1.9×10² | 1.26×10³±1.8×10² |
| 5.01×10⁻⁶±2.3×10⁻⁸ | 1.44×10³±1.3×10² | 1.25×10³±1.2×10² | 1.28×10³±1.2×10² | 1.29×10³±1.2×10² | 1.21×10³±1.1×10² | 1.46×10³±1.3×10² |
| 6.31×10⁻⁶±2.9×10⁻⁸ | 1.30×10³±1.0×10² | 1.23×10³±9.6×10¹ | 1.19×10³±9.3×10¹ | 1.22×10³±9.5×10¹ | 1.28×10³±9.9×10¹ | 1.34×10³±1.0×10² |
| 7.94×10⁻⁶±3.7×10⁻⁸ | 1.02×10³±2.2×10² | 1.12×10³±2.4×10² | 9.69×10²±2.1×10² | 9.98×10²±2.2×10² | 1.08×10³±2.4×10² | 1.25×10³±2.7×10² |
| 1.00×10⁻⁵±4.7×10⁻⁸ | 1.15×10³±8.4×10¹ | 9.53×10²±7.3×10¹ | 9.30×10²±7.2×10¹ | 1.03×10³±7.7×10¹ | 1.06×10³±7.9×10¹ | 1.22×10³±8.5×10¹ |
| 1.26×10⁻⁵±6.4×10⁻⁸ | 9.95×10²±1.1×10² | 1.04×10³±1.1×10² | 9.82×10²±1.0×10² | 8.62×10²±9.2×10¹ | 9.12×10²±9.7×10¹ | 9.31×10²±9.8×10¹ |
| 1.58×10⁻⁵±8.6×10⁻⁸ | 6.91×10²±1.1×10² | 6.65×10²±1.1×10² | 6.02×10²±9.8×10¹ | 6.94×10²±1.1×10² | 6.79×10²±1.1×10² | 7.28×10²±1.2×10² |
| 2.00×10⁻⁵±1.2×10⁻⁷ | 7.02×10²±1.4×10² | 6.60×10²±1.3×10² | 5.88×10²±1.1×10² | 5.38×10²±1.1×10² | 6.39×10²±1.2×10² | 5.92×10²±1.2×10² |
| 2.51×10⁻⁵±1.6×10⁻⁷ | 7.07×10²±6.5×10¹ | 7.09×10²±6.5×10¹ | 6.89×10²±6.3×10¹ | 6.97×10²±6.3×10¹ | 5.97×10²±5.8×10¹ | 7.75×10²±6.8×10¹ |
| 3.16×10⁻⁵±2.1×10⁻⁷ | 4.97×10²±1.1×10² | 4.14×10²±9.5×10¹ | 5.62×10²±1.3×10² | 4.39×10²±1.0×10² | 4.91×10²±1.1×10² | 4.43×10²±1.0×10² |
| 3.98×10⁻⁵±2.7×10⁻⁷ | 5.22×10²±6.4×10¹ | 4.59×10²±5.7×10¹ | 4.70×10²±5.7×10¹ | 4.97×10²±6.0×10¹ | 4.34×10²±5.4×10¹ | 4.11×10²±5.2×10¹ |
| 5.01×10⁻⁵±3.6×10⁻⁷ | 5.03×10²±5.0×10¹ | 4.81×10²±4.8×10¹ | 4.31×10²±4.3×10¹ | 5.11×10²±5.0×10¹ | 4.68×10²±4.6×10¹ | 4.40×10²±4.4×10¹ |
| 6.31×10⁻⁵±4.8×10⁻⁷ | 3.49×10²±4.2×10¹ | 3.21×10²±3.9×10¹ | 3.36×10²±4.1×10¹ | 3.67×10²±4.3×10¹ | 3.24×10²±3.9×10¹ | 3.27×10²±4.0×10¹ |
| 7.94×10⁻⁵±6.4×10⁻⁷ | 3.54×10²±4.3×10¹ | 2.89×10²±3.7×10¹ | 3.37×10²±4.1×10¹ | 3.16×10²±3.9×10¹ | 3.29×10²±4.0×10¹ | 2.66×10²±3.4×10¹ |
| 1.00×10⁻⁴±8.4×10⁻⁷ | 4.25×10²±4.6×10¹ | 3.10×10²±3.5×10¹ | 3.01×10²±3.4×10¹ | 2.89×10²±3.2×10¹ | 2.51×10²±2.9×10¹ | 2.70×10²±3.1×10¹ |
| 1.26×10⁻⁴±1.0×10⁻⁶ | 3.88×10²±5.3×10¹ | 2.56×10²±4.0×10¹ | 3.14×10²±4.4×10¹ | 2.62×10²±3.8×10¹ | 2.75×10²±4.0×10¹ | 2.58×10²±3.9×10¹ |
| 1.58×10⁻⁴±1.3×10⁻⁶ | 3.26×10²±3.7×10¹ | 1.96×10²±2.6×10¹ | 2.22×10²±2.8×10¹ | 2.40×10²±2.9×10¹ | 2.09×10²±2.6×10¹ | 2.49×10²±3.0×10¹ |

| $E_n$ (MeV) | 78.8° | 90.3° | 101.2° | 112.0° | 122.7° | (cont.) |
|---|---|---|---|---|---|---|
| $2.00\times10^{-4}\pm1.6\times10^{-6}$ | $3.06\times10^2\pm3.8\times10^1$ | $1.77\times10^2\pm2.7\times10^1$ | $2.13\times10^2\pm2.9\times10^1$ | $2.12\times10^2\pm2.8\times10^1$ | $1.73\times10^2\pm2.5\times10^1$ | $1.74\times10^2\pm2.5\times10^1$ |
| $2.51\times10^{-4}\pm2.0\times10^{-6}$ | $2.87\times10^2\pm3.5\times10^1$ | $1.76\times10^2\pm2.5\times10^1$ | $2.03\times10^2\pm2.7\times10^1$ | $2.24\times10^2\pm2.8\times10^1$ | $2.07\times10^2\pm2.7\times10^1$ | $2.17\times10^2\pm2.9\times10^1$ |
| $3.16\times10^{-4}\pm2.5\times10^{-6}$ | $2.40\times10^2\pm3.0\times10^1$ | $1.13\times10^2\pm1.9\times10^1$ | $1.33\times10^2\pm2.0\times10^1$ | $1.44\times10^2\pm2.1\times10^1$ | $1.78\times10^2\pm2.4\times10^1$ | $1.16\times10^2\pm1.9\times10^1$ |
| $3.98\times10^{-4}\pm3.1\times10^{-6}$ | $2.74\times10^2\pm3.1\times10^1$ | $1.53\times10^2\pm2.1\times10^1$ | $1.35\times10^2\pm1.9\times10^1$ | $1.42\times10^2\pm1.9\times10^1$ | $1.53\times10^2\pm2.1\times10^1$ | $1.68\times10^2\pm2.3\times10^1$ |
| $5.01\times10^{-4}\pm3.9\times10^{-6}$ | $2.25\times10^2\pm2.3\times10^1$ | $1.51\times10^2\pm1.9\times10^1$ | $1.29\times10^2\pm1.6\times10^1$ | $1.38\times10^2\pm1.5\times10^1$ | $1.32\times10^2\pm1.6\times10^1$ | $1.58\times10^2\pm1.8\times10^1$ |
| $6.31\times10^{-4}\pm4.8\times10^{-6}$ | $1.82\times10^2\pm1.8\times10^1$ | $9.09\times10^1\pm1.3\times10^1$ | $1.18\times10^2\pm1.4\times10^1$ | $1.09\times10^2\pm1.3\times10^1$ | $1.00\times10^2\pm1.2\times10^1$ | $1.31\times10^2\pm1.5\times10^1$ |
| $7.94\times10^{-4}\pm6.1\times10^{-6}$ | $1.71\times10^2\pm2.0\times10^1$ | $8.67\times10^1\pm1.3\times10^1$ | $8.53\times10^1\pm1.2\times10^1$ | $1.32\times10^2\pm1.6\times10^1$ | $1.01\times10^2\pm1.4\times10^1$ | $1.17\times10^2\pm1.6\times10^1$ |
| $1.00\times10^{-3}\pm7.7\times10^{-6}$ | $8.99\times10^1\pm1.2\times10^1$ | $7.08\times10^1\pm1.0\times10^1$ | $9.50\times10^1\pm1.1\times10^1$ | $1.06\times10^2\pm1.2\times10^1$ | $7.27\times10^1\pm9.8\times10^0$ | $9.54\times10^1\pm1.2\times10^1$ |
| $1.26\times10^{-3}\pm9.8\times10^{-6}$ | $4.62\times10^1\pm8.1\times10^0$ | $7.76\times10^1\pm1.0\times10^1$ | $8.09\times10^1\pm1.0\times10^1$ | $8.16\times10^1\pm9.7\times10^0$ | $7.37\times10^1\pm9.6\times10^0$ | $9.38\times10^1\pm1.2\times10^1$ |
| $1.58\times10^{-3}\pm1.3\times10^{-5}$ | $7.58\times10^1\pm9.5\times10^0$ | $5.97\times10^1\pm8.7\times10^0$ | $7.29\times10^1\pm8.6\times10^0$ | $7.98\times10^1\pm9.2\times10^0$ | $7.24\times10^1\pm8.9\times10^0$ | $8.72\times10^1\pm1.0\times10^1$ |
| $2.00\times10^{-3}\pm1.6\times10^{-5}$ | $7.46\times10^1\pm9.4\times10^0$ | $7.33\times10^1\pm9.2\times10^0$ | $8.29\times10^1\pm9.4\times10^0$ | $7.05\times10^1\pm8.4\times10^0$ | $9.39\times10^1\pm9.9\times10^0$ | $7.08\times10^1\pm9.5\times10^0$ |
| $2.51\times10^{-3}\pm2.1\times10^{-5}$ | $6.67\times10^1\pm1.1\times10^1$ | $5.51\times10^1\pm9.1\times10^0$ | $7.05\times10^1\pm9.9\times10^0$ | $6.87\times10^1\pm9.4\times10^0$ | $7.15\times10^1\pm1.0\times10^1$ | $3.60\times10^1\pm7.4\times10^0$ |
| $3.16\times10^{-3}\pm2.7\times10^{-5}$ | $4.64\times10^1\pm7.0\times10^0$ | $6.07\times10^1\pm8.3\times10^0$ | $5.93\times10^1\pm7.6\times10^0$ | $5.55\times10^1\pm7.2\times10^0$ | $5.37\times10^1\pm7.2\times10^0$ | $4.34\times10^1\pm7.1\times10^0$ |
| $3.98\times10^{-3}\pm3.6\times10^{-5}$ | $4.85\times10^1\pm7.3\times10^0$ | $4.70\times10^1\pm6.6\times10^0$ | $4.51\times10^1\pm6.4\times10^0$ | $5.18\times10^1\pm6.6\times10^0$ | $6.05\times10^1\pm7.4\times10^0$ | $4.67\times10^1\pm6.9\times10^0$ |
| $5.01\times10^{-3}\pm4.8\times10^{-5}$ | $3.24\times10^1\pm5.1\times10^0$ | $5.43\times10^1\pm6.7\times10^0$ | $5.22\times10^1\pm6.1\times10^0$ | $4.20\times10^1\pm5.2\times10^0$ | $4.14\times10^1\pm5.5\times10^0$ | $4.62\times10^1\pm6.4\times10^0$ |
| $6.31\times10^{-3}\pm6.4\times10^{-5}$ | $3.96\times10^1\pm5.8\times10^0$ | $3.95\times10^1\pm5.6\times10^0$ | $3.52\times10^1\pm5.1\times10^0$ | $4.08\times10^1\pm5.3\times10^0$ | $3.86\times10^1\pm5.4\times10^0$ | $4.34\times10^1\pm6.0\times10^0$ |
| $7.94\times10^{-3}\pm8.6\times10^{-5}$ | $1.95\times10^1\pm4.0\times10^0$ | $4.51\times10^1\pm6.1\times10^0$ | $3.54\times10^1\pm5.4\times10^0$ | $3.00\times10^1\pm4.4\times10^0$ | $3.43\times10^1\pm5.0\times10^0$ | $3.76\times10^1\pm5.8\times10^0$ |
| $1.00\times10^{-2}\pm1.2\times10^{-4}$ | $3.86\times10^1\pm5.3\times10^0$ | $3.84\times10^1\pm5.2\times10^0$ | $2.67\times10^1\pm4.0\times10^0$ | $3.44\times10^1\pm4.5\times10^0$ | $2.83\times10^1\pm4.3\times10^0$ | $3.12\times10^1\pm4.7\times10^0$ |
| $1.26\times10^{-2}\pm1.6\times10^{-4}$ | $1.78\times10^1\pm3.3\times10^0$ | $4.00\times10^1\pm4.7\times10^0$ | $2.72\times10^1\pm3.7\times10^0$ | $3.24\times10^1\pm3.9\times10^0$ | $2.88\times10^1\pm3.9\times10^0$ | $2.36\times10^1\pm3.6\times10^0$ |
| $1.58\times10^{-2}\pm2.2\times10^{-4}$ | $2.90\times10^1\pm4.0\times10^0$ | $2.95\times10^1\pm3.8\times10^0$ | $3.04\times10^1\pm3.8\times10^0$ | $3.39\times10^1\pm4.0\times10^0$ | $2.91\times10^1\pm3.7\times10^0$ | $2.47\times10^1\pm3.6\times10^0$ |
| $2.00\times10^{-2}\pm1.2\times10^{-4}$ | $2.41\times10^1\pm3.3\times10^0$ | $2.50\times10^1\pm3.3\times10^0$ | $2.52\times10^1\pm3.2\times10^0$ | $2.24\times10^1\pm2.9\times10^0$ | $3.08\times10^1\pm3.6\times10^0$ | $2.50\times10^1\pm3.5\times10^0$ |
| $2.51\times10^{-2}\pm1.5\times10^{-4}$ | $2.15\times10^1\pm2.5\times10^0$ | $2.02\times10^1\pm2.4\times10^0$ | $2.39\times10^1\pm2.6\times10^0$ | $2.21\times10^1\pm2.4\times10^0$ | $2.08\times10^1\pm2.4\times10^0$ | $1.36\times10^1\pm2.0\times10^0$ |
| $3.16\times10^{-2}\pm1.9\times10^{-4}$ | $2.32\times10^1\pm3.4\times10^0$ | $1.44\times10^1\pm2.4\times10^0$ | $2.03\times10^1\pm2.9\times10^0$ | $1.82\times10^1\pm2.8\times10^0$ | $2.45\times10^1\pm3.5\times10^0$ | $1.91\times10^1\pm3.2\times10^0$ |
| $3.98\times10^{-2}\pm2.4\times10^{-4}$ | $1.65\times10^1\pm2.1\times10^0$ | $1.87\times10^1\pm2.4\times10^0$ | $1.72\times10^1\pm2.1\times10^0$ | $1.75\times10^1\pm2.1\times10^0$ | $2.30\times10^1\pm2.6\times10^0$ | $1.58\times10^1\pm2.3\times10^0$ |
| $5.01\times10^{-2}\pm3.0\times10^{-4}$ | $1.39\times10^1\pm1.7\times10^0$ | $1.62\times10^1\pm1.8\times10^0$ | $1.86\times10^1\pm1.9\times10^0$ | $1.64\times10^1\pm1.7\times10^0$ | $1.73\times10^1\pm1.8\times10^0$ | $1.57\times10^1\pm1.9\times10^0$ |
| $6.31\times10^{-2}\pm3.8\times10^{-4}$ | $1.63\times10^1\pm1.5\times10^0$ | $1.55\times10^1\pm1.5\times10^0$ | $1.62\times10^1\pm1.4\times10^0$ | $1.58\times10^1\pm1.4\times10^0$ | $1.60\times10^1\pm1.5\times10^0$ | $1.40\times10^1\pm1.4\times10^0$ |
| $7.94\times10^{-2}\pm4.8\times10^{-4}$ | $1.29\times10^1\pm1.3\times10^0$ | $1.39\times10^1\pm1.3\times10^0$ | $1.69\times10^1\pm1.5\times10^0$ | $1.36\times10^1\pm1.3\times10^0$ | $1.39\times10^1\pm1.4\times10^0$ | $1.13\times10^1\pm1.3\times10^0$ |
| $1.00\times10^{-1}\pm6.1\times10^{-4}$ | $1.50\times10^1\pm8.0\times10^{-1}$ | $1.56\times10^1\pm8.2\times10^{-1}$ | $1.55\times10^1\pm8.0\times10^{-1}$ | $1.37\times10^1\pm7.5\times10^{-1}$ | $1.32\times10^1\pm7.4\times10^{-1}$ | $1.21\times10^1\pm7.4\times10^{-1}$ |
| $2.00\times10^{-1}\pm1.3\times10^{-3}$ | $1.37\times10^1\pm7.0\times10^{-1}$ | $1.35\times10^1\pm6.3\times10^{-1}$ | $1.38\times10^1\pm6.4\times10^{-1}$ | $1.25\times10^1\pm6.0\times10^{-1}$ | $1.22\times10^1\pm6.1\times10^{-1}$ | $1.23\times10^1\pm6.2\times10^{-1}$ |
| $3.00\times10^{-1}\pm2.0\times10^{-3}$ | $1.04\times10^1\pm5.3\times10^{-1}$ | $1.14\times10^1\pm5.6\times10^{-1}$ | $1.07\times10^1\pm5.2\times10^{-1}$ | $1.07\times10^1\pm5.2\times10^{-1}$ | $1.09\times10^1\pm5.8\times10^{-1}$ | $1.00\times10^1\pm5.4\times10^{-1}$ |
| $4.00\times10^{-1}\pm2.9\times10^{-3}$ | $1.21\times10^1\pm7.4\times10^{-1}$ | $1.10\times10^1\pm5.8\times10^{-1}$ | $1.08\times10^1\pm6.1\times10^{-1}$ | $1.12\times10^1\pm6.0\times10^{-1}$ | $1.08\times10^1\pm5.8\times10^{-1}$ | $1.05\times10^1\pm5.8\times10^{-1}$ |
| $5.00\times10^{-1}\pm3.9\times10^{-3}$ | $1.37\times10^1\pm1.0\times10^0$ | $1.37\times10^1\pm8.9\times10^{-1}$ | $1.36\times10^1\pm8.4\times10^{-1}$ | $1.27\times10^1\pm8.6\times10^{-1}$ | $1.21\times10^1\pm7.6\times10^{-1}$ | $1.23\times10^1\pm7.9\times10^{-1}$ |
| $6.00\times10^{-1}\pm5.0\times10^{-3}$ | $1.17\times10^1\pm9.4\times10^{-1}$ | $1.22\times10^1\pm8.7\times10^{-1}$ | $1.28\times10^1\pm1.0\times10^0$ | $9.44\times10^0\pm7.5\times10^{-1}$ | $1.01\times10^1\pm7.3\times10^{-1}$ | $9.99\times10^0\pm7.7\times10^{-1}$ |
| $7.00\times10^{-1}\pm6.1\times10^{-3}$ | $1.32\times10^1\pm2.9\times10^0$ | $1.15\times10^1\pm1.6\times10^0$ | $1.66\times10^1\pm2.3\times10^0$ | $8.47\times10^0\pm1.4\times10^0$ | $8.95\times10^0\pm1.5\times10^0$ | $7.84\times10^0\pm7.1\times10^{-1}$ |
| $8.00\times10^{-1}\pm7.4\times10^{-3}$ | $1.17\times10^1\pm2.1\times10^0$ | $1.19\times10^1\pm4.4\times10^0$ | $1.12\times10^1\pm1.6\times10^0$ | $1.32\times10^1\pm3.3\times10^0$ | $9.69\times10^0\pm2.0\times10^0$ | $8.43\times10^0\pm1.4\times10^0$ |
| $9.00\times10^{-1}\pm8.7\times10^{-3}$ | $1.60\times10^1\pm4.5\times10^0$ | $1.21\times10^1\pm3.3\times10^0$ | $9.81\times10^0\pm3.2\times10^0$ | $9.84\times10^0\pm2.8\times10^0$ | $8.74\times10^0\pm8.7\times10^{-1}$ | $9.00\times10^0\pm9.7\times10^{-1}$ |

**Table III (b).** The differential cross sections of the $^{10}$B($n$, $\alpha_0$)$^7$Li reaction in the laboratory reference system.

| $E_n$ (MeV) | $\sigma^0_{E\_bin,\theta}$ (mb/sr) | | | | |
|---|---|---|---|---|---|
| | 78.8° | 90.3° | 101.2° | 112.0° | 122.7° |
| $1.00\times10^{-6}\pm4.2\times10^{-9}$ | $3.08\times10^3\pm1.6\times10^2$ | $3.17\times10^3\pm1.7\times10^2$ | $3.02\times10^3\pm1.6\times10^2$ | $3.18\times10^3\pm1.6\times10^2$ | $3.20\times10^3\pm1.8\times10^2$ |
| $1.26\times10^{-6}\pm5.4\times10^{-9}$ | $2.79\times10^3\pm1.3\times10^2$ | $2.74\times10^3\pm1.3\times10^2$ | $2.65\times10^3\pm1.3\times10^2$ | $2.49\times10^3\pm1.2\times10^2$ | $2.92\times10^3\pm1.4\times10^2$ |
| $1.58\times10^{-6}\pm6.8\times10^{-9}$ | $2.15\times10^3\pm1.1\times10^2$ | $1.96\times10^3\pm1.0\times10^2$ | $2.14\times10^3\pm1.1\times10^2$ | $1.98\times10^3\pm9.9\times10^1$ | $2.16\times10^3\pm1.1\times10^2$ |
| $2.00\times10^{-6}\pm8.7\times10^{-9}$ | $1.97\times10^3\pm1.1\times10^2$ | $1.85\times10^3\pm1.0\times10^2$ | $1.84\times10^3\pm1.0\times10^2$ | $1.98\times10^3\pm1.1\times10^2$ | $1.90\times10^3\pm1.1\times10^2$ |
| $2.51\times10^{-6}\pm1.1\times10^{-8}$ | $1.64\times10^3\pm1.1\times10^2$ | $1.70\times10^3\pm1.1\times10^2$ | $1.57\times10^3\pm1.0\times10^2$ | $1.67\times10^3\pm1.1\times10^2$ | $1.80\times10^3\pm1.2\times10^2$ |
| $3.16\times10^{-6}\pm1.4\times10^{-8}$ | $1.95\times10^3\pm1.3\times10^2$ | $2.00\times10^3\pm1.3\times10^2$ | $1.89\times10^3\pm1.2\times10^2$ | $1.73\times10^3\pm1.1\times10^2$ | $1.94\times10^3\pm1.3\times10^2$ |
| $3.98\times10^{-6}\pm1.8\times10^{-8}$ | $1.35\times10^3\pm2.0\times10^2$ | $1.58\times10^3\pm2.2\times10^2$ | $1.31\times10^3\pm1.9\times10^2$ | $1.30\times10^3\pm1.8\times10^2$ | $1.35\times10^3\pm2.0\times10^2$ |
| $5.01\times10^{-6}\pm2.3\times10^{-8}$ | $1.25\times10^3\pm1.2\times10^2$ | $1.22\times10^3\pm1.1\times10^2$ | $1.29\times10^3\pm1.2\times10^2$ | $1.29\times10^3\pm1.2\times10^2$ | $1.28\times10^3\pm1.2\times10^2$ |
| $6.31\times10^{-6}\pm2.9\times10^{-8}$ | $1.25\times10^3\pm9.7\times10^1$ | $1.24\times10^3\pm9.5\times10^1$ | $1.37\times10^3\pm1.0\times10^2$ | $1.19\times10^3\pm9.2\times10^1$ | $1.35\times10^3\pm1.1\times10^2$ |
| $7.94\times10^{-6}\pm3.7\times10^{-8}$ | $1.01\times10^{3}\pm2.2\times10^{2}$ | $1.03\times10^{3}\pm2.3\times10^{2}$ | $1.17\times10^{3}\pm2.5\times10^{2}$ | $9.94\times10^{2}\pm2.2\times10^{2}$ | $1.05\times10^{3}\pm2.3\times10^{2}$ |
| $1.00\times10^{-5}\pm4.7\times10^{-8}$ | $1.04\times10^{3}\pm7.8\times10^{1}$ | $1.05\times10^{3}\pm7.6\times10^{1}$ | $1.02\times10^{3}\pm7.6\times10^{1}$ | $1.03\times10^{3}\pm7.5\times10^{1}$ | $1.10\times10^{3}\pm8.4\times10^{1}$ |
| $1.26\times10^{-5}\pm6.4\times10^{-8}$ | $9.36\times10^{2}\pm1.0\times10^{2}$ | $9.84\times10^{2}\pm1.0\times10^{2}$ | $8.60\times10^{2}\pm9.2\times10^{1}$ | $9.22\times10^{2}\pm9.7\times10^{1}$ | $1.07\times10^{3}\pm1.1\times10^{2}$ |
| $1.58\times10^{-5}\pm8.6\times10^{-8}$ | $7.08\times10^{2}\pm1.1\times10^{2}$ | $7.38\times10^{2}\pm1.2\times10^{2}$ | $6.15\times10^{2}\pm1.0\times10^{2}$ | $7.51\times10^{2}\pm1.2\times10^{2}$ | $6.81\times10^{2}\pm1.1\times10^{2}$ |
| $2.00\times10^{-5}\pm1.2\times10^{-7}$ | $5.65\times10^{2}\pm1.1\times10^{2}$ | $5.38\times10^{2}\pm1.0\times10^{2}$ | $5.75\times10^{2}\pm1.1\times10^{2}$ | $5.45\times10^{2}\pm1.1\times10^{2}$ | $6.38\times10^{2}\pm1.2\times10^{2}$ |
| $2.51\times10^{-5}\pm1.6\times10^{-7}$ | $6.21\times10^{2}\pm6.0\times10^{1}$ | $6.97\times10^{2}\pm6.2\times10^{1}$ | $6.01\times10^{2}\pm5.6\times10^{1}$ | $6.01\times10^{2}\pm5.5\times10^{1}$ | $7.75\times10^{2}\pm7.1\times10^{1}$ |
| $3.16\times10^{-5}\pm2.1\times10^{-7}$ | $4.54\times10^{2}\pm1.0\times10^{2}$ | $4.15\times10^{2}\pm9.4\times10^{1}$ | $4.40\times10^{2}\pm1.0\times10^{2}$ | $4.32\times10^{2}\pm9.8\times10^{1}$ | $4.89\times10^{2}\pm1.1\times10^{2}$ |
| $3.98\times10^{-5}\pm2.7\times10^{-7}$ | $4.07\times10^{2}\pm5.3\times10^{1}$ | $5.27\times10^{2}\pm6.1\times10^{1}$ | $5.02\times10^{2}\pm5.9\times10^{1}$ | $4.75\times10^{2}\pm5.6\times10^{1}$ | $4.90\times10^{2}\pm6.1\times10^{1}$ |
| $5.01\times10^{-5}\pm3.6\times10^{-7}$ | $4.11\times10^{2}\pm4.3\times10^{1}$ | $4.81\times10^{2}\pm4.7\times10^{1}$ | $4.72\times10^{2}\pm4.6\times10^{1}$ | $4.91\times10^{2}\pm4.7\times10^{1}$ | $4.35\times10^{2}\pm4.4\times10^{1}$ |
| $6.31\times10^{-5}\pm4.8\times10^{-7}$ | $2.79\times10^{2}\pm3.5\times10^{1}$ | $3.16\times10^{2}\pm3.8\times10^{1}$ | $3.69\times10^{2}\pm4.3\times10^{1}$ | $3.17\times10^{2}\pm3.8\times10^{1}$ | $3.49\times10^{2}\pm4.2\times10^{1}$ |
| $7.94\times10^{-5}\pm6.4\times10^{-7}$ | $3.04\times10^{2}\pm3.9\times10^{1}$ | $2.87\times10^{2}\pm3.5\times10^{1}$ | $3.46\times10^{2}\pm4.1\times10^{1}$ | $2.66\times10^{2}\pm3.3\times10^{1}$ | $3.34\times10^{2}\pm4.1\times10^{1}$ |
| $1.00\times10^{-4}\pm8.4\times10^{-7}$ | $2.68\times10^{2}\pm3.2\times10^{1}$ | $2.73\times10^{2}\pm3.0\times10^{1}$ | $3.08\times10^{2}\pm3.3\times10^{1}$ | $3.35\times10^{2}\pm3.5\times10^{1}$ | $3.17\times10^{2}\pm3.5\times10^{1}$ |
| $1.26\times10^{-4}\pm1.0\times10^{-6}$ | $2.53\times10^{2}\pm3.9\times10^{1}$ | $2.63\times10^{2}\pm3.6\times10^{1}$ | $3.42\times10^{2}\pm4.5\times10^{1}$ | $2.75\times10^{2}\pm3.8\times10^{1}$ | $2.35\times10^{2}\pm3.5\times10^{1}$ |
| $1.58\times10^{-4}\pm1.3\times10^{-6}$ | $2.30\times10^{2}\pm3.0\times10^{1}$ | $1.95\times10^{2}\pm2.3\times10^{1}$ | $2.70\times10^{2}\pm3.0\times10^{1}$ | $1.98\times10^{2}\pm2.4\times10^{1}$ | $2.26\times10^{2}\pm2.7\times10^{1}$ |
| $2.00\times10^{-4}\pm1.6\times10^{-6}$ | $2.29\times10^{2}\pm3.2\times10^{1}$ | $1.76\times10^{2}\pm2.3\times10^{1}$ | $1.77\times10^{2}\pm2.4\times10^{1}$ | $1.91\times10^{2}\pm2.4\times10^{1}$ | $2.12\times10^{2}\pm2.8\times10^{1}$ |
| $2.51\times10^{-4}\pm2.0\times10^{-6}$ | $1.78\times10^{2}\pm2.7\times10^{1}$ | $1.77\times10^{2}\pm2.2\times10^{1}$ | $2.06\times10^{2}\pm2.6\times10^{1}$ | $1.62\times10^{2}\pm2.1\times10^{1}$ | $1.91\times10^{2}\pm2.5\times10^{1}$ |
| $3.16\times10^{-4}\pm2.5\times10^{-6}$ | $2.28\times10^{2}\pm3.1\times10^{1}$ | $1.24\times10^{2}\pm1.7\times10^{1}$ | $1.25\times10^{2}\pm1.7\times10^{1}$ | $1.66\times10^{2}\pm2.0\times10^{1}$ | $1.74\times10^{2}\pm2.2\times10^{1}$ |
| $3.98\times10^{-4}\pm3.1\times10^{-6}$ | $1.82\times10^{2}\pm2.6\times10^{1}$ | $1.41\times10^{2}\pm1.7\times10^{1}$ | $1.46\times10^{2}\pm1.8\times10^{1}$ | $1.55\times10^{2}\pm1.8\times10^{1}$ | $2.03\times10^{2}\pm2.3\times10^{1}$ |
| $5.01\times10^{-4}\pm3.9\times10^{-6}$ | $1.26\times10^{2}\pm1.7\times10^{1}$ | $1.09\times10^{2}\pm1.2\times10^{1}$ | $1.29\times10^{2}\pm1.4\times10^{1}$ | $1.32\times10^{2}\pm1.4\times10^{1}$ | $1.27\times10^{2}\pm1.5\times10^{1}$ |
| $6.31\times10^{-4}\pm4.8\times10^{-6}$ | $7.23\times10^{1}\pm1.2\times10^{1}$ | $8.99\times10^{1}\pm1.0\times10^{1}$ | $8.78\times10^{1}\pm1.0\times10^{1}$ | $1.08\times10^{2}\pm1.2\times10^{1}$ | $1.23\times10^{2}\pm1.3\times10^{1}$ |
| $7.94\times10^{-4}\pm6.1\times10^{-6}$ | $9.27\times10^{1}\pm1.5\times10^{1}$ | $1.21\times10^{2}\pm1.3\times10^{1}$ | $1.01\times10^{2}\pm1.2\times10^{1}$ | $1.21\times10^{2}\pm1.3\times10^{1}$ | $8.99\times10^{1}\pm1.2\times10^{1}$ |
| $1.00\times10^{-3}\pm7.7\times10^{-6}$ | $6.06\times10^{1}\pm1.0\times10^{1}$ | $9.54\times10^{1}\pm9.8\times10^{0}$ | $9.75\times10^{1}\pm1.0\times10^{1}$ | $9.22\times10^{1}\pm9.7\times10^{0}$ | $8.81\times10^{1}\pm1.0\times10^{1}$ |
| $1.26\times10^{-3}\pm9.8\times10^{-6}$ | $5.13\times10^{1}\pm9.4\times10^{0}$ | $8.21\times10^{1}\pm8.7\times10^{0}$ | $6.73\times10^{1}\pm7.9\times10^{0}$ | $1.05\times10^{2}\pm9.9\times10^{0}$ | $8.41\times10^{1}\pm9.8\times10^{0}$ |
| $1.58\times10^{-3}\pm1.3\times10^{-5}$ | $5.74\times10^{1}\pm9.2\times10^{0}$ | $7.19\times10^{1}\pm7.6\times10^{0}$ | $7.16\times10^{1}\pm7.7\times10^{0}$ | $5.43\times10^{1}\pm6.4\times10^{0}$ | $8.05\times10^{1}\pm8.9\times10^{0}$ |
| $2.00\times10^{-3}\pm1.6\times10^{-5}$ | $7.68\times10^{1}\pm1.1\times10^{1}$ | $7.01\times10^{1}\pm7.3\times10^{0}$ | $6.22\times10^{1}\pm6.9\times10^{0}$ | $6.46\times10^{1}\pm7.0\times10^{0}$ | $6.33\times10^{1}\pm7.5\times10^{0}$ |
| $2.51\times10^{-3}\pm2.1\times10^{-5}$ | $7.30\times10^{1}\pm1.2\times10^{1}$ | $6.14\times10^{1}\pm7.6\times10^{0}$ | $4.63\times10^{1}\pm6.7\times10^{0}$ | $7.00\times10^{1}\pm8.3\times10^{0}$ | $6.19\times10^{1}\pm8.4\times10^{0}$ |
| $3.16\times10^{-3}\pm2.7\times10^{-5}$ | $5.88\times10^{1}\pm9.0\times10^{0}$ | $5.12\times10^{1}\pm6.2\times10^{0}$ | $7.31\times10^{1}\pm7.3\times10^{0}$ | $5.61\times10^{1}\pm6.2\times10^{0}$ | $5.25\times10^{1}\pm6.6\times10^{0}$ |
| $3.98\times10^{-3}\pm3.6\times10^{-5}$ | $5.01\times10^{1}\pm8.1\times10^{0}$ | $5.60\times10^{1}\pm6.1\times10^{0}$ | $3.76\times10^{1}\pm4.9\times10^{0}$ | $3.91\times10^{1}\pm5.0\times10^{0}$ | $4.42\times10^{1}\pm5.9\times10^{0}$ |
| $5.01\times10^{-3}\pm4.8\times10^{-5}$ | $4.41\times10^{1}\pm6.4\times10^{0}$ | $4.04\times10^{1}\pm4.6\times10^{0}$ | $3.94\times10^{1}\pm4.5\times10^{0}$ | $4.46\times10^{1}\pm4.8\times10^{0}$ | $4.24\times10^{1}\pm5.1\times10^{0}$ |
| $6.31\times10^{-3}\pm6.4\times10^{-5}$ | $2.28\times10^{1}\pm5.0\times10^{0}$ | $3.08\times10^{1}\pm4.0\times10^{0}$ | $3.82\times10^{1}\pm4.6\times10^{0}$ | $4.21\times10^{1}\pm4.8\times10^{0}$ | $2.98\times10^{1}\pm4.3\times10^{0}$ |
| $7.94\times10^{-3}\pm8.6\times10^{-5}$ | $2.51\times10^{1}\pm5.2\times10^{0}$ | $3.06\times10^{1}\pm4.1\times10^{0}$ | $3.33\times10^{1}\pm4.3\times10^{0}$ | $3.05\times10^{1}\pm4.0\times10^{0}$ | $3.19\times10^{1}\pm4.5\times10^{0}$ |
| $1.00\times10^{-2}\pm1.2\times10^{-4}$ | $2.79\times10^{1}\pm5.0\times10^{0}$ | $2.47\times10^{1}\pm3.3\times10^{0}$ | $2.67\times10^{1}\pm3.6\times10^{0}$ | $3.43\times10^{1}\pm4.1\times10^{0}$ | $3.23\times10^{1}\pm4.3\times10^{0}$ |
| $1.26\times10^{-2}\pm1.6\times10^{-4}$ | $3.80\times10^{1}\pm5.2\times10^{0}$ | $2.47\times10^{1}\pm3.1\times10^{0}$ | $3.00\times10^{1}\pm3.5\times10^{0}$ | $2.39\times10^{1}\pm3.0\times10^{0}$ | $2.30\times10^{1}\pm3.3\times10^{0}$ |
| $1.58\times10^{-2}\pm2.2\times10^{-4}$ | $2.46\times10^{1}\pm3.9\times10^{0}$ | $2.08\times10^{1}\pm2.7\times10^{0}$ | $2.22\times10^{1}\pm2.9\times10^{0}$ | $2.49\times10^{1}\pm2.9\times10^{0}$ | $1.96\times10^{1}\pm2.9\times10^{0}$ |
| $2.00\times10^{-2}\pm1.2\times10^{-4}$ | $2.20\times10^{1}\pm3.5\times10^{0}$ | $1.81\times10^{1}\pm2.3\times10^{0}$ | $2.55\times10^{1}\pm2.8\times10^{0}$ | $1.37\times10^{1}\pm2.0\times10^{0}$ | $2.31\times10^{1}\pm2.9\times10^{0}$ |
| $2.51\times10^{-2}\pm1.5\times10^{-4}$ | $2.22\times10^{1}\pm2.8\times10^{0}$ | $1.77\times10^{1}\pm2.0\times10^{0}$ | $1.85\times10^{1}\pm2.1\times10^{0}$ | $1.49\times10^{1}\pm1.8\times10^{0}$ | $1.92\times10^{1}\pm2.3\times10^{0}$ |
| $3.16\times10^{-2}\pm1.9\times10^{-4}$ | $1.27\times10^{1}\pm2.8\times10^{0}$ | $1.60\times10^{1}\pm2.2\times10^{0}$ | $1.66\times10^{1}\pm2.3\times10^{0}$ | $1.58\times10^{1}\pm2.2\times10^{0}$ | $1.55\times10^{1}\pm2.4\times10^{0}$ |
| $3.98\times10^{-2}\pm2.4\times10^{-4}$ | $1.45\times10^{1}\pm2.3\times10^{0}$ | $1.65\times10^{1}\pm1.9\times10^{0}$ | $1.51\times10^{1}\pm1.9\times10^{0}$ | $1.47\times10^{1}\pm1.8\times10^{0}$ | $1.29\times10^{1}\pm1.8\times10^{0}$ |
| $5.01\times10^{-2}\pm3.0\times10^{-4}$ | $1.42\times10^{1}\pm1.8\times10^{0}$ | $1.43\times10^{1}\pm1.5\times10^{0}$ | $1.28\times10^{1}\pm1.4\times10^{0}$ | $1.24\times10^{1}\pm1.4\times10^{0}$ | $1.01\times10^{1}\pm1.4\times10^{0}$ |
| $6.31\times10^{-2}\pm3.8\times10^{-4}$ | $1.13\times10^{1}\pm1.4\times10^{0}$ | $1.26\times10^{1}\pm1.2\times10^{0}$ | $1.22\times10^{1}\pm1.2\times10^{0}$ | $1.15\times10^{1}\pm1.1\times10^{0}$ | $1.03\times10^{1}\pm1.2\times10^{0}$ |
| $7.94\times10^{-2}\pm4.8\times10^{-4}$ | $1.27\times10^{1}\pm1.5\times10^{0}$ | $1.06\times10^{1}\pm1.1\times10^{0}$ | $1.15\times10^{1}\pm1.2\times10^{0}$ | $9.33\times10^{0}\pm1.0\times10^{0}$ | $7.94\times10^{0}\pm1.0\times10^{0}$ |
| $1.00\times10^{-1}\pm6.1\times10^{-4}$ | $1.39\times10^{1}\pm8.3\times10^{-1}$ | $1.13\times10^{1}\pm6.5\times10^{-1}$ | $9.79\times10^{0}\pm6.0\times10^{-1}$ | $9.35\times10^{0}\pm5.8\times10^{-1}$ | $8.94\times10^{0}\pm6.1\times10^{-1}$ |
| $2.00\times10^{-1}\pm1.3\times10^{-3}$ | $1.24\times10^{1}\pm6.5\times10^{-1}$ | $1.03\times10^{1}\pm5.6\times10^{-1}$ | $9.39\times10^{0}\pm5.0\times10^{-1}$ | $9.63\times10^{0}\pm5.1\times10^{-1}$ | $9.25\times10^{0}\pm5.3\times10^{-1}$ |
| $3.00\times10^{-1}\pm2.0\times10^{-3}$ | $1.07\times10^{1}\pm5.9\times10^{-1}$ | $1.03\times10^{1}\pm5.4\times10^{-1}$ | $1.14\times10^{1}\pm5.9\times10^{-1}$ | $1.09\times10^{1}\pm5.5\times10^{-1}$ | $1.20\times10^{1}\pm6.5\times10^{-1}$ |
| $4.00\times10^{-1}\pm2.9\times10^{-3}$ | $1.23\times10^{1}\pm7.0\times10^{-1}$ | $1.41\times10^{1}\pm7.4\times10^{-1}$ | $1.30\times10^{1}\pm7.2\times10^{-1}$ | $1.44\times10^{1}\pm7.6\times10^{-1}$ | $1.64\times10^{1}\pm9.2\times10^{-1}$ |
| $5.00\times10^{-1}\pm3.9\times10^{-3}$ | $1.26\times10^{1}\pm8.3\times10^{-1}$ | $1.49\times10^{1}\pm8.9\times10^{-1}$ | $1.53\times10^{1}\pm9.4\times10^{-1}$ | $1.50\times10^{1}\pm8.8\times10^{-1}$ | $1.76\times10^{1}\pm1.1\times10^{0}$ |
| $6.00\times10^{-1}\pm5.0\times10^{-3}$ | $1.11\times10^{1}\pm8.6\times10^{-1}$ | $1.07\times10^{1}\pm7.5\times10^{-1}$ | $1.12\times10^{1}\pm8.3\times10^{-1}$ | $1.18\times10^{1}\pm8.2\times10^{-1}$ | $1.27\times10^{1}\pm9.7\times10^{-1}$ |
| $7.00\times10^{-1}\pm6.1\times10^{-3}$ | $8.62\times10^{0}\pm7.2\times10^{-1}$ | $7.71\times10^{0}\pm6.5\times10^{-1}$ | $9.75\times10^{0}\pm7.8\times10^{-1}$ | $8.16\times10^{0}\pm6.9\times10^{-1}$ | $9.48\times10^{0}\pm8.5\times10^{-1}$ |
| $8.00\times10^{-1}\pm7.4\times10^{-3}$ | $8.82\times10^{0}\pm9.1\times10^{-1}$ | $3.63\times10^{0}\pm4.7\times10^{-1}$ | $7.83\times10^{0}\pm7.3\times10^{-1}$ | $6.62\times10^{0}\pm7.6\times10^{-1}$ | $8.23\times10^{0}\pm8.9\times10^{-1}$ |
| $9.00\times10^{-1}\pm8.7\times10^{-3}$ | $9.67\times10^{0}\pm1.0\times10^{0}$ | $6.85\times10^{0}\pm7.5\times10^{-1}$ | $7.52\times10^{0}\pm8.2\times10^{-1}$ | $6.02\times10^{0}\pm6.8\times10^{-1}$ | $6.42\times10^{0}\pm7.9\times10^{-1}$ |

**Table III (c).** The differential and angle-integrated cross sections of the $^{10}$B$(n, \alpha_0)^{7}$Li reaction in the laboratory reference system.

| $E_n$ (MeV) | $\sigma^{0}_{E\_bin,\theta}$ (mb/sr) | | | | $\sigma^{0}_{E\_bin}$ (mb) |
|---|---|---|---|---|---|
| | 133.2° | 143.5° | 153.1° | 160.8° | |
| $1.00\times10^{-6}\pm4.2\times10^{-9}$ | $3.21\times10^{3}\pm1.6\times10^{2}$ | $3.04\times10^{3}\pm1.6\times10^{2}$ | $3.17\times10^{3}\pm1.7\times10^{2}$ | $3.12\times10^{3}\pm1.6\times10^{2}$ | $3.93\times10^{4}\pm1.8\times10^{3}$ |
| $1.26\times10^{-6}\pm5.4\times10^{-9}$ | $2.84\times10^{3}\pm1.4\times10^{2}$ | $2.44\times10^{3}\pm1.2\times10^{2}$ | $2.63\times10^{3}\pm1.3\times10^{2}$ | $2.66\times10^{3}\pm1.4\times10^{2}$ | $3.34\times10^{4}\pm1.1\times10^{3}$ |
| $1.58\times10^{-6}\pm6.8\times10^{-9}$ | $2.15\times10^{3}\pm1.1\times10^{2}$ | $1.98\times10^{3}\pm1.0\times10^{2}$ | $1.90\times10^{3}\pm1.0\times10^{2}$ | $2.33\times10^{3}\pm1.2\times10^{2}$ | $2.63\times10^{4}\pm8.6\times10^{2}$ |
| $2.00\times10^{-6}\pm8.7\times10^{-9}$ | $1.97\times10^{3}\pm1.1\times10^{2}$ | $2.02\times10^{3}\pm1.1\times10^{2}$ | $1.80\times10^{3}\pm1.0\times10^{2}$ | $1.78\times10^{3}\pm1.1\times10^{2}$ | $2.38\times10^{4}\pm8.7\times10^{2}$ |
| $2.51\times10^{-6}\pm1.1\times10^{-8}$ | $1.70\times10^{3}\pm1.1\times10^{2}$ | $1.61\times10^{3}\pm1.1\times10^{2}$ | $1.50\times10^{3}\pm1.0\times10^{2}$ | $1.54\times10^{3}\pm1.1\times10^{2}$ | $2.08\times10^{4}\pm1.0\times10^{3}$ |
| $3.16\times10^{-6}\pm1.4\times10^{-8}$ | $1.63\times10^{3}\pm1.2\times10^{2}$ | $1.89\times10^{3}\pm1.3\times10^{2}$ | $1.96\times10^{3}\pm1.3\times10^{2}$ | $1.66\times10^{3}\pm1.2\times10^{2}$ | $2.34\times10^{4}\pm9.3\times10^{2}$ |
| $3.98\times10^{-6}\pm1.8\times10^{-8}$ | $1.31\times10^{3}\pm1.9\times10^{2}$ | $1.26\times10^{3}\pm1.9\times10^{2}$ | $1.44\times10^{3}\pm2.1\times10^{2}$ | $1.21\times10^{3}\pm1.8\times10^{2}$ | $1.69\times10^{4}\pm2.0\times10^{3}$ |
| $5.01\times10^{-6}\pm2.3\times10^{-8}$ | $1.35\times10^{3}\pm1.3\times10^{2}$ | $1.37\times10^{3}\pm1.3\times10^{2}$ | $1.25\times10^{3}\pm1.2\times10^{2}$ | $1.38\times10^{3}\pm1.3\times10^{2}$ | $1.64\times10^{4}\pm1.2\times10^{3}$ |
| $6.31\times10^{-6}\pm2.9\times10^{-8}$ | $1.32\times10^{3}\pm1.0\times10^{2}$ | $1.26\times10^{3}\pm9.9\times10^{1}$ | $1.21\times10^{3}\pm9.6\times10^{1}$ | $1.11\times10^{3}\pm9.1\times10^{1}$ | $1.58\times10^{4}\pm8.9\times10^{2}$ |
| $7.94\times10^{-6}\pm3.7\times10^{-8}$ | $1.12\times10^{3}\pm2.4\times10^{2}$ | $1.20\times10^{3}\pm2.6\times10^{2}$ | $1.11\times10^{3}\pm2.4\times10^{2}$ | $9.62\times10^{2}\pm2.1\times10^{2}$ | $1.35\times10^{4}\pm2.8\times10^{3}$ |
| $1.00\times10^{-5}\pm4.7\times10^{-8}$ | $1.07\times10^{3}\pm8.3\times10^{1}$ | $9.91\times10^{2}\pm7.7\times10^{1}$ | $1.01\times10^{3}\pm7.7\times10^{1}$ | $1.00\times10^{3}\pm7.9\times10^{1}$ | $1.31\times10^{4}\pm5.6\times10^{2}$ |
| $1.26\times10^{-5}\pm6.4\times10^{-8}$ | $8.93\times10^{2}\pm9.7\times10^{1}$ | $9.37\times10^{2}\pm1.0\times10^{2}$ | $8.79\times10^{2}\pm9.6\times10^{1}$ | $8.71\times10^{2}\pm9.6\times10^{1}$ | $1.18\times10^{4}\pm1.0\times10^{3}$ |
| $1.58\times10^{-5}\pm8.6\times10^{-8}$ | $7.94\times10^{2}\pm1.3\times10^{2}$ | $7.44\times10^{2}\pm1.2\times10^{2}$ | $6.86\times10^{2}\pm1.1\times10^{2}$ | $7.59\times10^{2}\pm1.2\times10^{2}$ | $8.82\times10^{3}\pm1.3\times10^{3}$ |
| $2.00\times10^{-5}\pm1.2\times10^{-7}$ | $6.26\times10^{2}\pm1.2\times10^{2}$ | $6.30\times10^{2}\pm1.2\times10^{2}$ | $6.32\times10^{2}\pm1.2\times10^{2}$ | $4.67\times10^{2}\pm9.4\times10^{1}$ | $7.48\times10^{3}\pm1.4\times10^{3}$ |
| $2.51\times10^{-5}\pm1.6\times10^{-7}$ | $5.93\times10^{2}\pm5.8\times10^{1}$ | $6.70\times10^{2}\pm6.2\times10^{1}$ | $5.73\times10^{2}\pm5.7\times10^{1}$ | $6.22\times10^{2}\pm6.2\times10^{1}$ | $8.32\times10^{3}\pm4.4\times10^{2}$ |
| $3.16\times10^{-5}\pm2.1\times10^{-7}$ | $4.30\times10^{2}\pm9.8\times10^{1}$ | $5.09\times10^{2}\pm1.2\times10^{2}$ | $4.29\times10^{2}\pm9.8\times10^{1}$ | $4.03\times10^{2}\pm9.3\times10^{1}$ | $5.74\times10^{3}\pm1.2\times10^{3}$ |
| $3.98\times10^{-5}\pm2.7\times10^{-7}$ | $4.38\times10^{2}\pm5.5\times10^{1}$ | $4.82\times10^{2}\pm5.9\times10^{1}$ | $4.54\times10^{2}\pm5.7\times10^{1}$ | $5.01\times10^{2}\pm6.3\times10^{1}$ | $5.92\times10^{3}\pm5.5\times10^{2}$ |
| $5.01\times10^{-5}\pm3.6\times10^{-7}$ | $4.78\times10^{2}\pm4.7\times10^{1}$ | $4.67\times10^{2}\pm4.6\times10^{1}$ | $4.59\times10^{2}\pm4.6\times10^{1}$ | $4.54\times10^{2}\pm4.8\times10^{1}$ | $5.85\times10^{3}\pm4.3\times10^{2}$ |
| $6.31\times10^{-5}\pm4.8\times10^{-7}$ | $2.98\times10^{2}\pm3.7\times10^{1}$ | $3.24\times10^{2}\pm3.9\times10^{1}$ | $3.49\times10^{2}\pm4.2\times10^{1}$ | $2.70\times10^{2}\pm3.5\times10^{1}$ | $4.10\times10^{3}\pm4.2\times10^{2}$ |
| $7.94\times10^{-5}\pm6.4\times10^{-7}$ | $3.21\times10^{2}\pm3.9\times10^{1}$ | $3.06\times10^{2}\pm3.8\times10^{1}$ | $3.38\times10^{2}\pm4.1\times10^{1}$ | $3.26\times10^{2}\pm4.2\times10^{1}$ | $3.96\times10^{3}\pm3.8\times10^{2}$ |
| $1.00\times10^{-4}\pm8.4\times10^{-7}$ | $2.91\times10^{2}\pm3.2\times10^{1}$ | $3.18\times10^{2}\pm3.5\times10^{1}$ | $2.70\times10^{2}\pm3.2\times10^{1}$ | $3.10\times10^{2}\pm3.8\times10^{1}$ | $3.80\times10^{3}\pm3.0\times10^{2}$ |
| $1.26\times10^{-4}\pm1.0\times10^{-6}$ | $3.32\times10^{2}\pm4.6\times10^{1}$ | $2.28\times10^{2}\pm3.5\times10^{1}$ | $2.86\times10^{2}\pm4.1\times10^{1}$ | $2.94\times10^{2}\pm4.6\times10^{1}$ | $3.57\times10^{3}\pm3.8\times10^{2}$ |
| $1.58\times10^{-4}\pm1.3\times10^{-6}$ | $2.04\times10^{2}\pm2.5\times10^{1}$ | $2.14\times10^{2}\pm2.6\times10^{1}$ | $2.15\times10^{2}\pm2.7\times10^{1}$ | $1.53\times10^{2}\pm2.3\times10^{1}$ | $2.80\times10^{3}\pm2.2\times10^{2}$ |
| $2.00\times10^{-4}\pm1.6\times10^{-6}$ | $2.09\times10^{2}\pm2.7\times10^{1}$ | $1.44\times10^{2}\pm2.1\times10^{1}$ | $1.55\times10^{2}\pm2.4\times10^{1}$ | $2.32\times10^{2}\pm3.3\times10^{1}$ | $2.50\times10^{3}\pm2.1\times10^{2}$ |
| $2.51\times10^{-4}\pm2.0\times10^{-6}$ | $1.97\times10^{2}\pm2.5\times10^{1}$ | $1.68\times10^{2}\pm2.3\times10^{1}$ | $1.70\times10^{2}\pm2.4\times10^{1}$ | $1.14\times10^{2}\pm2.0\times10^{1}$ | $2.41\times10^{3}\pm2.1\times10^{2}$ |
| $3.16\times10^{-4}\pm2.5\times10^{-6}$ | $1.58\times10^{2}\pm2.1\times10^{1}$ | $1.40\times10^{2}\pm1.9\times10^{1}$ | $1.48\times10^{2}\pm2.1\times10^{1}$ | $6.25\times10^{1}\pm1.5\times10^{1}$ | $1.88\times10^{3}\pm1.4\times10^{2}$ |
| $3.98\times10^{-4}\pm3.1\times10^{-6}$ | $1.71\times10^{2}\pm2.1\times10^{1}$ | $1.26\times10^{2}\pm1.7\times10^{1}$ | $1.59\times10^{2}\pm2.2\times10^{1}$ | $6.70\times10^{1}\pm1.4\times10^{1}$ | $1.99\times10^{3}\pm1.4\times10^{2}$ |
| $5.01\times10^{-4}\pm3.9\times10^{-6}$ | $1.47\times10^{2}\pm1.6\times10^{1}$ | $1.26\times10^{2}\pm1.4\times10^{1}$ | $1.10\times10^{2}\pm1.4\times10^{1}$ | $1.26\times10^{2}\pm1.7\times10^{1}$ | $1.73\times10^{3}\pm9.5\times10^{1}$ |
| $6.31\times10^{-4}\pm4.8\times10^{-6}$ | $1.03\times10^{2}\pm1.2\times10^{1}$ | $1.10\times10^{2}\pm1.3\times10^{1}$ | $1.29\times10^{2}\pm1.5\times10^{1}$ | $1.03\times10^{2}\pm1.5\times10^{1}$ | $1.39\times10^{3}\pm6.8\times10^{1}$ |
| $7.94\times10^{-4}\pm6.1\times10^{-6}$ | $1.05\times10^{2}\pm1.3\times10^{1}$ | $8.94\times10^{1}\pm1.2\times10^{1}$ | $9.83\times10^{1}\pm1.3\times10^{1}$ | $8.12\times10^{1}\pm1.5\times10^{1}$ | $1.33\times10^{3}\pm7.5\times10^{1}$ |
| $1.00\times10^{-3}\pm7.7\times10^{-6}$ | $9.14\times10^{1}\pm1.0\times10^{1}$ | $7.97\times10^{1}\pm9.8\times10^{0}$ | $6.22\times10^{1}\pm9.1\times10^{0}$ | $9.34\times10^{1}\pm1.3\times10^{1}$ | $1.08\times10^{3}\pm5.3\times10^{1}$ |
| $1.26\times10^{-3}\pm9.8\times10^{-6}$ | $8.83\times10^{1}\pm9.9\times10^{0}$ | $8.44\times10^{1}\pm9.8\times10^{0}$ | $8.38\times10^{1}\pm1.0\times10^{1}$ | $6.91\times10^{1}\pm1.1\times10^{1}$ | $9.92\times10^{2}\pm4.2\times10^{1}$ |
| $1.58\times10^{-3}\pm1.3\times10^{-5}$ | $7.96\times10^{1}\pm8.6\times10^{0}$ | $5.08\times10^{1}\pm6.9\times10^{0}$ | $5.20\times10^{1}\pm7.4\times10^{0}$ | $4.83\times10^{1}\pm8.6\times10^{0}$ | $8.66\times10^{2}\pm3.7\times10^{1}$ |
| $2.00\times10^{-3}\pm1.6\times10^{-5}$ | $5.56\times10^{1}\pm7.1\times10^{0}$ | $5.72\times10^{1}\pm7.4\times10^{0}$ | $5.95\times10^{1}\pm7.6\times10^{0}$ | $4.05\times10^{1}\pm7.4\times10^{0}$ | $8.64\times10^{2}\pm3.5\times10^{1}$ |
| $2.51\times10^{-3}\pm2.1\times10^{-5}$ | $5.02\times10^{1}\pm7.4\times10^{0}$ | $4.87\times10^{1}\pm7.6\times10^{0}$ | $3.18\times10^{1}\pm6.4\times10^{0}$ | $2.84\times10^{1}\pm6.8\times10^{0}$ | $7.23\times10^{2}\pm4.8\times10^{1}$ |
| $3.16\times10^{-3}\pm2.7\times10^{-5}$ | $5.10\times10^{1}\pm6.4\times10^{0}$ | $6.66\times10^{1}\pm7.5\times10^{0}$ | $4.88\times10^{1}\pm7.1\times10^{0}$ | $4.74\times10^{1}\pm7.8\times10^{0}$ | $6.98\times10^{2}\pm3.2\times10^{1}$ |
| $3.98\times10^{-3}\pm3.6\times10^{-5}$ | $5.19\times10^{1}\pm6.3\times10^{0}$ | $3.86\times10^{1}\pm5.5\times10^{0}$ | $3.44\times10^{1}\pm5.4\times10^{0}$ | $4.30\times10^{1}\pm7.1\times10^{0}$ | $5.89\times10^{2}\pm2.6\times10^{1}$ |
| $5.01\times10^{-3}\pm4.8\times10^{-5}$ | $2.91\times10^{1}\pm4.2\times10^{0}$ | $4.05\times10^{1}\pm5.2\times10^{0}$ | $2.72\times10^{1}\pm4.4\times10^{0}$ | $3.17\times10^{1}\pm5.7\times10^{0}$ | $5.18\times10^{2}\pm2.3\times10^{1}$ |
| $6.31\times10^{-3}\pm6.4\times10^{-5}$ | $3.97\times10^{1}\pm5.0\times10^{0}$ | $3.65\times10^{1}\pm4.9\times10^{0}$ | $3.03\times10^{1}\pm5.1\times10^{0}$ | $1.46\times10^{1}\pm3.8\times10^{0}$ | $4.42\times10^{2}\pm2.1\times10^{1}$ |
| $7.94\times10^{-3}\pm8.6\times10^{-5}$ | $3.07\times10^{1}\pm4.3\times10^{0}$ | $1.76\times10^{1}\pm3.4\times10^{0}$ | $2.63\times10^{1}\pm4.5\times10^{0}$ | $2.03\times10^{1}\pm4.8\times10^{0}$ | $3.84\times10^{2}\pm2.0\times10^{1}$ |

| | | | | | | |
|---|---|---|---|---|---|---|
| $1.00\times10^{-2}\pm1.2\times10^{-4}$ | $3.17\times10^{1}\pm4.3\times10^{0}$ | $2.49\times10^{1}\pm3.8\times10^{0}$ | $2.50\times10^{1}\pm4.0\times10^{0}$ | $2.26\times10^{1}\pm4.5\times10^{0}$ | $3.73\times10^{2}\pm2.2\times10^{1}$ | |
| $1.26\times10^{-2}\pm1.6\times10^{-4}$ | $2.20\times10^{1}\pm3.1\times10^{0}$ | $2.83\times10^{1}\pm3.7\times10^{0}$ | $2.42\times10^{1}\pm3.5\times10^{0}$ | $2.25\times10^{1}\pm4.0\times10^{0}$ | $3.44\times10^{2}\pm1.5\times10^{1}$ | |
| $1.58\times10^{-2}\pm2.2\times10^{-4}$ | $1.91\times10^{1}\pm2.8\times10^{0}$ | $2.63\times10^{1}\pm3.4\times10^{0}$ | $2.24\times10^{1}\pm3.4\times10^{0}$ | $2.36\times10^{1}\pm4.2\times10^{0}$ | $3.12\times10^{2}\pm1.4\times10^{1}$ | |
| $2.00\times10^{-2}\pm1.2\times10^{-4}$ | $2.28\times10^{1}\pm2.8\times10^{0}$ | $2.17\times10^{1}\pm2.8\times10^{0}$ | $2.34\times10^{1}\pm3.2\times10^{0}$ | $1.19\times10^{1}\pm2.6\times10^{0}$ | $2.81\times10^{2}\pm1.3\times10^{1}$ | |
| $2.51\times10^{-2}\pm1.5\times10^{-4}$ | $2.03\times10^{1}\pm2.3\times10^{0}$ | $1.74\times10^{1}\pm2.2\times10^{0}$ | $1.17\times10^{1}\pm1.8\times10^{0}$ | $9.49\times10^{0}\pm2.0\times10^{0}$ | $2.31\times10^{2}\pm1.0\times10^{1}$ | |
| $3.16\times10^{-2}\pm1.9\times10^{-4}$ | $1.47\times10^{1}\pm2.4\times10^{0}$ | $1.41\times10^{1}\pm2.4\times10^{0}$ | $1.41\times10^{1}\pm2.7\times10^{0}$ | $8.50\times10^{0}\pm2.5\times10^{0}$ | $2.09\times10^{2}\pm1.1\times10^{1}$ | |
| $3.98\times10^{-2}\pm2.4\times10^{-4}$ | $1.14\times10^{1}\pm1.7\times10^{0}$ | $1.22\times10^{1}\pm1.8\times10^{0}$ | $9.80\times10^{0}\pm1.7\times10^{0}$ | $1.10\times10^{1}\pm2.3\times10^{0}$ | $1.93\times10^{2}\pm9.3\times10^{0}$ | |
| $5.01\times10^{-2}\pm3.0\times10^{-4}$ | $1.32\times10^{1}\pm1.5\times10^{0}$ | $9.89\times10^{0}\pm1.3\times10^{0}$ | $1.24\times10^{1}\pm1.6\times10^{0}$ | $1.05\times10^{1}\pm1.8\times10^{0}$ | $1.75\times10^{2}\pm7.6\times10^{0}$ | |
| $6.31\times10^{-2}\pm3.8\times10^{-4}$ | $1.00\times10^{1}\pm1.1\times10^{0}$ | $1.01\times10^{1}\pm1.2\times10^{0}$ | $7.35\times10^{0}\pm1.0\times10^{0}$ | $8.92\times10^{0}\pm1.3\times10^{0}$ | $1.58\times10^{2}\pm5.8\times10^{0}$ | |
| $7.94\times10^{-2}\pm4.8\times10^{-4}$ | $7.46\times10^{0}\pm1.0\times10^{0}$ | $7.96\times10^{0}\pm1.0\times10^{0}$ | $6.84\times10^{0}\pm1.0\times10^{0}$ | $7.83\times10^{0}\pm1.3\times10^{0}$ | $1.39\times10^{2}\pm5.3\times10^{0}$ | |
| $1.00\times10^{-1}\pm6.1\times10^{-4}$ | $9.14\times10^{0}\pm6.2\times10^{-1}$ | $8.33\times10^{0}\pm5.9\times10^{-1}$ | $8.59\times10^{0}\pm6.3\times10^{-1}$ | $8.52\times10^{0}\pm7.0\times10^{-1}$ | $1.44\times10^{2}\pm4.3\times10^{0}$ | |
| $2.00\times10^{-1}\pm1.3\times10^{-3}$ | $9.58\times10^{0}\pm5.4\times10^{-1}$ | $9.30\times10^{0}\pm5.4\times10^{-1}$ | $9.61\times10^{0}\pm5.8\times10^{-1}$ | $8.42\times10^{0}\pm6.1\times10^{-1}$ | $1.38\times10^{2}\pm4.2\times10^{0}$ | |
| $3.00\times10^{-1}\pm2.0\times10^{-3}$ | $1.33\times10^{1}\pm6.9\times10^{-1}$ | $1.31\times10^{1}\pm7.2\times10^{-1}$ | $1.27\times10^{1}\pm7.4\times10^{-1}$ | $1.38\times10^{1}\pm9.1\times10^{-1}$ | $1.42\times10^{2}\pm4.0\times10^{0}$ | |
| $4.00\times10^{-1}\pm2.9\times10^{-3}$ | $1.74\times10^{1}\pm9.9\times10^{-1}$ | $1.82\times10^{1}\pm1.1\times10^{0}$ | $1.87\times10^{1}\pm1.2\times10^{0}$ | $1.77\times10^{1}\pm1.4\times10^{0}$ | $1.71\times10^{2}\pm5.1\times10^{0}$ | |
| $5.00\times10^{-1}\pm3.9\times10^{-3}$ | $1.91\times10^{1}\pm1.2\times10^{0}$ | $1.83\times10^{1}\pm1.2\times10^{0}$ | $1.81\times10^{1}\pm1.4\times10^{0}$ | $1.81\times10^{1}\pm1.7\times10^{0}$ | $1.87\times10^{2}\pm7.4\times10^{0}$ | |
| $6.00\times10^{-1}\pm5.0\times10^{-3}$ | $1.24\times10^{1}\pm9.7\times10^{-1}$ | $1.30\times10^{1}\pm1.1\times10^{0}$ | $1.23\times10^{1}\pm1.2\times10^{0}$ | $1.38\times10^{1}\pm1.7\times10^{0}$ | $1.44\times10^{2}\pm6.5\times10^{0}$ | |
| $7.00\times10^{-1}\pm6.1\times10^{-3}$ | $9.50\times10^{0}\pm8.9\times10^{-1}$ | $8.28\times10^{0}\pm8.5\times10^{-1}$ | $9.05\times10^{0}\pm1.0\times10^{0}$ | $1.04\times10^{1}\pm1.6\times10^{0}$ | $1.18\times10^{2}\pm5.3\times10^{0}$ | |
| $8.00\times10^{-1}\pm7.4\times10^{-3}$ | $8.20\times10^{0}\pm8.6\times10^{-1}$ | $7.56\times10^{0}\pm8.8\times10^{-1}$ | $7.67\times10^{0}\pm1.2\times10^{0}$ | $5.80\times10^{0}\pm1.4\times10^{0}$ | $1.05\times10^{2}\pm6.3\times10^{0}$ | |
| $9.00\times10^{-1}\pm8.7\times10^{-3}$ | $7.97\times10^{0}\pm9.5\times10^{-1}$ | $8.59\times10^{0}\pm1.1\times10^{0}$ | $5.91\times10^{0}\pm1.1\times10^{0}$ | $7.22\times10^{0}\pm1.6\times10^{0}$ | $1.06\times10^{2}\pm6.6\times10^{0}$ | |

**Table IV (a).** The differential cross sections of the $^{10}$B(n, $\alpha_1$)$^7$Li reaction in the laboratory reference system.

| $E_n$ (MeV) | $\sigma^{1}_{E\_\text{bin},\theta}$ (mb/sr) | | | | | |
|---|---|---|---|---|---|---|
| | 19.2° | 26.9° | 36.5° | 46.7° | 57.3° | 68.0° |
| $1.00\times10^{-6}\pm4.2\times10^{-9}$ | $4.67\times10^{4}\pm2.4\times10^{3}$ | $4.69\times10^{4}\pm2.1\times10^{3}$ | $4.74\times10^{4}\pm2.1\times10^{3}$ | $4.65\times10^{4}\pm2.1\times10^{3}$ | $4.75\times10^{4}\pm2.0\times10^{3}$ | $4.70\times10^{4}\pm2.0\times10^{3}$ |
| $1.26\times10^{-6}\pm5.4\times10^{-9}$ | $3.86\times10^{4}\pm1.2\times10^{3}$ | $3.92\times10^{4}\pm1.2\times10^{3}$ | $3.97\times10^{4}\pm1.2\times10^{3}$ | $3.97\times10^{4}\pm1.2\times10^{3}$ | $3.96\times10^{4}\pm1.2\times10^{3}$ | $3.89\times10^{4}\pm1.2\times10^{3}$ |
| $1.58\times10^{-6}\pm6.8\times10^{-9}$ | $3.10\times10^{4}\pm1.2\times10^{3}$ | $3.03\times10^{4}\pm9.8\times10^{2}$ | $3.07\times10^{4}\pm9.9\times10^{2}$ | $3.06\times10^{4}\pm9.9\times10^{2}$ | $3.09\times10^{4}\pm9.9\times10^{2}$ | $3.05\times10^{4}\pm9.8\times10^{2}$ |
| $2.00\times10^{-6}\pm8.7\times10^{-9}$ | $2.88\times10^{4}\pm1.2\times10^{3}$ | $2.91\times10^{4}\pm1.1\times10^{3}$ | $2.89\times10^{4}\pm1.0\times10^{3}$ | $2.87\times10^{4}\pm1.1\times10^{3}$ | $2.94\times10^{4}\pm1.1\times10^{3}$ | $2.94\times10^{4}\pm1.1\times10^{3}$ |
| $2.51\times10^{-6}\pm1.1\times10^{-8}$ | $2.47\times10^{4}\pm1.2\times10^{3}$ | $2.49\times10^{4}\pm1.2\times10^{3}$ | $2.46\times10^{4}\pm1.2\times10^{3}$ | $2.50\times10^{4}\pm1.2\times10^{3}$ | $2.43\times10^{4}\pm1.2\times10^{3}$ | $2.44\times10^{4}\pm1.2\times10^{3}$ |
| $3.16\times10^{-6}\pm1.4\times10^{-8}$ | $2.75\times10^{4}\pm1.1\times10^{3}$ | $2.77\times10^{4}\pm1.1\times10^{3}$ | $2.75\times10^{4}\pm1.1\times10^{3}$ | $2.86\times10^{4}\pm1.2\times10^{3}$ | $2.78\times10^{4}\pm1.1\times10^{3}$ | $2.82\times10^{4}\pm1.1\times10^{3}$ |
| $3.98\times10^{-6}\pm1.8\times10^{-8}$ | $1.94\times10^{4}\pm2.4\times10^{3}$ | $2.04\times10^{4}\pm2.5\times10^{3}$ | $1.99\times10^{4}\pm2.4\times10^{3}$ | $1.99\times10^{4}\pm2.4\times10^{3}$ | $2.00\times10^{4}\pm2.4\times10^{3}$ | $1.97\times10^{4}\pm2.4\times10^{3}$ |
| $5.01\times10^{-6}\pm2.3\times10^{-8}$ | $1.93\times10^{4}\pm1.5\times10^{3}$ | $1.95\times10^{4}\pm1.5\times10^{3}$ | $1.94\times10^{4}\pm1.5\times10^{3}$ | $1.99\times10^{4}\pm1.5\times10^{3}$ | $1.95\times10^{4}\pm1.5\times10^{3}$ | $1.99\times10^{4}\pm1.5\times10^{3}$ |
| $6.31\times10^{-6}\pm2.9\times10^{-8}$ | $1.90\times10^{4}\pm1.1\times10^{3}$ | $1.99\times10^{4}\pm1.1\times10^{3}$ | $1.91\times10^{4}\pm1.1\times10^{3}$ | $1.91\times10^{4}\pm1.1\times10^{3}$ | $1.94\times10^{4}\pm1.1\times10^{3}$ | $1.92\times10^{4}\pm1.1\times10^{3}$ |
| $7.94\times10^{-6}\pm3.7\times10^{-8}$ | $1.59\times10^{4}\pm3.3\times10^{3}$ | $1.58\times10^{4}\pm3.3\times10^{3}$ | $1.61\times10^{4}\pm3.4\times10^{3}$ | $1.63\times10^{4}\pm3.4\times10^{3}$ | $1.64\times10^{4}\pm3.4\times10^{3}$ | $1.63\times10^{4}\pm3.4\times10^{3}$ |
| $1.00\times10^{-5}\pm4.7\times10^{-8}$ | $1.56\times10^{4}\pm7.5\times10^{2}$ | $1.52\times10^{4}\pm6.8\times10^{2}$ | $1.57\times10^{4}\pm7.0\times10^{2}$ | $1.54\times10^{4}\pm6.9\times10^{2}$ | $1.52\times10^{4}\pm6.8\times10^{2}$ | $1.52\times10^{4}\pm6.8\times10^{2}$ |
| $1.26\times10^{-5}\pm6.4\times10^{-8}$ | $1.35\times10^{4}\pm1.2\times10^{3}$ | $1.40\times10^{4}\pm1.2\times10^{3}$ | $1.34\times10^{4}\pm1.2\times10^{3}$ | $1.36\times10^{4}\pm1.2\times10^{3}$ | $1.36\times10^{4}\pm1.2\times10^{3}$ | $1.36\times10^{4}\pm1.2\times10^{3}$ |
| $1.58\times10^{-5}\pm8.6\times10^{-8}$ | $1.05\times10^{4}\pm1.6\times10^{3}$ | $1.08\times10^{4}\pm1.6\times10^{3}$ | $1.06\times10^{4}\pm1.6\times10^{3}$ | $1.06\times10^{4}\pm1.6\times10^{3}$ | $1.09\times10^{4}\pm1.6\times10^{3}$ | $1.05\times10^{4}\pm1.6\times10^{3}$ |
| $2.00\times10^{-5}\pm1.2\times10^{-7}$ | $8.80\times10^{3}\pm1.6\times10^{3}$ | $8.73\times10^{3}\pm1.6\times10^{3}$ | $8.87\times10^{3}\pm1.6\times10^{3}$ | $8.88\times10^{3}\pm1.6\times10^{3}$ | $8.82\times10^{3}\pm1.6\times10^{3}$ | $8.59\times10^{3}\pm1.6\times10^{3}$ |
| $2.51\times10^{-5}\pm1.6\times10^{-7}$ | $9.68\times10^{3}\pm5.5\times10^{2}$ | $9.58\times10^{3}\pm5.5\times10^{2}$ | $9.98\times10^{3}\pm5.6\times10^{2}$ | $9.78\times10^{3}\pm5.5\times10^{2}$ | $9.63\times10^{3}\pm5.5\times10^{2}$ | $9.52\times10^{3}\pm5.4\times10^{2}$ |
| $3.16\times10^{-5}\pm2.1\times10^{-7}$ | $6.80\times10^{3}\pm1.5\times10^{3}$ | $6.93\times10^{3}\pm1.5\times10^{3}$ | $6.84\times10^{3}\pm1.5\times10^{3}$ | $7.09\times10^{3}\pm1.5\times10^{3}$ | $7.00\times10^{3}\pm1.5\times10^{3}$ | $6.75\times10^{3}\pm1.5\times10^{3}$ |
| $3.98\times10^{-5}\pm2.7\times10^{-7}$ | $7.21\times10^{3}\pm6.9\times10^{2}$ | $6.90\times10^{3}\pm6.6\times10^{2}$ | $7.08\times10^{3}\pm6.8\times10^{2}$ | $6.95\times10^{3}\pm6.6\times10^{2}$ | $7.16\times10^{3}\pm6.8\times10^{2}$ | $7.07\times10^{3}\pm6.8\times10^{2}$ |
| $5.01\times10^{-5}\pm3.6\times10^{-7}$ | $6.73\times10^{3}\pm5.3\times10^{2}$ | $6.74\times10^{3}\pm5.1\times10^{2}$ | $6.57\times10^{3}\pm5.0\times10^{2}$ | $6.41\times10^{3}\pm4.9\times10^{2}$ | $6.78\times10^{3}\pm5.1\times10^{2}$ | $6.63\times10^{3}\pm5.0\times10^{2}$ |
| $6.31\times10^{-5}\pm4.8\times10^{-7}$ | $5.05\times10^{3}\pm5.2\times10^{2}$ | $5.10\times10^{3}\pm5.2\times10^{2}$ | $5.04\times10^{3}\pm5.2\times10^{2}$ | $4.97\times10^{3}\pm5.1\times10^{2}$ | $5.05\times10^{3}\pm5.2\times10^{2}$ | $4.76\times10^{3}\pm4.9\times10^{2}$ |
| $7.94\times10^{-5}\pm6.4\times10^{-7}$ | $5.04\times10^{3}\pm5.0\times10^{2}$ | $4.95\times10^{3}\pm4.9\times10^{2}$ | $4.82\times10^{3}\pm4.8\times10^{2}$ | $4.86\times10^{3}\pm4.8\times10^{2}$ | $4.67\times10^{3}\pm4.6\times10^{2}$ | $4.84\times10^{3}\pm4.8\times10^{2}$ |
| $1.00\times10^{-4}\pm8.4\times10^{-7}$ | $4.33\times10^{3}\pm3.7\times10^{2}$ | $4.48\times10^{3}\pm3.7\times10^{2}$ | $4.32\times10^{3}\pm3.6\times10^{2}$ | $4.51\times10^{3}\pm3.7\times10^{2}$ | $4.49\times10^{3}\pm3.7\times10^{2}$ | $4.28\times10^{3}\pm3.5\times10^{2}$ |
| $1.26\times10^{-4}\pm1.0\times10^{-6}$ | $4.09\times10^{3}\pm4.6\times10^{2}$ | $3.74\times10^{3}\pm4.3\times10^{2}$ | $3.91\times10^{3}\pm4.3\times10^{2}$ | $4.06\times10^{3}\pm4.5\times10^{2}$ | $4.13\times10^{3}\pm4.6\times10^{2}$ | $3.87\times10^{3}\pm4.3\times10^{2}$ |
| $1.58\times10^{-4}\pm1.3\times10^{-6}$ | $3.40\times10^{3}\pm3.0\times10^{2}$ | $3.52\times10^{3}\pm3.0\times10^{2}$ | $3.35\times10^{3}\pm2.8\times10^{2}$ | $3.38\times10^{3}\pm2.9\times10^{2}$ | $3.65\times10^{3}\pm3.1\times10^{2}$ | $3.36\times10^{3}\pm2.8\times10^{2}$ |
| $2.00\times10^{-4}\pm1.6\times10^{-6}$ | $3.06\times10^{3}\pm2.9\times10^{2}$ | $2.89\times10^{3}\pm2.8\times10^{2}$ | $2.87\times10^{3}\pm2.7\times10^{2}$ | $2.93\times10^{3}\pm2.7\times10^{2}$ | $2.97\times10^{3}\pm2.7\times10^{2}$ | $2.87\times10^{3}\pm2.7\times10^{2}$ |
| $2.51\times10^{-4}\pm2.0\times10^{-6}$ | $2.79\times10^{3}\pm2.6\times10^{2}$ | $2.70\times10^{3}\pm2.6\times10^{2}$ | $2.70\times10^{3}\pm2.6\times10^{2}$ | $2.70\times10^{3}\pm2.5\times10^{2}$ | $2.64\times10^{3}\pm2.5\times10^{2}$ | $2.64\times10^{3}\pm2.5\times10^{2}$ |
| $3.16\times10^{-4}\pm2.5\times10^{-6}$ | $2.54\times10^{3}\pm2.0\times10^{2}$ | $2.30\times10^{3}\pm2.0\times10^{2}$ | $2.33\times10^{3}\pm1.9\times10^{2}$ | $2.15\times10^{3}\pm1.8\times10^{2}$ | $2.37\times10^{3}\pm1.9\times10^{2}$ | $2.25\times10^{3}\pm1.9\times10^{2}$ |
| $3.98\times10^{-4}\pm3.1\times10^{-6}$ | $2.36\times10^{3}\pm1.9\times10^{2}$ | $2.34\times10^{3}\pm1.9\times10^{2}$ | $2.14\times10^{3}\pm1.8\times10^{2}$ | $2.26\times10^{3}\pm1.8\times10^{2}$ | $2.25\times10^{3}\pm1.8\times10^{2}$ | $2.10\times10^{3}\pm1.7\times10^{2}$ |

| $E_n$ (MeV) | | | | | | |
|---|---|---|---|---|---|---|
| 5.01×10⁻⁴±3.9×10⁻⁶ | 2.06×10³±1.3×10² | 1.89×10³±1.3×10² | 1.91×10³±1.2×10² | 2.16×10³±1.3×10² | 2.05×10³±1.3×10² | 1.95×10³±1.2×10² |
| 6.31×10⁻⁴±4.8×10⁻⁶ | 1.80×10³±1.0×10² | 1.65×10³±1.0×10² | 1.75×10³±1.0×10² | 1.70×10³±9.7×10¹ | 1.79×10³±1.0×10² | 1.60×10³±9.9×10¹ |
| 7.94×10⁻⁴±6.1×10⁻⁶ | 1.62×10³±1.2×10² | 1.54×10³±1.1×10² | 1.58×10³±1.1×10² | 1.51×10³±9.9×10¹ | 1.50×10³±1.1×10² | 1.51×10³±1.0×10² |
| 1.00×10⁻³±7.7×10⁻⁶ | 1.31×10³±8.3×10¹ | 1.33×10³±8.1×10¹ | 1.33×10³±7.8×10¹ | 1.38×10³±7.9×10¹ | 1.31×10³±7.8×10¹ | 1.15×10³±7.5×10¹ |
| 1.26×10⁻³±9.8×10⁻⁶ | 1.20×10³±7.2×10¹ | 1.11×10³±6.7×10¹ | 1.12×10³±6.3×10¹ | 1.23×10³±6.4×10¹ | 1.17×10³±6.5×10¹ | 1.04×10³±6.5×10¹ |
| 1.58×10⁻³±1.3×10⁻⁵ | 1.09×10³±6.2×10¹ | 1.02×10³±6.6×10¹ | 1.10×10³±5.7×10¹ | 1.03×10³±5.6×10¹ | 1.06×10³±5.7×10¹ | 9.97×10²±5.9×10¹ |
| 2.00×10⁻³±1.6×10⁻⁵ | 9.89×10²±5.7×10¹ | 9.65×10²±5.5×10¹ | 9.43×10²±5.2×10¹ | 9.59×10²±5.2×10¹ | 9.42×10²±5.1×10¹ | 8.70×10²±5.6×10¹ |
| 2.51×10⁻³±2.1×10⁻⁵ | 8.05×10²±6.9×10¹ | 8.27×10²±6.8×10¹ | 8.51×10²±6.7×10¹ | 8.87×10²±6.8×10¹ | 8.32×10²±6.8×10¹ | 7.62×10²±6.8×10¹ |
| 3.16×10⁻³±2.7×10⁻⁵ | 8.28×10²±5.2×10¹ | 7.68×10²±4.9×10¹ | 7.37×10²±4.4×10¹ | 7.80×10²±4.5×10¹ | 7.37×10²±4.5×10¹ | 6.95×10²±4.9×10¹ |
| 3.98×10⁻³±3.6×10⁻⁵ | 6.56×10²±4.5×10¹ | 7.24×10²±4.6×10¹ | 7.17×10²±4.2×10¹ | 7.02×10²±4.0×10¹ | 6.99×10²±4.1×10¹ | 6.39×10²±4.3×10¹ |
| 5.01×10⁻³±4.8×10⁻⁵ | 5.57×10²±3.6×10¹ | 5.53×10²±3.5×10¹ | 5.87×10²±3.4×10¹ | 5.90×10²±3.3×10¹ | 5.79×10²±3.4×10¹ | 4.98×10²±3.5×10¹ |
| 6.31×10⁻³±6.4×10⁻⁵ | 5.65×10²±3.7×10¹ | 5.54×10²±3.5×10¹ | 5.28×10²±3.3×10¹ | 5.50×10²±3.3×10¹ | 5.27×10²±3.3×10¹ | 5.10×10²±3.3×10¹ |
| 7.94×10⁻³±8.6×10⁻⁵ | 4.91×10²±3.8×10¹ | 4.95×10²±3.4×10¹ | 4.71×10²±3.3×10¹ | 5.05×10²±3.2×10¹ | 4.94×10²±3.3×10¹ | 4.44×10²±3.3×10¹ |
| 1.00×10⁻²±1.2×10⁻⁴ | 4.75×10²±3.4×10¹ | 4.51×10²±3.3×10¹ | 4.77×10²±3.3×10¹ | 4.66×10²±3.2×10¹ | 4.40×10²±3.2×10¹ | 4.40×10²±3.3×10¹ |
| 1.26×10⁻²±1.6×10⁻⁴ | 3.84×10²±2.6×10¹ | 3.93×10²±2.4×10¹ | 4.19×10²±2.4×10¹ | 4.17×10²±2.3×10¹ | 3.93×10²±2.4×10¹ | 3.95×10²±2.5×10¹ |
| 1.58×10⁻²±2.2×10⁻⁴ | 3.52×10²±2.3×10¹ | 3.54×10²±2.2×10¹ | 3.69×10²±2.2×10¹ | 3.47×10²±2.1×10¹ | 3.65×10²±2.2×10¹ | 3.30×10²±2.2×10¹ |
| 2.00×10⁻²±1.2×10⁻⁴ | 3.37×10²±2.3×10¹ | 3.31×10²±2.3×10¹ | 3.37×10²±2.2×10¹ | 3.27×10²±2.1×10¹ | 3.30×10²±2.2×10¹ | 2.85×10²±2.2×10¹ |
| 2.51×10⁻²±1.5×10⁻⁴ | 3.23×10²±1.8×10¹ | 3.00×10²±1.8×10¹ | 2.98×10²±1.7×10¹ | 3.12×10²±1.7×10¹ | 2.97×10²±1.8×10¹ | 2.64×10²±1.8×10¹ |
| 3.16×10⁻²±1.9×10⁻⁴ | 2.47×10²±2.0×10¹ | 2.54×10²±1.9×10¹ | 2.41×10²±1.8×10¹ | 2.27×10²±1.8×10¹ | 2.39×10²±1.9×10¹ | 2.14×10²±2.0×10¹ |
| 3.98×10⁻²±2.4×10⁻⁴ | 2.49×10²±1.6×10¹ | 2.24×10²±1.5×10¹ | 2.35×10²±1.5×10¹ | 2.27×10²±1.5×10¹ | 2.33×10²±1.5×10¹ | 2.06×10²±1.6×10¹ |
| 5.01×10⁻²±3.0×10⁻⁴ | 2.16×10²±1.4×10¹ | 2.05×10²±1.3×10¹ | 2.16×10²±1.2×10¹ | 2.14×10²±1.2×10¹ | 2.01×10²±1.2×10¹ | 1.86×10²±1.2×10¹ |
| 6.31×10⁻²±3.8×10⁻⁴ | 1.98×10²±9.9×10⁰ | 1.92×10²±9.7×10⁰ | 2.01×10²±9.6×10⁰ | 1.95×10²±9.6×10⁰ | 1.87×10²±9.5×10⁰ | 1.78×10²±9.6×10⁰ |
| 7.94×10⁻²±4.8×10⁻⁴ | 1.78×10²±9.3×10⁰ | 1.83×10²±9.2×10⁰ | 1.76×10²±8.9×10⁰ | 1.73×10²±8.9×10⁰ | 1.76×10²±9.2×10⁰ | 1.55×10²±8.9×10⁰ |
| 1.00×10⁻¹±6.1×10⁻⁴ | 1.71×10²±6.6×10⁰ | 1.69×10²±5.5×10⁰ | 1.64×10²±5.4×10⁰ | 1.61×10²±5.4×10⁰ | 1.59×10²±5.4×10⁰ | 1.40×10²±5.2×10⁰ |
| 2.00×10⁻¹±1.3×10⁻³ | 1.29×10²±5.5×10⁰ | 1.24×10²±4.6×10⁰ | 1.22×10²±4.7×10⁰ | 1.18×10²±4.4×10⁰ | 1.15×10²±3.8×10⁰ | 1.00×10²±3.6×10⁰ |
| 3.00×10⁻¹±2.0×10⁻³ | 1.01×10²±4.0×10⁰ | 9.62×10¹±3.8×10⁰ | 9.48×10¹±3.4×10⁰ | 9.18×10¹±3.6×10⁰ | 8.73×10¹±2.9×10⁰ | 7.37×10¹±2.6×10⁰ |
| 4.00×10⁻¹±2.9×10⁻³ | 1.04×10²±4.9×10⁰ | 1.00×10²±4.4×10⁰ | 9.75×10¹±3.6×10⁰ | 8.56×10¹±3.9×10⁰ | 7.95×10¹±3.5×10⁰ | 6.33×10¹±2.8×10⁰ |
| 5.00×10⁻¹±3.9×10⁻³ | 1.10×10²±6.7×10⁰ | 1.03×10²±6.0×10⁰ | 9.64×10¹±5.5×10⁰ | 8.59×10¹±5.3×10⁰ | 7.48×10¹±4.2×10⁰ | 5.83×10¹±3.5×10⁰ |
| 6.00×10⁻¹±5.0×10⁻³ | 7.16×10¹±5.1×10⁰ | 6.78×10¹±4.4×10⁰ | 5.95×10¹±4.3×10⁰ | 5.68×10¹±3.9×10⁰ | 5.03×10¹±3.3×10⁰ | 3.83×10¹±2.8×10⁰ |
| 7.00×10⁻¹±6.1×10⁻³ | 3.61×10¹±5.1×10⁰ | 3.94×10¹±4.1×10⁰ | 3.12×10¹±4.1×10⁰ | 3.51×10¹±3.6×10⁰ | 3.07×10¹±3.4×10⁰ | 2.31×10¹±2.0×10⁰ |
| 8.00×10⁻¹±7.4×10⁻³ | 2.77×10¹±3.7×10⁰ | 2.61×10¹±5.0×10⁰ | 2.49×10¹±3.0×10⁰ | 1.88×10¹±3.7×10⁰ | 1.66×10¹±2.9×10⁰ | 1.48×10¹±2.2×10⁰ |
| 9.00×10⁻¹±8.7×10⁻³ | 2.04×10¹±4.5×10⁰ | 2.07×10¹±4.0×10⁰ | 1.85×10¹±3.7×10⁰ | 1.70×10¹±3.3×10⁰ | 1.63×10¹±1.7×10⁰ | 1.25×10¹±1.5×10⁰ |

**Table IV (b).** The differential cross sections of the $^{10}$B(n, $\alpha_1$)$^7$Li reaction in the laboratory reference system.

| $E_n$ (MeV) | $\sigma^1_{E\_bin,\theta}$ (mb/sr) | | | | |
|---|---|---|---|---|---|
| | 78.8° | 90.3° | 101.2° | 112.0° | 122.7° |
| 1.00×10⁻⁶±4.2×10⁻⁹ | 4.65×10⁴±2.0×10³ | 4.77×10⁴±2.0×10³ | 4.90×10⁴±2.3×10³ | 4.77×10⁴±2.0×10³ | 4.84×10⁴±2.2×10³ |
| 1.26×10⁻⁶±5.4×10⁻⁹ | 3.96×10⁴±1.2×10³ | 4.12×10⁴±1.3×10³ | 4.07×10⁴±1.3×10³ | 4.07×10⁴±1.2×10³ | 4.04×10⁴±1.3×10³ |
| 1.58×10⁻⁶±6.8×10⁻⁹ | 3.12×10⁴±1.0×10³ | 3.19×10⁴±1.0×10³ | 3.17×10⁴±1.0×10³ | 3.15×10⁴±1.0×10³ | 3.12×10⁴±1.0×10³ |
| 2.00×10⁻⁶±8.7×10⁻⁹ | 2.88×10⁴±1.0×10³ | 2.98×10⁴±1.1×10³ | 2.96×10⁴±1.1×10³ | 2.99×10⁴±1.1×10³ | 2.99×10⁴±1.1×10³ |
| 2.51×10⁻⁶±1.1×10⁻⁸ | 2.53×10⁴±1.2×10³ | 2.53×10⁴±1.3×10³ | 2.61×10⁴±1.3×10³ | 2.52×10⁴±1.2×10³ | 2.59×10⁴±1.3×10³ |
| 3.16×10⁻⁶±1.4×10⁻⁸ | 2.79×10⁴±1.1×10³ | 2.91×10⁴±1.2×10³ | 2.89×10⁴±1.2×10³ | 2.93×10⁴±1.2×10³ | 2.83×10⁴±1.2×10³ |
| 3.98×10⁻⁶±1.8×10⁻⁸ | 2.02×10⁴±2.5×10³ | 2.09×10⁴±2.5×10³ | 2.06×10⁴±2.5×10³ | 2.09×10⁴±2.5×10³ | 2.07×10⁴±2.5×10³ |
| 5.01×10⁻⁶±2.3×10⁻⁸ | 1.93×10⁴±1.5×10³ | 1.98×10⁴±1.5×10³ | 1.98×10⁴±1.5×10³ | 2.00×10⁴±1.5×10³ | 2.00×10⁴±1.5×10³ |
| 6.31×10⁻⁶±2.9×10⁻⁸ | 1.96×10⁴±1.1×10³ | 1.98×10⁴±1.1×10³ | 1.98×10⁴±1.1×10³ | 1.97×10⁴±1.1×10³ | 2.03×10⁴±1.2×10³ |
| 7.94×10⁻⁶±3.7×10⁻⁸ | 1.64×10⁴±3.4×10³ | 1.64×10⁴±3.4×10³ | 1.70×10⁴±3.6×10³ | 1.66×10⁴±3.5×10³ | 1.66×10⁴±3.5×10³ |
| 1.00×10⁻⁵±4.7×10⁻⁸ | 1.58×10⁴±7.0×10² | 1.58×10⁴±6.9×10² | 1.59×10⁴±7.0×10² | 1.55×10⁴±6.9×10² | 1.61×10⁴±7.3×10² |
| 1.26×10⁻⁵±6.4×10⁻⁸ | 1.37×10⁴±1.2×10³ | 1.36×10⁴±1.2×10³ | 1.37×10⁴±1.2×10³ | 1.41×10⁴±1.3×10³ | 1.42×10⁴±1.3×10³ |
| $1.58\times10^{-5}\pm8.6\times10^{-8}$ | $1.06\times10^{4}\pm1.6\times10^{3}$ | $1.09\times10^{4}\pm1.6\times10^{3}$ | $1.10\times10^{4}\pm1.6\times10^{3}$ | $1.09\times10^{4}\pm1.6\times10^{3}$ | $1.14\times10^{4}\pm1.7\times10^{3}$ |
| $2.00\times10^{-5}\pm1.2\times10^{-7}$ | $8.72\times10^{3}\pm1.6\times10^{3}$ | $9.24\times10^{3}\pm1.7\times10^{3}$ | $8.99\times10^{3}\pm1.6\times10^{3}$ | $8.68\times10^{3}\pm1.6\times10^{3}$ | $9.10\times10^{3}\pm1.7\times10^{3}$ |
| $2.51\times10^{-5}\pm1.6\times10^{-7}$ | $1.00\times10^{4}\pm5.7\times10^{2}$ | $1.01\times10^{4}\pm5.6\times10^{2}$ | $1.01\times10^{4}\pm5.6\times10^{2}$ | $1.00\times10^{4}\pm5.6\times10^{2}$ | $9.92\times10^{3}\pm5.7\times10^{2}$ |
| $3.16\times10^{-5}\pm2.1\times10^{-7}$ | $6.86\times10^{3}\pm1.5\times10^{3}$ | $7.00\times10^{3}\pm1.5\times10^{3}$ | $6.95\times10^{3}\pm1.5\times10^{3}$ | $7.16\times10^{3}\pm1.6\times10^{3}$ | $6.99\times10^{3}\pm1.5\times10^{3}$ |
| $3.98\times10^{-5}\pm2.7\times10^{-7}$ | $6.95\times10^{3}\pm6.7\times10^{2}$ | $7.39\times10^{3}\pm7.0\times10^{2}$ | $7.35\times10^{3}\pm7.0\times10^{2}$ | $7.26\times10^{3}\pm6.9\times10^{2}$ | $7.06\times10^{3}\pm6.8\times10^{2}$ |
| $5.01\times10^{-5}\pm3.6\times10^{-7}$ | $6.68\times10^{3}\pm5.1\times10^{2}$ | $6.77\times10^{3}\pm5.1\times10^{2}$ | $6.89\times10^{3}\pm5.2\times10^{2}$ | $6.82\times10^{3}\pm5.1\times10^{2}$ | $6.79\times10^{3}\pm5.1\times10^{2}$ |
| $6.31\times10^{-5}\pm4.8\times10^{-7}$ | $5.04\times10^{3}\pm5.2\times10^{2}$ | $5.21\times10^{3}\pm5.3\times10^{2}$ | $5.25\times10^{3}\pm5.3\times10^{2}$ | $5.15\times10^{3}\pm5.2\times10^{2}$ | $5.18\times10^{3}\pm5.3\times10^{2}$ |
| $7.94\times10^{-5}\pm6.4\times10^{-7}$ | $4.85\times10^{3}\pm4.8\times10^{2}$ | $4.93\times10^{3}\pm4.8\times10^{2}$ | $4.90\times10^{3}\pm4.8\times10^{2}$ | $4.95\times10^{3}\pm4.9\times10^{2}$ | $4.96\times10^{3}\pm4.9\times10^{2}$ |
| $1.00\times10^{-4}\pm8.4\times10^{-7}$ | $4.28\times10^{3}\pm3.6\times10^{2}$ | $4.42\times10^{3}\pm3.6\times10^{2}$ | $4.44\times10^{3}\pm3.6\times10^{2}$ | $4.48\times10^{3}\pm3.6\times10^{2}$ | $4.46\times10^{3}\pm3.7\times10^{2}$ |
| $1.26\times10^{-4}\pm1.0\times10^{-6}$ | $3.94\times10^{3}\pm4.4\times10^{2}$ | $4.09\times10^{3}\pm4.4\times10^{2}$ | $4.18\times10^{3}\pm4.5\times10^{2}$ | $3.96\times10^{3}\pm4.3\times10^{2}$ | $4.25\times10^{3}\pm4.7\times10^{2}$ |
| $1.58\times10^{-4}\pm1.3\times10^{-6}$ | $3.31\times10^{3}\pm2.9\times10^{2}$ | $3.41\times10^{3}\pm2.8\times10^{2}$ | $3.66\times10^{3}\pm3.0\times10^{2}$ | $3.45\times10^{3}\pm2.8\times10^{2}$ | $3.31\times10^{3}\pm2.8\times10^{2}$ |
| $2.00\times10^{-4}\pm1.6\times10^{-6}$ | $3.01\times10^{3}\pm2.8\times10^{2}$ | $3.07\times10^{3}\pm2.7\times10^{2}$ | $3.10\times10^{3}\pm2.8\times10^{2}$ | $3.15\times10^{3}\pm2.8\times10^{2}$ | $2.95\times10^{3}\pm2.7\times10^{2}$ |
| $2.51\times10^{-4}\pm2.0\times10^{-6}$ | $2.64\times10^{3}\pm2.6\times10^{2}$ | $2.84\times10^{3}\pm2.6\times10^{2}$ | $2.67\times10^{3}\pm2.5\times10^{2}$ | $2.96\times10^{3}\pm2.7\times10^{2}$ | $2.89\times10^{3}\pm2.7\times10^{2}$ |
| $3.16\times10^{-4}\pm2.5\times10^{-6}$ | $2.35\times10^{3}\pm2.0\times10^{2}$ | $2.44\times10^{3}\pm1.8\times10^{2}$ | $2.36\times10^{3}\pm1.8\times10^{2}$ | $2.32\times10^{3}\pm1.8\times10^{2}$ | $2.55\times10^{3}\pm1.9\times10^{2}$ |
| $3.98\times10^{-4}\pm3.1\times10^{-6}$ | $2.17\times10^{3}\pm1.8\times10^{2}$ | $2.33\times10^{3}\pm1.7\times10^{2}$ | $2.37\times10^{3}\pm1.8\times10^{2}$ | $2.29\times10^{3}\pm1.7\times10^{2}$ | $2.31\times10^{3}\pm1.8\times10^{2}$ |
| $5.01\times10^{-4}\pm3.9\times10^{-6}$ | $1.93\times10^{3}\pm1.3\times10^{2}$ | $2.06\times10^{3}\pm1.2\times10^{2}$ | $2.11\times10^{3}\pm1.2\times10^{2}$ | $2.00\times10^{3}\pm1.2\times10^{2}$ | $2.04\times10^{3}\pm1.2\times10^{2}$ |
| $6.31\times10^{-4}\pm4.8\times10^{-6}$ | $1.65\times10^{3}\pm1.1\times10^{2}$ | $1.79\times10^{3}\pm9.3\times10^{1}$ | $1.72\times10^{3}\pm9.2\times10^{1}$ | $1.71\times10^{3}\pm9.0\times10^{1}$ | $1.81\times10^{3}\pm9.8\times10^{1}$ |
| $7.94\times10^{-4}\pm6.1\times10^{-6}$ | $1.42\times10^{3}\pm1.1\times10^{2}$ | $1.55\times10^{3}\pm9.5\times10^{1}$ | $1.62\times10^{3}\pm9.9\times10^{1}$ | $1.59\times10^{3}\pm9.6\times10^{1}$ | $1.56\times10^{3}\pm1.0\times10^{2}$ |
| $1.00\times10^{-3}\pm7.7\times10^{-6}$ | $1.22\times10^{3}\pm8.4\times10^{1}$ | $1.36\times10^{3}\pm7.1\times10^{1}$ | $1.39\times10^{3}\pm7.3\times10^{1}$ | $1.34\times10^{3}\pm7.1\times10^{1}$ | $1.33\times10^{3}\pm7.5\times10^{1}$ |
| $1.26\times10^{-3}\pm9.8\times10^{-6}$ | $1.09\times10^{3}\pm7.5\times10^{1}$ | $1.25\times10^{3}\pm6.0\times10^{1}$ | $1.23\times10^{3}\pm5.9\times10^{1}$ | $1.21\times10^{3}\pm5.6\times10^{1}$ | $1.17\times10^{3}\pm6.2\times10^{1}$ |
| $1.58\times10^{-3}\pm1.3\times10^{-5}$ | $1.00\times10^{3}\pm6.5\times10^{1}$ | $1.11\times10^{3}\pm5.3\times10^{1}$ | $1.07\times10^{3}\pm5.2\times10^{1}$ | $1.09\times10^{3}\pm5.1\times10^{1}$ | $1.09\times10^{3}\pm5.6\times10^{1}$ |
| $2.00\times10^{-3}\pm1.6\times10^{-5}$ | $8.97\times10^{2}\pm6.0\times10^{1}$ | $9.58\times10^{2}\pm4.6\times10^{1}$ | $9.82\times10^{2}\pm4.7\times10^{1}$ | $9.24\times10^{2}\pm4.5\times10^{1}$ | $9.82\times10^{2}\pm5.0\times10^{1}$ |
| $2.51\times10^{-3}\pm2.1\times10^{-5}$ | $8.53\times10^{2}\pm7.3\times10^{1}$ | $8.94\times10^{2}\pm6.1\times10^{1}$ | $9.01\times10^{2}\pm6.3\times10^{1}$ | $8.83\times10^{2}\pm6.1\times10^{1}$ | $8.62\times10^{2}\pm6.3\times10^{1}$ |
| $3.16\times10^{-3}\pm2.7\times10^{-5}$ | $7.32\times10^{2}\pm5.2\times10^{1}$ | $7.54\times10^{2}\pm4.1\times10^{1}$ | $7.94\times10^{2}\pm4.0\times10^{1}$ | $8.27\times10^{2}\pm4.1\times10^{1}$ | $8.29\times10^{2}\pm4.5\times10^{1}$ |
| $3.98\times10^{-3}\pm3.6\times10^{-5}$ | $6.18\times10^{2}\pm4.6\times10^{1}$ | $6.61\times10^{2}\pm3.6\times10^{1}$ | $7.26\times10^{2}\pm3.7\times10^{1}$ | $6.75\times10^{2}\pm3.5\times10^{1}$ | $6.76\times10^{2}\pm3.9\times10^{1}$ |
| $5.01\times10^{-3}\pm4.8\times10^{-5}$ | $5.65\times10^{2}\pm3.8\times10^{1}$ | $5.64\times10^{2}\pm3.0\times10^{1}$ | $6.17\times10^{2}\pm3.1\times10^{1}$ | $5.73\times10^{2}\pm2.9\times10^{1}$ | $5.99\times10^{2}\pm3.3\times10^{1}$ |
| $6.31\times10^{-3}\pm6.4\times10^{-5}$ | $5.01\times10^{2}\pm4.1\times10^{1}$ | $5.60\times10^{2}\pm3.0\times10^{1}$ | $5.63\times10^{2}\pm3.0\times10^{1}$ | $5.44\times10^{2}\pm2.9\times10^{1}$ | $5.30\times10^{2}\pm3.1\times10^{1}$ |
| $7.94\times10^{-3}\pm8.6\times10^{-5}$ | $4.44\times10^{2}\pm3.8\times10^{1}$ | $4.92\times10^{2}\pm3.0\times10^{1}$ | $4.94\times10^{2}\pm2.9\times10^{1}$ | $4.83\times10^{2}\pm2.8\times10^{1}$ | $4.86\times10^{2}\pm3.1\times10^{1}$ |
| $1.00\times10^{-2}\pm1.2\times10^{-4}$ | $3.99\times10^{2}\pm3.4\times10^{1}$ | $4.79\times10^{2}\pm3.0\times10^{1}$ | $4.69\times10^{2}\pm3.0\times10^{1}$ | $4.41\times10^{2}\pm2.8\times10^{1}$ | $4.47\times10^{2}\pm3.0\times10^{1}$ |
| $1.26\times10^{-2}\pm1.6\times10^{-4}$ | $3.89\times10^{2}\pm2.6\times10^{1}$ | $4.14\times10^{2}\pm2.2\times10^{1}$ | $3.82\times10^{2}\pm2.1\times10^{1}$ | $3.96\times10^{2}\pm2.1\times10^{1}$ | $3.84\times10^{2}\pm2.2\times10^{1}$ |
| $1.58\times10^{-2}\pm2.2\times10^{-4}$ | $3.59\times10^{2}\pm2.5\times10^{1}$ | $3.46\times10^{2}\pm1.9\times10^{1}$ | $3.54\times10^{2}\pm1.9\times10^{1}$ | $3.53\times10^{2}\pm1.9\times10^{1}$ | $3.44\times10^{2}\pm2.0\times10^{1}$ |
| $2.00\times10^{-2}\pm1.2\times10^{-4}$ | $3.18\times10^{2}\pm2.5\times10^{1}$ | $3.34\times10^{2}\pm2.0\times10^{1}$ | $3.45\times10^{2}\pm2.0\times10^{1}$ | $3.13\times10^{2}\pm1.9\times10^{1}$ | $3.25\times10^{2}\pm2.1\times10^{1}$ |
| $2.51\times10^{-2}\pm1.5\times10^{-4}$ | $2.89\times10^{2}\pm2.0\times10^{1}$ | $2.75\times10^{2}\pm1.5\times10^{1}$ | $2.78\times10^{2}\pm1.6\times10^{1}$ | $2.76\times10^{2}\pm1.5\times10^{1}$ | $2.78\times10^{2}\pm1.7\times10^{1}$ |
| $3.16\times10^{-2}\pm1.9\times10^{-4}$ | $2.27\times10^{2}\pm2.2\times10^{1}$ | $2.44\times10^{2}\pm1.7\times10^{1}$ | $2.31\times10^{2}\pm1.6\times10^{1}$ | $2.35\times10^{2}\pm1.6\times10^{1}$ | $2.41\times10^{2}\pm1.8\times10^{1}$ |
| $3.98\times10^{-2}\pm2.4\times10^{-4}$ | $2.15\times10^{2}\pm1.8\times10^{1}$ | $2.30\times10^{2}\pm1.4\times10^{1}$ | $2.22\times10^{2}\pm1.4\times10^{1}$ | $2.11\times10^{2}\pm1.3\times10^{1}$ | $2.19\times10^{2}\pm1.5\times10^{1}$ |
| $5.01\times10^{-2}\pm3.0\times10^{-4}$ | $1.98\times10^{2}\pm1.3\times10^{1}$ | $1.94\times10^{2}\pm1.1\times10^{1}$ | $1.99\times10^{2}\pm1.1\times10^{1}$ | $1.89\times10^{2}\pm1.1\times10^{1}$ | $1.91\times10^{2}\pm1.2\times10^{1}$ |
| $6.31\times10^{-2}\pm3.8\times10^{-4}$ | $1.66\times10^{2}\pm1.0\times10^{1}$ | $1.68\times10^{2}\pm8.4\times10^{0}$ | $1.79\times10^{2}\pm8.8\times10^{0}$ | $1.68\times10^{2}\pm8.3\times10^{0}$ | $1.67\times10^{2}\pm9.2\times10^{0}$ |
| $7.94\times10^{-2}\pm4.8\times10^{-4}$ | $1.52\times10^{2}\pm9.6\times10^{0}$ | $1.58\times10^{2}\pm8.1\times10^{0}$ | $1.53\times10^{2}\pm8.0\times10^{0}$ | $1.55\times10^{2}\pm8.0\times10^{0}$ | $1.49\times10^{2}\pm8.7\times10^{0}$ |
| $1.00\times10^{-1}\pm6.1\times10^{-4}$ | $1.47\times10^{2}\pm5.4\times10^{0}$ | $1.47\times10^{2}\pm5.0\times10^{0}$ | $1.39\times10^{2}\pm5.0\times10^{0}$ | $1.38\times10^{2}\pm4.7\times10^{0}$ | $1.35\times10^{2}\pm5.0\times10^{0}$ |
| $2.00\times10^{-1}\pm1.3\times10^{-3}$ | $1.00\times10^{2}\pm3.5\times10^{0}$ | $9.65\times10^{1}\pm3.3\times10^{0}$ | $9.50\times10^{1}\pm3.3\times10^{0}$ | $8.90\times10^{1}\pm3.1\times10^{0}$ | $9.01\times10^{1}\pm3.3\times10^{0}$ |
| $3.00\times10^{-1}\pm2.0\times10^{-3}$ | $7.38\times10^{1}\pm2.7\times10^{0}$ | $6.62\times10^{1}\pm2.8\times10^{0}$ | $6.63\times10^{1}\pm2.4\times10^{0}$ | $5.97\times10^{1}\pm2.2\times10^{0}$ | $5.92\times10^{1}\pm2.4\times10^{0}$ |
| $4.00\times10^{-1}\pm2.9\times10^{-3}$ | $5.73\times10^{1}\pm2.5\times10^{0}$ | $4.92\times10^{1}\pm2.5\times10^{0}$ | $4.40\times10^{1}\pm2.0\times10^{0}$ | $3.74\times10^{1}\pm1.7\times10^{0}$ | $3.44\times10^{1}\pm1.7\times10^{0}$ |
| $5.00\times10^{-1}\pm3.9\times10^{-3}$ | $5.19\times10^{1}\pm3.2\times10^{0}$ | $3.74\times10^{1}\pm2.4\times10^{0}$ | $3.00\times10^{1}\pm1.7\times10^{0}$ | $2.39\times10^{1}\pm1.3\times10^{0}$ | $1.78\times10^{1}\pm1.1\times10^{0}$ |
| $6.00\times10^{-1}\pm5.0\times10^{-3}$ | $3.49\times10^{1}\pm2.6\times10^{0}$ | $2.48\times10^{1}\pm1.7\times10^{0}$ | $2.14\times10^{1}\pm1.4\times10^{0}$ | $1.63\times10^{1}\pm1.1\times10^{0}$ | $1.29\times10^{1}\pm9.8\times10^{-1}$ |
| $7.00\times10^{-1}\pm6.1\times10^{-3}$ | $2.22\times10^{1}\pm1.7\times10^{0}$ | $1.66\times10^{1}\pm1.4\times10^{0}$ | $1.44\times10^{1}\pm1.2\times10^{0}$ | $1.20\times10^{1}\pm9.4\times10^{-1}$ | $1.03\times10^{1}\pm9.2\times10^{-1}$ |
| $8.00\times10^{-1}\pm7.4\times10^{-3}$ | $1.34\times10^{1}\pm1.4\times10^{0}$ | $1.38\times10^{1}\pm1.4\times10^{0}$ | $1.12\times10^{1}\pm1.0\times10^{0}$ | $1.05\times10^{1}\pm1.1\times10^{0}$ | $8.27\times10^{0}\pm9.6\times10^{-1}$ |
| $9.00\times10^{-1}\pm8.7\times10^{-3}$ | $1.14\times10^{1}\pm1.2\times10^{0}$ | $8.55\times10^{0}\pm1.0\times10^{0}$ | $8.52\times10^{0}\pm9.4\times10^{-1}$ | $9.06\times10^{0}\pm9.5\times10^{-1}$ | $9.52\times10^{0}\pm1.1\times10^{0}$ |

**Table IV (c).** The differential and angle-integrated cross sections of the $^{10}$B$(n, \alpha_1)^{7}$Li reaction in the laboratory reference system.

| $E_n$ (MeV) | $\sigma^1_{E\_bin,\theta}$ (mb/sr) | | | | $\sigma^1_{E\_bin}$ (mb) |
|---|---|---|---|---|---|
| | 133.2° | 143.5° | 153.1° | 160.8° | |
| $1.00\times10^{-6}\pm4.2\times10^{-9}$ | $4.84\times10^{4}\pm2.0\times10^{3}$ | $4.86\times10^{4}\pm2.1\times10^{3}$ | $4.79\times10^{4}\pm2.1\times10^{3}$ | $4.81\times10^{4}\pm2.2\times10^{3}$ | $5.98\times10^{5}\pm2.5\times10^{4}$ |
| $1.26\times10^{-6}\pm5.4\times10^{-9}$ | $4.08\times10^{4}\pm1.3\times10^{3}$ | $4.04\times10^{4}\pm1.3\times10^{3}$ | $4.07\times10^{4}\pm1.3\times10^{3}$ | $4.03\times10^{4}\pm1.3\times10^{3}$ | $5.03\times10^{5}\pm1.3\times10^{4}$ |
| $1.58\times10^{-6}\pm6.8\times10^{-9}$ | $3.21\times10^{4}\pm1.0\times10^{3}$ | $3.14\times10^{4}\pm1.0\times10^{3}$ | $3.16\times10^{4}\pm1.0\times10^{3}$ | $3.16\times10^{4}\pm1.0\times10^{3}$ | $3.92\times10^{5}\pm1.1\times10^{4}$ |
| $2.00\times10^{-6}\pm8.7\times10^{-9}$ | $3.04\times10^{4}\pm1.1\times10^{3}$ | $3.01\times10^{4}\pm1.1\times10^{3}$ | $3.00\times10^{4}\pm1.1\times10^{3}$ | $2.99\times10^{4}\pm1.1\times10^{3}$ | $3.71\times10^{5}\pm1.2\times10^{4}$ |
| $2.51\times10^{-6}\pm1.1\times10^{-8}$ | $2.56\times10^{4}\pm1.3\times10^{3}$ | $2.52\times10^{4}\pm1.3\times10^{3}$ | $2.56\times10^{4}\pm1.3\times10^{3}$ | $2.55\times10^{4}\pm1.3\times10^{3}$ | $3.16\times10^{5}\pm1.4\times10^{4}$ |
| $3.16\times10^{-6}\pm1.4\times10^{-8}$ | $2.93\times10^{4}\pm1.2\times10^{3}$ | $2.88\times10^{4}\pm1.2\times10^{3}$ | $2.84\times10^{4}\pm1.2\times10^{3}$ | $2.80\times10^{4}\pm1.2\times10^{3}$ | $3.56\times10^{5}\pm1.2\times10^{4}$ |
| $3.98\times10^{-6}\pm1.8\times10^{-8}$ | $2.03\times10^{4}\pm2.5\times10^{3}$ | $2.13\times10^{4}\pm2.6\times10^{3}$ | $2.07\times10^{4}\pm2.5\times10^{3}$ | $2.09\times10^{4}\pm2.5\times10^{3}$ | $2.56\times10^{5}\pm3.0\times10^{4}$ |
| $5.01\times10^{-6}\pm2.3\times10^{-8}$ | $1.98\times10^{4}\pm1.5\times10^{3}$ | $2.01\times10^{4}\pm1.5\times10^{3}$ | $2.09\times10^{4}\pm1.6\times10^{3}$ | $1.97\times10^{4}\pm1.5\times10^{3}$ | $2.49\times10^{5}\pm1.8\times10^{4}$ |
| $6.31\times10^{-6}\pm2.9\times10^{-8}$ | $1.98\times10^{4}\pm1.1\times10^{3}$ | $1.98\times10^{4}\pm1.1\times10^{3}$ | $1.96\times10^{4}\pm1.1\times10^{3}$ | $1.98\times10^{4}\pm1.1\times10^{3}$ | $2.46\times10^{5}\pm1.3\times10^{4}$ |
| $7.94\times10^{-6}\pm3.7\times10^{-8}$ | $1.62\times10^{4}\pm3.4\times10^{3}$ | $1.66\times10^{4}\pm3.5\times10^{3}$ | $1.65\times10^{4}\pm3.5\times10^{3}$ | $1.67\times10^{4}\pm3.5\times10^{3}$ | $2.06\times10^{5}\pm4.3\times10^{4}$ |
| $1.00\times10^{-5}\pm4.7\times10^{-8}$ | $1.51\times10^{4}\pm6.9\times10^{2}$ | $1.58\times10^{4}\pm7.2\times10^{2}$ | $1.64\times10^{4}\pm7.3\times10^{2}$ | $1.62\times10^{4}\pm7.4\times10^{2}$ | $1.97\times10^{5}\pm7.2\times10^{3}$ |
| $1.26\times10^{-5}\pm6.4\times10^{-8}$ | $1.40\times10^{4}\pm1.3\times10^{3}$ | $1.40\times10^{4}\pm1.3\times10^{3}$ | $1.39\times10^{4}\pm1.2\times10^{3}$ | $1.34\times10^{4}\pm1.2\times10^{3}$ | $1.73\times10^{5}\pm1.5\times10^{4}$ |
| $1.58\times10^{-5}\pm8.6\times10^{-8}$ | $1.08\times10^{4}\pm1.6\times10^{3}$ | $1.10\times10^{4}\pm1.7\times10^{3}$ | $1.09\times10^{4}\pm1.6\times10^{3}$ | $1.05\times10^{4}\pm1.6\times10^{3}$ | $1.36\times10^{5}\pm2.0\times10^{4}$ |
| $2.00\times10^{-5}\pm1.2\times10^{-7}$ | $8.73\times10^{3}\pm1.6\times10^{3}$ | $9.04\times10^{3}\pm1.6\times10^{3}$ | $8.84\times10^{3}\pm1.6\times10^{3}$ | $8.72\times10^{3}\pm1.6\times10^{3}$ | $1.11\times10^{5}\pm2.0\times10^{4}$ |
| $2.51\times10^{-5}\pm1.6\times10^{-7}$ | $1.01\times10^{4}\pm5.7\times10^{2}$ | $1.00\times10^{4}\pm5.7\times10^{2}$ | $1.01\times10^{4}\pm5.8\times10^{2}$ | $1.00\times10^{4}\pm5.9\times10^{2}$ | $1.24\times10^{5}\pm5.8\times10^{3}$ |
| $3.16\times10^{-5}\pm2.1\times10^{-7}$ | $7.17\times10^{3}\pm1.6\times10^{3}$ | $7.15\times10^{3}\pm1.6\times10^{3}$ | $6.96\times10^{3}\pm1.5\times10^{3}$ | $6.95\times10^{3}\pm1.5\times10^{3}$ | $8.76\times10^{4}\pm1.9\times10^{4}$ |
| $3.98\times10^{-5}\pm2.7\times10^{-7}$ | $7.36\times10^{3}\pm7.0\times10^{2}$ | $7.33\times10^{3}\pm7.0\times10^{2}$ | $7.14\times10^{3}\pm6.9\times10^{2}$ | $6.87\times10^{3}\pm6.7\times10^{2}$ | $8.97\times10^{4}\pm8.1\times10^{3}$ |
| $5.01\times10^{-5}\pm3.6\times10^{-7}$ | $6.95\times10^{3}\pm5.2\times10^{2}$ | $6.79\times10^{3}\pm5.1\times10^{2}$ | $6.84\times10^{3}\pm5.2\times10^{2}$ | $6.61\times10^{3}\pm5.1\times10^{2}$ | $8.46\times10^{4}\pm5.9\times10^{3}$ |
| $6.31\times10^{-5}\pm4.8\times10^{-7}$ | $5.00\times10^{3}\pm5.1\times10^{2}$ | $5.06\times10^{3}\pm5.2\times10^{2}$ | $5.21\times10^{3}\pm5.3\times10^{2}$ | $5.16\times10^{3}\pm5.3\times10^{2}$ | $6.39\times10^{4}\pm6.3\times10^{3}$ |
| $7.94\times10^{-5}\pm6.4\times10^{-7}$ | $5.09\times10^{3}\pm5.0\times10^{2}$ | $4.95\times10^{3}\pm4.9\times10^{2}$ | $5.06\times10^{3}\pm5.0\times10^{2}$ | $4.79\times10^{3}\pm4.8\times10^{2}$ | $6.17\times10^{4}\pm5.8\times10^{3}$ |
| $1.00\times10^{-4}\pm8.4\times10^{-7}$ | $4.30\times10^{3}\pm3.5\times10^{2}$ | $4.55\times10^{3}\pm3.7\times10^{2}$ | $4.31\times10^{3}\pm3.6\times10^{2}$ | $4.02\times10^{3}\pm3.5\times10^{2}$ | $5.50\times10^{4}\pm4.1\times10^{3}$ |
| $1.26\times10^{-4}\pm1.0\times10^{-6}$ | $4.01\times10^{3}\pm4.4\times10^{2}$ | $3.97\times10^{3}\pm4.4\times10^{2}$ | $3.85\times10^{3}\pm4.3\times10^{2}$ | $3.71\times10^{3}\pm4.3\times10^{2}$ | $5.01\times10^{4}\pm5.1\times10^{3}$ |
| $1.58\times10^{-4}\pm1.3\times10^{-6}$ | $3.52\times10^{3}\pm2.9\times10^{2}$ | $3.59\times10^{3}\pm3.0\times10^{2}$ | $3.50\times10^{3}\pm2.9\times10^{2}$ | $3.54\times10^{3}\pm3.1\times10^{2}$ | $4.35\times10^{4}\pm3.3\times10^{3}$ |
| $2.00\times10^{-4}\pm1.6\times10^{-6}$ | $3.13\times10^{3}\pm2.8\times10^{2}$ | $3.05\times10^{3}\pm2.8\times10^{2}$ | $2.90\times10^{3}\pm2.7\times10^{2}$ | $3.01\times10^{3}\pm2.9\times10^{2}$ | $3.77\times10^{4}\pm3.0\times10^{3}$ |
| $2.51\times10^{-4}\pm2.0\times10^{-6}$ | $2.81\times10^{3}\pm2.6\times10^{2}$ | $2.71\times10^{3}\pm2.5\times10^{2}$ | $2.79\times10^{3}\pm2.7\times10^{2}$ | $2.64\times10^{3}\pm2.6\times10^{2}$ | $3.45\times10^{4}\pm2.9\times10^{3}$ |
| $3.16\times10^{-4}\pm2.5\times10^{-6}$ | $2.37\times10^{3}\pm1.8\times10^{2}$ | $2.46\times10^{3}\pm1.9\times10^{2}$ | $2.46\times10^{3}\pm2.0\times10^{2}$ | $2.15\times10^{3}\pm2.0\times10^{2}$ | $2.97\times10^{4}\pm1.9\times10^{3}$ |
| $3.98\times10^{-4}\pm3.1\times10^{-6}$ | $2.32\times10^{3}\pm1.8\times10^{2}$ | $2.23\times10^{3}\pm1.7\times10^{2}$ | $2.19\times10^{3}\pm1.8\times10^{2}$ | $2.23\times10^{3}\pm2.0\times10^{2}$ | $2.84\times10^{4}\pm1.8\times10^{3}$ |
| $5.01\times10^{-4}\pm3.9\times10^{-6}$ | $2.08\times10^{3}\pm1.3\times10^{2}$ | $2.07\times10^{3}\pm1.3\times10^{2}$ | $2.09\times10^{3}\pm1.3\times10^{2}$ | $2.05\times10^{3}\pm1.4\times10^{2}$ | $2.55\times10^{4}\pm1.2\times10^{3}$ |
| $6.31\times10^{-4}\pm4.8\times10^{-6}$ | $1.91\times10^{3}\pm1.0\times10^{2}$ | $1.82\times10^{3}\pm9.9\times10^{1}$ | $1.73\times10^{3}\pm1.0\times10^{2}$ | $1.67\times10^{3}\pm1.1\times10^{2}$ | $2.19\times10^{4}\pm8.3\times10^{2}$ |
| $7.94\times10^{-4}\pm6.1\times10^{-6}$ | $1.60\times10^{3}\pm1.0\times10^{2}$ | $1.56\times10^{3}\pm1.0\times10^{2}$ | $1.55\times10^{3}\pm1.0\times10^{2}$ | $1.39\times10^{3}\pm1.1\times10^{2}$ | $1.94\times10^{4}\pm8.7\times10^{2}$ |
| $1.00\times10^{-3}\pm7.7\times10^{-6}$ | $1.37\times10^{3}\pm7.6\times10^{1}$ | $1.31\times10^{3}\pm7.6\times10^{1}$ | $1.26\times10^{3}\pm7.7\times10^{1}$ | $1.31\times10^{3}\pm8.8\times10^{1}$ | $1.65\times10^{4}\pm6.1\times10^{2}$ |
| $1.26\times10^{-3}\pm9.8\times10^{-6}$ | $1.16\times10^{3}\pm6.0\times10^{1}$ | $1.19\times10^{3}\pm6.2\times10^{1}$ | $1.16\times10^{3}\pm6.4\times10^{1}$ | $1.14\times10^{3}\pm7.5\times10^{1}$ | $1.47\times10^{4}\pm3.9\times10^{2}$ |
| $1.58\times10^{-3}\pm1.3\times10^{-5}$ | $1.05\times10^{3}\pm5.4\times10^{1}$ | $1.10\times10^{3}\pm5.7\times10^{1}$ | $1.02\times10^{3}\pm5.7\times10^{1}$ | $9.48\times10^{2}\pm6.5\times10^{1}$ | $1.33\times10^{4}\pm3.6\times10^{2}$ |
| $2.00\times10^{-3}\pm1.6\times10^{-5}$ | $9.48\times10^{2}\pm5.0\times10^{1}$ | $9.40\times10^{2}\pm5.1\times10^{1}$ | $9.69\times10^{2}\pm5.2\times10^{1}$ | $9.50\times10^{2}\pm6.1\times10^{1}$ | $1.19\times10^{4}\pm2.9\times10^{2}$ |
| $2.51\times10^{-3}\pm2.1\times10^{-5}$ | $8.72\times10^{2}\pm6.4\times10^{1}$ | $7.91\times10^{2}\pm6.1\times10^{1}$ | $7.97\times10^{2}\pm6.5\times10^{1}$ | $8.51\times10^{2}\pm7.7\times10^{1}$ | $1.07\times10^{4}\pm5.8\times10^{2}$ |
| $3.16\times10^{-3}\pm2.7\times10^{-5}$ | $8.15\times10^{2}\pm4.3\times10^{1}$ | $8.05\times10^{2}\pm4.4\times10^{1}$ | $7.12\times10^{2}\pm4.5\times10^{1}$ | $7.70\times10^{2}\pm5.4\times10^{1}$ | $9.68\times10^{3}\pm2.8\times10^{2}$ |
| $3.98\times10^{-3}\pm3.6\times10^{-5}$ | $7.02\times10^{2}\pm3.9\times10^{1}$ | $6.56\times10^{2}\pm3.8\times10^{1}$ | $7.07\times10^{2}\pm4.2\times10^{1}$ | $6.97\times10^{2}\pm4.7\times10^{1}$ | $8.55\times10^{3}\pm2.3\times10^{2}$ |
| $5.01\times10^{-3}\pm4.8\times10^{-5}$ | $5.76\times10^{2}\pm3.2\times10^{1}$ | $5.76\times10^{2}\pm3.3\times10^{1}$ | $5.64\times10^{2}\pm3.5\times10^{1}$ | $5.33\times10^{2}\pm4.0\times10^{1}$ | $7.17\times10^{3}\pm2.1\times10^{2}$ |
| $6.31\times10^{-3}\pm6.4\times10^{-5}$ | $5.35\times10^{2}\pm3.1\times10^{1}$ | $5.52\times10^{2}\pm3.3\times10^{1}$ | $4.45\times10^{2}\pm3.2\times10^{1}$ | $5.04\times10^{2}\pm3.9\times10^{1}$ | $6.70\times10^{3}\pm2.0\times10^{2}$ |
| $7.94\times10^{-3}\pm8.6\times10^{-5}$ | $4.92\times10^{2}\pm3.1\times10^{1}$ | $4.62\times10^{2}\pm3.1\times10^{1}$ | $4.62\times10^{2}\pm3.3\times10^{1}$ | $4.38\times10^{2}\pm4.1\times10^{1}$ | $6.00\times10^{3}\pm2.1\times10^{2}$ |
| $1.00\times10^{-2}\pm1.2\times10^{-4}$ | $4.24\times10^{2}\pm2.9\times10^{1}$ | $4.32\times10^{2}\pm3.0\times10^{1}$ | $4.49\times10^{2}\pm3.2\times10^{1}$ | $4.20\times10^{2}\pm3.6\times10^{1}$ | $5.62\times10^{3}\pm2.5\times10^{2}$ |
| $1.26\times10^{-2}\pm1.6\times10^{-4}$ | $3.88\times10^{2}\pm2.2\times10^{1}$ | $3.83\times10^{2}\pm2.3\times10^{1}$ | $4.08\times10^{2}\pm2.5\times10^{1}$ | $3.80\times10^{2}\pm2.8\times10^{1}$ | $4.97\times10^{3}\pm1.3\times10^{2}$ |
| $1.58\times10^{-2}\pm2.2\times10^{-4}$ | $3.52\times10^{2}\pm2.0\times10^{1}$ | $3.52\times10^{2}\pm2.1\times10^{1}$ | $3.41\times10^{2}\pm2.3\times10^{1}$ | $3.10\times10^{2}\pm2.5\times10^{1}$ | $4.39\times10^{3}\pm1.2\times10^{2}$ |
| $2.00\times10^{-2}\pm1.2\times10^{-4}$ | $3.27\times10^{2}\pm2.0\times10^{1}$ | $3.13\times10^{2}\pm2.1\times10^{1}$ | $3.14\times10^{2}\pm2.2\times10^{1}$ | $2.80\times10^{2}\pm2.5\times10^{1}$ | $4.04\times10^{3}\pm1.3\times10^{2}$ |
| $2.51\times10^{-2}\pm1.5\times10^{-4}$ | $2.77\times10^{2}\pm1.6\times10^{1}$ | $2.65\times10^{2}\pm1.6\times10^{1}$ | $2.78\times10^{2}\pm1.7\times10^{1}$ | $2.35\times10^{2}\pm2.0\times10^{1}$ | $3.55\times10^{3}\pm1.1\times10^{2}$ |
| $3.16\times10^{-2}\pm1.9\times10^{-4}$ | $2.30\times10^{2}\pm1.8\times10^{1}$ | $2.12\times10^{2}\pm1.7\times10^{1}$ | $1.96\times10^{2}\pm1.9\times10^{1}$ | $1.91\times10^{2}\pm2.3\times10^{1}$ | $2.89\times10^{3}\pm9.2\times10^{1}$ |
| $3.98\times10^{-2}\pm2.4\times10^{-4}$ | $2.09\times10^{2}\pm1.4\times10^{1}$ | $2.02\times10^{2}\pm1.5\times10^{1}$ | $1.88\times10^{2}\pm1.5\times10^{1}$ | $1.77\times10^{2}\pm1.8\times10^{1}$ | $2.73\times10^{3}\pm9.1\times10^{1}$ |
| $5.01\times10^{-2}\pm3.0\times10^{-4}$ | $1.80\times10^{2}\pm1.1\times10^{1}$ | $1.86\times10^{2}\pm1.2\times10^{1}$ | $1.78\times10^{2}\pm1.2\times10^{1}$ | $1.61\times10^{2}\pm1.3\times10^{1}$ | $2.44\times10^{3}\pm7.5\times10^{1}$ |
| $6.31\times10^{-2}\pm3.8\times10^{-4}$ | $1.62\times10^{2}\pm8.9\times10^{0}$ | $1.60\times10^{2}\pm8.9\times10^{0}$ | $1.52\times10^{2}\pm9.2\times10^{0}$ | $1.55\times10^{2}\pm1.0\times10^{1}$ | $2.19\times10^{3}\pm5.3\times10^{1}$ |
| $7.94\times10^{-2}\pm4.8\times10^{-4}$ | $1.42\times10^{2}\pm8.6\times10^{0}$ | $1.46\times10^{2}\pm8.6\times10^{0}$ | $1.39\times10^{2}\pm8.9\times10^{0}$ | $1.35\times10^{2}\pm1.0\times10^{1}$ | $1.98\times10^{3}\pm4.8\times10^{1}$ |
| $1.00\times10^{-1}\pm6.1\times10^{-4}$ | $1.31\times10^{2}\pm4.9\times10^{0}$ | $1.29\times10^{2}\pm5.0\times10^{0}$ | $1.25\times10^{2}\pm5.0\times10^{0}$ | $1.23\times10^{2}\pm5.3\times10^{0}$ | $1.82\times10^{3}\pm4.2\times10^{1}$ |
| $2.00\times10^{-1}\pm1.3\times10^{-3}$ | $8.59\times10^{1}\pm3.2\times10^{0}$ | $8.36\times10^{1}\pm3.1\times10^{0}$ | $7.93\times10^{1}\pm3.1\times10^{0}$ | $7.63\times10^{1}\pm4.0\times10^{0}$ | $1.25\times10^{3}\pm3.1\times10^{1}$ |
| $3.00\times10^{-1}\pm2.0\times10^{-3}$ | $5.42\times10^{1}\pm2.2\times10^{0}$ | $5.50\times10^{1}\pm2.3\times10^{0}$ | $5.30\times10^{1}\pm2.3\times10^{0}$ | $4.80\times10^{1}\pm2.9\times10^{0}$ | $8.94\times10^{2}\pm2.0\times10^{1}$ |
| $4.00\times10^{-1}\pm2.9\times10^{-3}$ | $2.93\times10^{1}\pm1.5\times10^{0}$ | $2.60\times10^{1}\pm1.5\times10^{0}$ | $2.52\times10^{1}\pm1.5\times10^{0}$ | $2.37\times10^{1}\pm1.8\times10^{0}$ | $6.97\times10^{2}\pm1.8\times10^{1}$ |
| $5.00\times10^{-1}\pm3.9\times10^{-3}$ | $1.38\times10^{1}\pm9.2\times10^{-1}$ | $1.07\times10^{1}\pm7.8\times10^{-1}$ | $8.55\times10^{0}\pm7.7\times10^{-1}$ | $7.01\times10^{0}\pm9.0\times10^{-1}$ | $5.83\times10^{2}\pm2.1\times10^{1}$ |
| $6.00\times10^{-1}\pm5.0\times10^{-3}$ | $1.05\times10^{1}\pm8.5\times10^{-1}$ | $9.82\times10^{0}\pm8.4\times10^{-1}$ | $6.29\times10^{0}\pm7.3\times10^{-1}$ | $5.03\times10^{0}\pm7.8\times10^{-1}$ | $3.90\times10^{2}\pm1.7\times10^{1}$ |
| $7.00\times10^{-1}\pm6.1\times10^{-3}$ | $9.68\times10^{0}\pm9.0\times10^{-1}$ | $8.56\times10^{0}\pm8.7\times10^{-1}$ | $1.00\times10^{1}\pm1.1\times10^{0}$ | $5.33\times10^{0}\pm9.5\times10^{-1}$ | $2.48\times10^{2}\pm1.0\times10^{1}$ |
| $8.00\times10^{-1}\pm7.4\times10^{-3}$ | $8.91\times10^{0}\pm9.1\times10^{-1}$ | $8.29\times10^{0}\pm9.4\times10^{-1}$ | $5.91\times10^{0}\pm9.4\times10^{-1}$ | $2.93\times10^{0}\pm8.1\times10^{-1}$ | $1.72\times10^{2}\pm9.0\times10^{0}$ |
| $9.00\times10^{-1}\pm8.7\times10^{-3}$ | $6.85\times10^{0}\pm8.6\times10^{-1}$ | $6.97\times10^{0}\pm9.4\times10^{-1}$ | $4.77\times10^{0}\pm9.7\times10^{-1}$ | $5.88\times10^{0}\pm1.4\times10^{0}$ | $1.44\times10^{2}\pm8.2\times10^{0}$ |